\documentclass[twocolumn]{aastex631}
\usepackage{multirow}
\usepackage[toc,page]{appendix}

\newcommand{\kms}{\hbox{km s$^{-1}$}}

\usepackage{CJK}

\shorttitle{The accretion bursts of HOPS 373}
\shortauthors{Yoon et al.}

\begin{document}
\begin{CJK*}{UTF8}{gbsn}

\title{Dissecting the different components of the modest accretion bursts \\
of the very young protostar HOPS 373}

\author{Sung-Yong Yoon}
\affiliation{The School of Space Research, Kyung Hee University, 1732 Deogyeong-daero, Giheung-gu, Yongin-si, Gyeonggi-do, Republic of Korea}
\affiliation{Korea Astronomy and Space Science Institute, 776, Daedeok-daero, Yuseong-gu, Daejeon, 34055, Republic of Korea}

\correspondingauthor{Gregory J. Herczeg, Jeong-Eun Lee}
\email{gherczeg1@gmail.com, jeongeun.lee@khu.ac.kr}

\author{Gregory J. Herczeg (沈雷歌)}
\affiliation{Kavli Institute for Astronomy and Astrophysics, Peking University, Yiheyuan 5, Haidian Qu, 100871 Beijing, China}
\affiliation{Department of Astronomy, Peking University, Yiheyuan 5, Haidian Qu, 100871 Beijing, China}

\author{Jeong-Eun Lee}
\affiliation{The School of Space Research, Kyung Hee University, 1732 Deogyeong-daero, Giheung-gu, Yongin-si, Gyeonggi-do, Republic of Korea}

\author{Ho-Gyu Lee}
\affiliation{Korea Astronomy and Space Science Institute, 776, Daedeok-daero, Yuseong-gu, Daejeon, 34055, Republic of Korea}

\author{Doug Johnstone}
\affiliation{NRC Herzberg Astronomy and Astrophysics, 5071 West Saanich Road, Victoria, BC, V9E 2E7, Canada}
\affiliation{Department of Physics and Astronomy, University of Victoria, 3800 Finnerty Road, Elliot Building, Victoria, BC, V8P 5C2, Canada}

\author{Watson Varricatt}
\affiliation{UKIRT Observatory, University of Hawaii, Institute for Astronomy, 640 N. Aohoku Place, Hilo, HI 96720, USA}

\author{John J. Tobin}
\affiliation{National Radio Astronomy Observatory, 520 Edgemont Rd., Charlottesville, VA 22903}

\author{Carlos Contreras Pe{\~n}a}
\affiliation{Physics and Astronomy, University of Exeter, Stocker Road, Exeter EX4 4QL, UK}
\affiliation{Centre for Astrophysics Research, University of Hertfordshire, Hatfield AL10 9AB, UK}

\author{Steve Mairs}
\affiliation{East Asian Observatory, 660 N. A'ohoku Place, Hilo, HI 96720, USA}

\author{Klaus Hodapp}
\affiliation{University of Hawaii, Institute for Astronomy, 640 N. Aohoku Place, Hilo, HI 96720, USA}

\author{P. Manoj}
\affiliation{Tata Institute of Fundamental Research, Homi Bhabha Road, Mumbai 400 005, India}

\author{Mayra Osorio}
\affiliation{Instituto de Astrofisica de Andalucia, CSIC, Glorieta de la Astronomia S/N, E-18008, Grenada, Spain}

\author{S. Thomas Megeath}
\affiliation{Ritter Astrophysical Research Center, Department of Physics and Astronomy, University of Toledo, W. Bancroft Street, Toledo, OH 43606 USA}

\author{the JCMT Transient Team}

\begin{abstract}
Observed changes in protostellar brightness can be complicated to interpret.  In our JCMT~Transient monitoring survey, we discovered that a young binary protostar, HOPS 373, is undergoing a modest $30\%$ brightness increase at 850 $\mu$m, caused by a factor of 1.8--3.3 enhancement in the accretion rate.  The initial burst occurred over a few months, with a sharp rise and then shallower decay.  A second rise occurred soon after the decay, and the source is still bright one year later.
The mid-IR emission, the small-scale CO outflow mapped with ALMA, and the location of variable maser emission indicate that the variability is associated with the SW component.
The near-infrared and NEOWISE $W1$ and $W2$ emission is located along the blueshifted CO outflow, spatially offset by $\sim3$ to $4^{\prime\prime}$ from the SW component.  The $K$-band emission imaged by UKIRT shows a compact H$_2$ emission source at
the edge of the outflow, with a tail tracing the outflow back to the
source.  The $W1$ emission, likely dominated by scattered light, brightens by 0.7 mag, consistent with expectations based on the sub-mm
lightcurve. The signal of continuum variability in $K$-band and $W2$ is
masked by stable H$_2$ emission, as seen in our Gemini/GNIRS spectrum, and
perhaps by CO emission.  
These differences in emission sources complicate infrared searches for variability of the youngest protostars.
\end{abstract}


\section{Introduction}

Accretion outbursts are thought to play an important role in the growth of the protostar and evolution of its disk. Historically, most accretion bursts were discovered with optical variability and are therefore identified during the later stages of protostellar evolution, after the star has already grown to near its final mass and has shed its envelope (see reviews by \citealt{hartmann96} and \citealt{audard14} and subsequent discoveries from Gaia by, e.g., \citealt{hillenbrand18} and \citealt{szegedi20}). However, accretion outbursts may play an even more important role at the youngest stages of stellar growth, as indicated by indirect probes such as outflow knots \citep[e.g.][]{reipurth89,plunkett15}, envelope chemistry \citep[e.g.][]{lee07,jorgensen15,hsieh19}, and by models of disk instabilities \citep[e.g.][]{bae14}.  A few accretion outbursts have been detected toward very young low-mass protostars \citep[e.g.][]{kospal07,safron15,kospal20}.

Over the past few years, several surveys at longer wavelengths have been designed to statistically evaluate accretion variability at earlier stages of protostellar evolution \citep[e.g.][]{scholz13,rebull14,antoniucci14,lucas17,Johnstone2018,fischer19,lee21, park21,zakri22}, when the star is still in its main growth phase and the disk is accreting from the envelope.  For protostars, sub-mm emission is produced by warm dust in the envelope (and disk), serving as a bolometer that can be converted into a luminosity \citep[e.g.][]{johnstone13,macfarlane19a,macfarlane19b,baek20}. 
However, the infrared emission from deeply embedded objects may be much more complicated, since this emission may escape only through optically-thin outflow cavities \citep[e.g.][]{lee10,herczeg12,tobin20hops370}.  In three previous cases, varying infrared emission has been resolved into scattered light nebulae surrounding embedded protostars \citep{muzerolle13,balog14,caratti15,cook19}.

\begin{figure*}[!t]
    \includegraphics[width=0.33\textwidth]{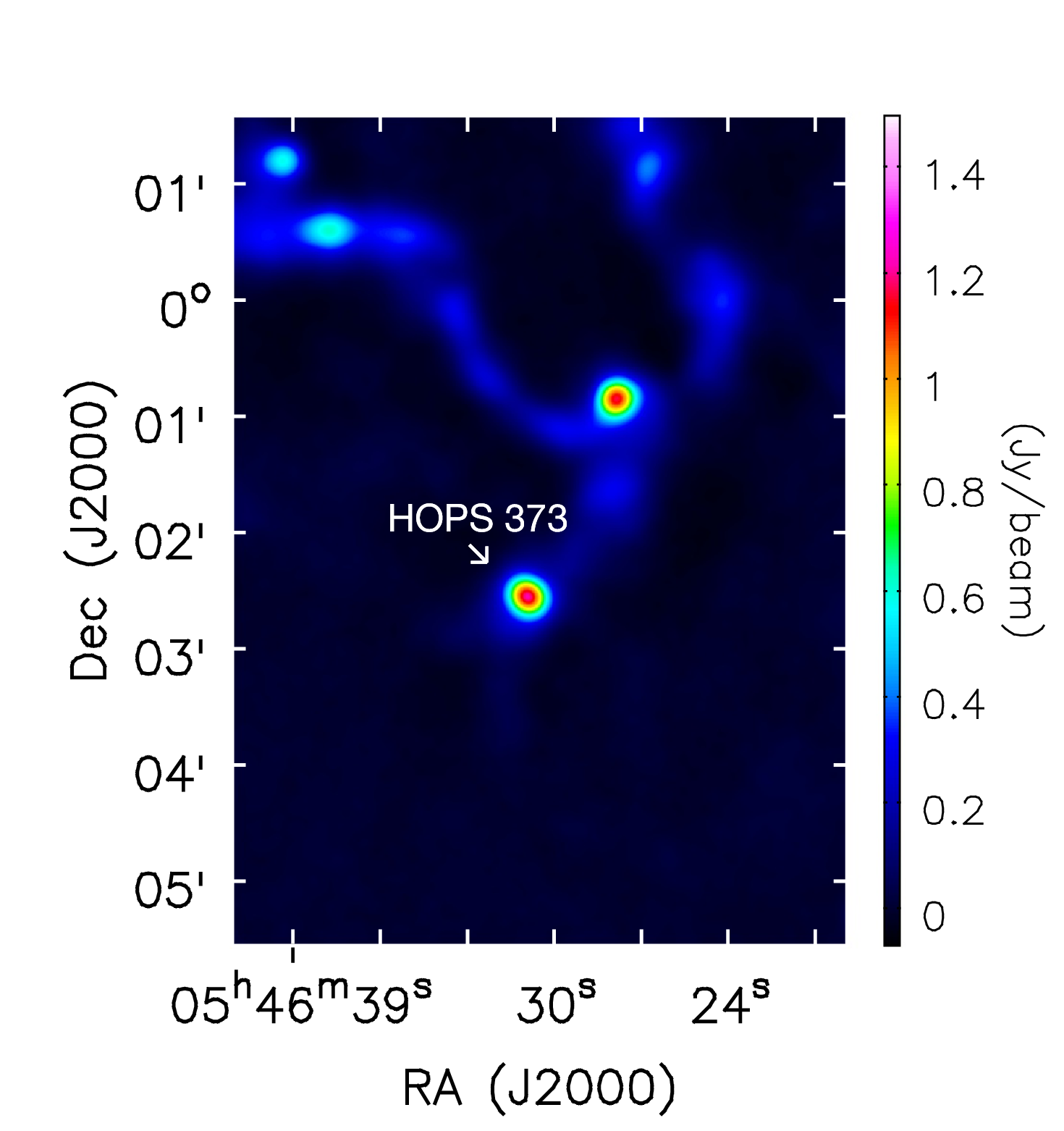}
    \includegraphics[width=0.33\textwidth]{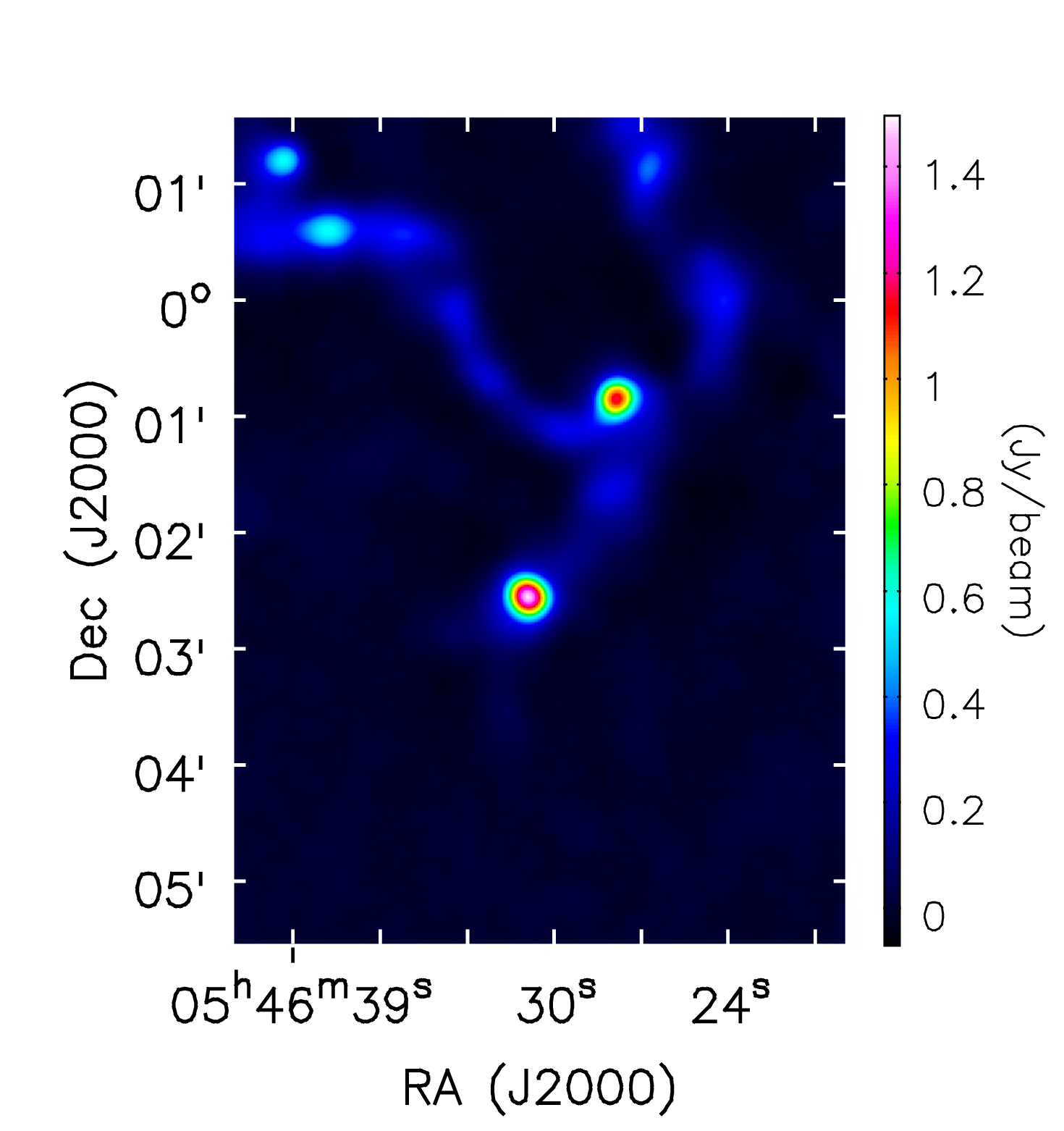}
    \includegraphics[width=0.33\textwidth]{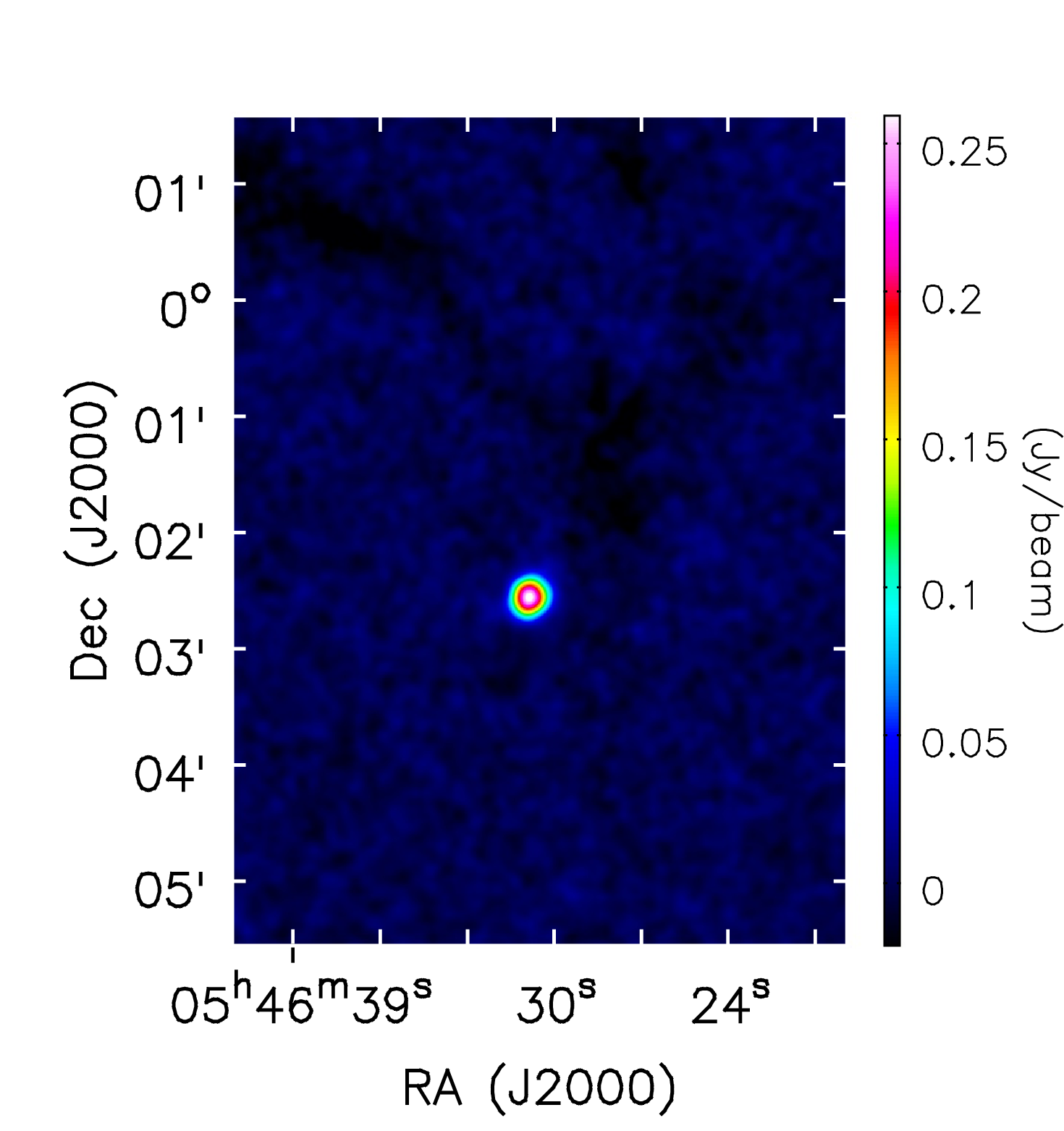}
    \caption{SCUBA-2 850 $\mu$m continuum images of HOPS 373 in the quiescent phase (left, coadded from 14 epochs in 2017), in the burst phase (center, coadded from 12 epochs in 2020--2021), and the residual (burst minus quiescent) image (right).
    \label{fig:scuba2}}
\end{figure*}

In this paper, we dissect the ongoing, modest accretion outburst of HOPS 373 as seen in near-IR, mid-IR, sub-mm, mm, and cm observations.  The central protostar has characteristics that are typical of a very young protostar driving a powerful outflow.  The protostellar envelope of HOPS 373 was initially identified in sub-mm mapping of NGC 2068 \citep[distance of 428 pc,][]{kounkel18} in the Orion B cloud complex \citep{phillips01,johnstone01,motte01}, following the discovery of a large CO outflow \citep{gibb00} and a nearby maser \citep{haschick83}.   From fits to the near-IR through mm spectral energy distribution, the protostar has a bolometric luminosity of $5.3-6.1$~L$_{\odot}$ and a bolometric temperature of $37$ K \citep{kang15,Furlan2016}.  The SED, with a total-to-submm luminosity ratio of 34, is red enough to be a PACS Bright Red Source \citep{Stutz2013} and consistent with an early stage of a Class 0 protostar.  HOPS 373 and its associated outflows are not detected in X-rays \citep{getman17}.  The central object was resolved with high-resolution mm interferometry into two distinct mm peaks separated by $3\farcs6$, or 1500 AU \citep{tobin15,tobin20}.  The source drives a powerful outflow.  Sub-mm CO emission from the large-scale outflow extends over $1$ arcmin to both the blue (southeast) and red (northwest) directions with a dynamical time of $10^4$ yr \citep{mitchell01,nagy20}. Far-IR spectra reveal excited molecular emission that likely traces the walls of an outflow cavity \citep{tobin16}, with far-IR CO emission that is the brightest of all PACS Bright Red Sources and amongst the brightest of all low-mass protostars  \citep{manoj16,karska18}.

The NGC 2068 star-forming region, including HOPS 373,
has been monitored in the sub-mm by the JCMT Transient Survey  since December 2015 \citep{herczeg2017}. Beginning in 2016, HOPS 373 had gotten fainter by 10\% \citep{Johnstone2018} and stayed steady in this lower luminosity state until a brief increase of 25\% in late 2019 \citep[]{lee21}, a pattern also seen at $4.5$~$\mu$m with NEOWISE \citep[]{contreras20,park21}.  We report that HOPS 373 has since returned to the bright state in the sub-mm, and analyze in detail the location of emission across the spectrum.  We combine the sub-mm lightcurve with near-IR spectroscopy and imaging to dissect how the source is seen across a range of diagnostics.  In Section~\ref{sec:obs}, we describe the array of observations that are used in this paper.  In Section~\ref{sec:lightcurves}, we describe the sub-mm and NEOWISE lightcurves and provide a physical interpretation for the sub-mm variability.  In Section~\ref{sec:morphology}, we describe the morphology of the emission sources.  In Section~\ref{sec:molecular}, we present Gemini/GNIRS spectra to demonstrate the dominance of H$_2$ emission in the $K$-band imaging.  In Section~\ref{sec:dissection}, we interpret these results and discuss their importance in on-going and future searches for variable protostars.

\section{Observations}
\label{sec:obs}

\subsection{Sub-mm Monitoring at 450 and 850 $\mu$m}

The JCMT-Transient survey \citep{herczeg2017} has been using SCUBA-2 \citep{Holland2013} on the James Clerk Maxwell Telescope (JCMT) to monitor every month emission at 450 and 850 $\mu$m from eight regions, including NGC 2068, beginning in December 2015.  We also include an earlier data point from the JCMT-Gould Belt Survey \citep{kirk2016} obtained from 2014 November 16-22, reported and re-analyzed by \citet{Mairs2017GBS}.  In November 2019, we discovered that HOPS 373 had brightened and increased our observational cadence of NGC 2068 to once every two weeks (Figure~\ref{fig:scuba2}).

\begin{table}[!b]
\begin{center}
\caption{Selected sub-mm peak brightness$^a$}
\begin{tabular}{ccc|ccc}
  \multirow{2}{*}{MJD} & 450$^b$ & 850$^b$ &  \multirow{2}{*}{MJD} & 450$^b$ & 850$^b$ \\
  & \multicolumn{2}{c|}{[Jy/beam]} & & \multicolumn{2}{c}{[Jy/beam]} \\
\hline
 57403.348 &  5.07 &  1.31 &  58850.285 &  5.32 &  1.32 \\
 57424.223 &  4.91 &  1.32 &  58872.391 &  5.14 &  1.27 \\
 57505.215 &  4.54 &  1.23 &  58915.297 &  4.99 &  1.26 \\
 57997.668 &  4.75 &  1.22 &  59072.633 &     &  1.42 \\
 58025.621 &  4.26 &  1.19 &  59087.645 &  6.21 &  1.47 \\
 58095.422 &  4.89 &  1.18 &  59106.723 &  6.09 &  1.48 \\
 58133.410 &  4.35 &  1.20 &  59155.480 &  6.07 &  1.46 \\
 58486.270 &  4.38 &  1.19 &  59180.371 &  6.22 &  1.43 \\
 58519.336 &  4.73 &  1.23 &  59256.199 &  5.96 &  1.44 \\
 58580.258 &     &  1.26 &  59275.273 &  6.08 &  1.46 \\
 58715.621 &     &  1.35 &  59300.223 &  6.47 &  1.52 \\
 58752.539 &     &  1.48 &  59321.219 &  6.69 &  1.57 \\
 58774.727 &  6.02 &  1.41 &  59454.633 &  6.31 &  1.45 \\
 58788.465 &  5.75 &  1.40 &  59485.562 &  5.81 &  1.48 \\
 58836.355 &  5.58 &  1.34 &  59541.398 &     &  1.41 \\
\hline
\multicolumn{6}{l}{$^a$Full table available online, selected points shown here}\\
\multicolumn{6}{l}{$^b$Error of 5\% at 450 $\mu$m, 2\% at 850 $\mu$m}\\
\end{tabular}
\label{tab:scuba2phot}
\end{center}
\end{table}

\begin{table}[!b]
\caption{Mid-IR Photometry from NEOWISE}
\begin{center}
\begin{tabular}{ccccc}
 \multirow{2}{*}{MJD} & $W1$ & Err & $W2$ & Err\\
 & [mag] & [mag] & [mag] & [mag] \\
\hline
55266.88$^a$ &  15.32 & 0.27 & 10.930 & 0.021\\
55458.14$^a$ & 15.49 & 0.38 & 10.933 & 0.026\\
56731.57 & 15.01 &  0.24 &10.765 &0.031 \\
56923.07 & 15.32 &  0.50 &10.802 & 0.040\\
57090.68 & 14.99 &  0.24  &10.741 & 0.029\\
57284.77 & 15.13 &  0.42 &10.745 & 0.036\\
57285.24 & 15.12 &  0.25 &10.772 & 0.024\\
57449.77 & 15.08 &  0.26 &10.768 & 0.038\\
57649.51 & 15.29 &  0.30 &10.867 & 0.037\\
57814.28 & 15.49 &  0.40 &10.942 & 0.036\\
58016.13 & 15.52 &  0.42 &10.954 & 0.040\\
58171.58 & 15.38  & 0.36 &10.983 & 0.035\\
58380.29 & 15.44 &  0.35 &10.966 & 0.032\\
58537.98 & 15.38 &  0.33 &10.905 & 0.031\\
58744.87 & 14.49  & 0.15 &10.602 & 0.033\\
58902.25 & 15.09 & 0.24 & 10.835 & 0.035 \\
59111.60 & 14.31 & 0.14 & 10.515 & 0.036 \\
\hline
\multicolumn{5}{l}{$^a$ALLWISE photometry from \citet{cutri14}}\\
\end{tabular}
\label{tab:wisephot}
\end{center}
\end{table}

\begin{table}[!b]
\begin{center}
\caption{Selected $K$-band Photometry$^a$}
\begin{tabular}{lccccc}
\multirow{2}{*}{MJD} & Brt & Err & Tail & Err & $t_{exp}$ \\
 & [mag] & [mag] & [mag] & [mag] & [s] \\
\hline
51093.310  & \multicolumn{2}{c}{$15.22\pm0.16$} & \multicolumn{3}{c}{summed 2MASS $K_S^b$}\\
55164.335   & \multicolumn{2}{c}{$15.037\pm0.029$} & \multicolumn{3}{c}{summed VISTA $K_S^c$}\\
   55984.308 &   15.74 &  0.02 &   16.26 &  0.03 &  72 \\
   56262.376 &   15.67 &  0.03 &   16.09 &  0.04 &  72 \\
   57990.635 &   15.67 &  0.02 &   16.32 &  0.03 &  72 \\
   58566.254 &   15.70 &  0.03 &   16.28 &  0.05 &  72 \\
   58812.650 &   15.67 &  0.02 &   16.25 &  0.03 &  72 \\
   58814.551 &   15.67 &  0.02 &   16.25 &  0.03 &  72 \\
   58818.513 &   15.71 &  0.03 &   16.28 &  0.04 &  72 \\
   59075.633 &   15.63 &  0.02 &   16.03 &  0.03 &  72 \\
   59103.641 &   15.61 &  0.02 &   16.11 &  0.02 &  72 \\
   59105.629 &   15.59 &  0.02 &   16.14 &  0.03 &  72 \\
   59100.576 &   15.62 &  0.02 &   16.15 &  0.02 &   360 \\
   59105.588 &   15.62 &  0.02 &   16.16 &  0.02 &   360 \\
   59114.569 &   15.61 &  0.02 &   16.11 &  0.02 &   360 \\
   59118.632 &   15.59 &  0.02 &   16.11 &  0.02 &   360 \\
   59122.538 &   15.59 &  0.02 &   16.14 &  0.02 &   360 \\
   59132.591 &   15.60 &  0.02 &   16.10 &  0.02 &   360 \\
   59136.594 &   15.57 &  0.02 &   16.07 &  0.02 &   360 \\
   59144.565 &   15.57 &  0.02 &   16.10 &  0.02 &   360 \\
   59161.540 &   15.59 &  0.02 &   16.07 &  0.02 &   360 \\
   59172.479 &   15.57 &  0.02 &   16.05 &  0.02 &   360 \\
   59229.241 &   15.55 &  0.03 &   16.05 &  0.03 &   360 \\
   59247.368 &   15.57 &  0.03 &   15.97 &  0.03 &   360 \\
   59257.290 &   15.62 &  0.02 &   16.08 &  0.02 &   360 \\
   59277.303 &   15.55 &  0.02 &   15.96 &  0.02 &   360 \\
   59295.228 &   15.49 &  0.02 &   15.85 &  0.02 &   360 \\
   59436.638 &   15.48 &  0.02 &   15.86 &  0.03 &  72 \\
   59451.640 &   15.51 &  0.02 &   15.87 &  0.02 &  72 \\
   59477.540 &   15.51 &  0.02 &   15.92 &  0.03 &  72 \\
\hline
\multicolumn{6}{l}{$^a$Full table available online, selected points shown here.}\\
\multicolumn{6}{l}{$^b$Includes bright compact object and tail.}\\
\multicolumn{6}{l}{$^c$Extended source extraction with $5\farcs7$ diameter aperture}\\
\multicolumn{6}{l}{~~from \citet{mcmahon13}.}\\
\end{tabular}
\label{tab:nearirphot}
\end{center}
\end{table}

JCMT Transient Survey observations occur only when the precipitable water vapor content is $<2.58$ mm (opacity of $<0.12$ at 225 GHZ), corresponding to JCMT weather bands 1--3.  Because of telluric absorption, the 450 $\mu$m imaging is only useable from the $\sim 50\%$ of our observations that occur during epochs with the lowest precipitable water vapor content.

A full description of the data, advanced reduction, relative alignment, and calibration process developed by the JCMT Transient Survey is provided by \citet{Mairs2017Cal} for the 850 $\mu$m data and by Mairs et al.~(in prep) for 450 $\mu$m data.  For HOPS 373, a bright sub-mm source, the relative flux calibration is typically accurate to 2\% at 850 $\mu$m and $\sim 5$\% at 450 $\mu$m.  The effective beam sizes are $14\farcs1$ and $9\farcs6$ at 850 and 450 $\mu$m, respectively \citep{dempsey13,mairs21}.

The absolute alignment of the SCUBA-2 images is uncertain by $\sim 2^{\prime\prime}$.  Six objects in the field are compact, bright enough for centroid, and have a well-identified {\it WISE} or {2MASS} counterpart.  The position of HOPS 373 is shifted to the average {\it WISE} position of those objects, with a standard deviation in the positional differences of $1\farcs4$.  However, possible systematics across the field\footnote[1]{Matching 2MASS and WISE coordinates with compact SCUBA-2 sources in the Ophiuchus star-forming region yields $\sim 2-3^{\prime\prime}$ trends across the image.  This trend is unexplained.} limit our confidence to $2^{\prime\prime}$.

\begin{figure*}
\begin{center}
    \includegraphics[width=0.4\textwidth]{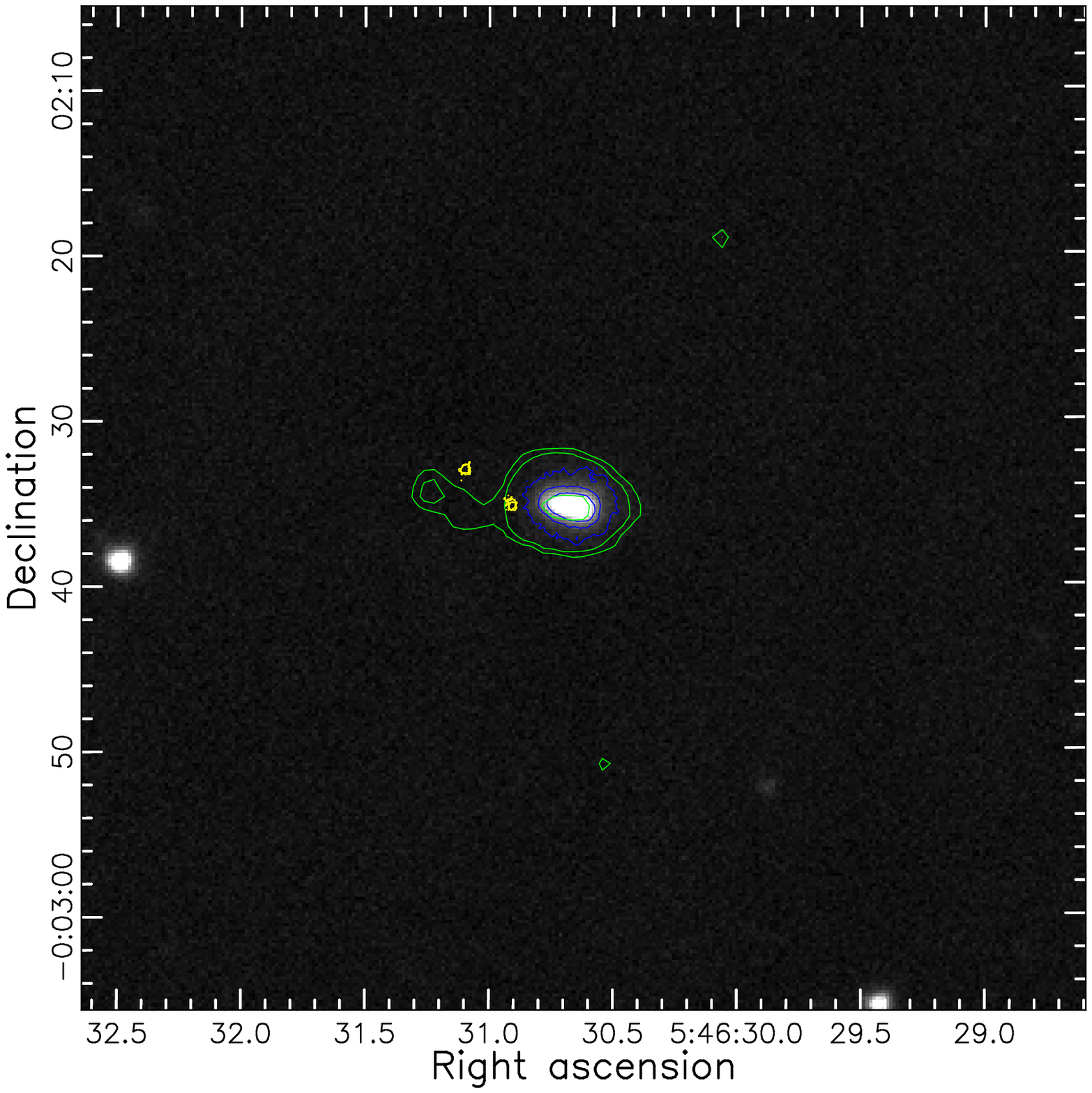}
    \includegraphics[width=0.4\textwidth]{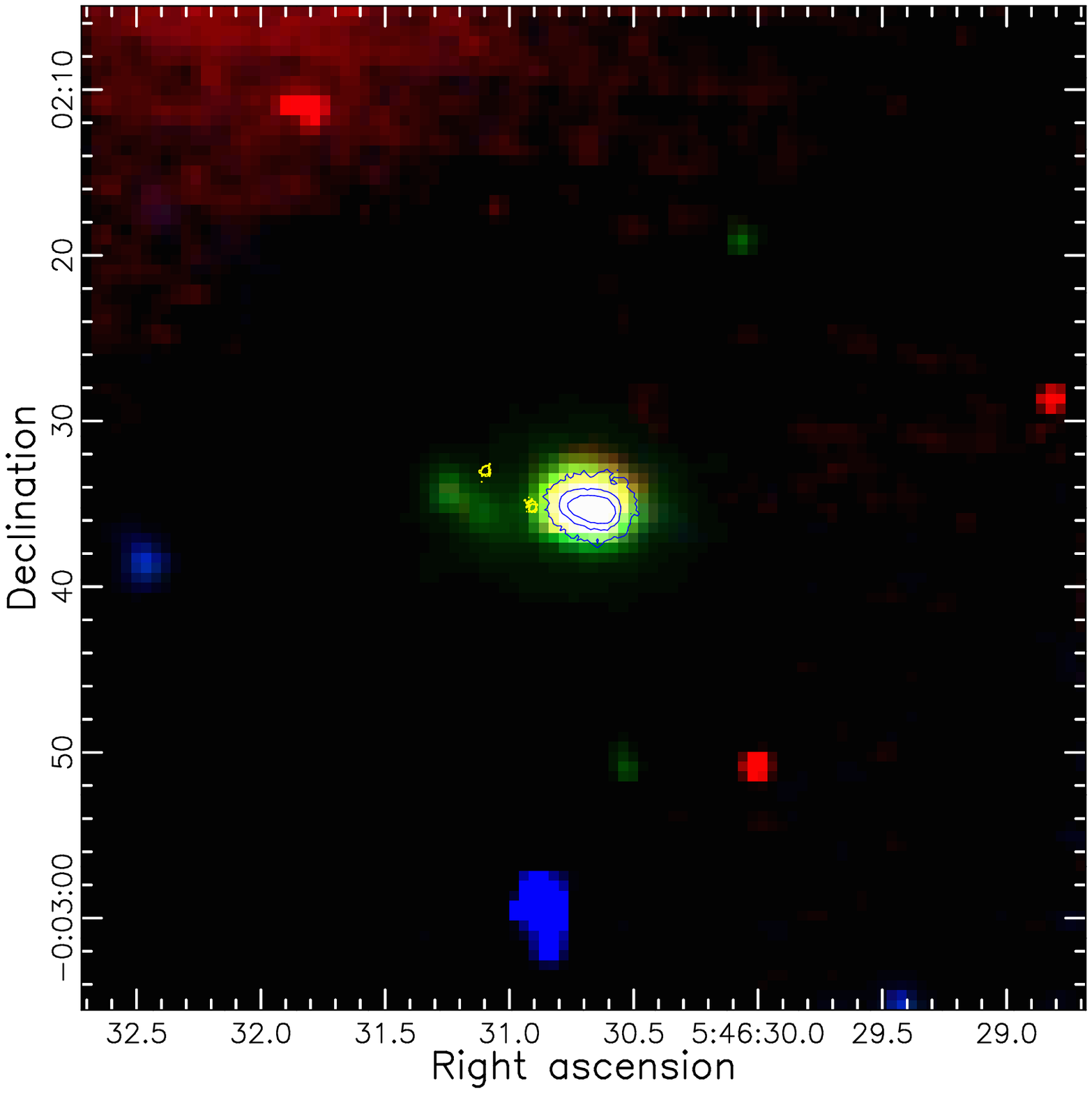}
        \end{center}
    \caption{A $1^\prime\times1^\prime$ field around HOPS 373 from (left) UKIRT $K$-band imaging and from (right) {\it Spitzer}/IRAC imaging in Band 1 (3.6\,$\mu$m; blue), Band 2 (4.5\,$\mu$m; green) and Band 4 (8\,$\mu$m; red).  Both images also show blue contours (99.9, 99.8, 99.5 percentiles) from UKIRT 2.122 $\mu$m H$_2$ imaging and the yellow contours (99.5, 99.0) from emission in the ALMA 890 $\mu$m continuum map; the left image shows green contours  (99.9, 99, 98.5) from Spitzer IRAC 4.5\,$\mu$m imaging.}
        \label{fig:HOPS373_UKIRT_K}
\end{figure*}

\subsection{Mid-IR Imaging from 3--24 $\mu$m}

The \textit{Wide-field Infrared Survey Explorer} (\textit{WISE})  \citep{wright10} surveyed the entire sky in four bands, $W1$ (3.4 $\mu$m), $W2$ (4.6 $\mu$m), $W3$ (12 $\mu$m), and $W4$ (22 $\mu$m), with the angular resolutions of 6\farcs1, 6\farcs4, 6\farcs5, and 12\arcsec, respectively, from January to September 2010.  After the depletion of hydrogen from the cryostat, the NEOWISE Post-Cryogenic Mission \citep{mainzer11,mainzer14} has been observing the full sky in $W1$ and $W2$ at a 6-month cadence since late 2013.  HOPS 373 has been observed in March and September each year from 2014--2020.

Each epoch consists of $\sim 10-20$ exposures taken over a few days.  Each set of exposures are averaged to produce a single photometric point per epoch \citep[following][]{contreras20}.  No significant variability is identified within each epoch.  

We identify consistent seasonal offsets in the NEOWISE position of HOPS 373, as observed during fainter epochs.  In March, the average position in $W1$ is located $0\farcs109\pm0\farcs018$ west and $0\farcs152\pm0\farcs016$ south of the average position in September; for $W2$ the September position is $0\farcs120\pm0\farcs016$ west and $0\farcs149\pm0\farcs016$ south.  The average standard deviation in the seasonal position in a filter is  $0\farcs024$ in each direction, which we adopt for the uncertainty in individual measurements.  We are not aware of the reason behind these seasonal differences or whether they are common.

The field of HOPS 373 was observed multiple times using {\it Spitzer} space telescope and the Infrared Array Camera \citep{fazio04} in the IRAC bands 1, 2, 3 and 4, centered at 3.6, 4.5, 5.8 and 8 $\mu$m, respectively. We obtained and averaged IRAC images from three AORs (4105472, 4105728 and 4105984) from the {\it Spitzer} Heritage Archive.
Figure~\ref{fig:HOPS373_UKIRT_K}-right shows a color composite image in a $1^{\prime}\times1^{\prime}$ field
extracted from the averaged images in Band 1 (blue), Band 2 (green) and Band 4 (red). The near-IR counterpart of HOPS 373 has excess IRAC Band 2 emission, consistent with classification as an extended green object \citep[EGO;][]{cyganowski08}.

The {\it Spitzer}/MIPS scan map observing mode covered the NGC 2068 region on 2004 March 15. We obtained the MIPS images at 24 $\mu$m, with the AOR of 4320256 from the {\it Spitzer} Heritage Archive. 

\subsection{Far-IR Imaging from 70--160 $\mu$m}

The {\it Herschel Space Observatory} \citep{pilbratt10} surveyed protostars in Orion star forming regions \citep{Stutz2013} with PACS 70 $\mu$m and 160 $\mu$m photometry. The PACS 70 $\mu$m and 160 $\mu$m data of HOPS 373 were acquired over 8\arcmin $\times$ 8\arcmin~ field size with beam sizes of 5\farcs6 and 10\farcs7, respectively. We obtained the PACS images of HOPS 373 observed on 2010 Sep 28 and 2011 Mar 6 from {\it Herschel} science archive with the observation ID of 1342205216, 1342205217, 1342215363, and 13422115364 \citep[][]{Fischer20}.

\subsection{Monitoring at $2.2$ $\mu$m and imaging from 1--2.5 $\mu$m}

We monitored HOPS 373 at near-IR wavelengths using the Wide Field Camera \citep[WFCAM;][]{casali07} on the 3.8-m UKIRT telescope from February 2012 to March 2021. WFCAM employs four 2048$\times$2048 HgCdTe Hawaii\,{\sc{ii}} arrays at an image scale of 0\farcs4~pixel$^{-1}$. 
The object was dithered to nine positions separated by a few arcsecs, with 2$\times$2 microsteps at each dither position to achieve an image scale of 0\farcs2~pixel$^{-1}$ in the final mosaics. The observations were obtained using the $J$, $H$ and $K$ MKO filters centered at 1.25, 1.65 and 2.2\,$\mu$m respectively, and in a narrow-band filter centered at the wavelength of the H$_2$ 1--0~S(1) line at 2.1218\,$\mu$m.  The monitoring comprised of a set of shallow observations (1 s$\times$2 coadds per frame; 72 s integration per mosaic) in $J$, $H$ and $K$ filters from 2012 to 2021 and observations with deeper integration in $K$ (2 s$\times$5 coadds per frame; 360 s integration per mosaic) from September 2020 to March 2021.  
The magnitudes are derived using the average zero points estimated from a set of isolated point sources, with absolute scaling from images obtained on a clear, photometric night.  Data points are discarded if the the standard deviation of the zero points on the derived on field stars is larger than 0.03 mag.

We observed HOPS 373 in the H$_2$ line on eight epochs.  The dither and microstep patter were similar to that in K, but with the per frame exposure time of 40s. This gives a total integration time of 1440 sec for the mosaic from each epoch.  The H$_2$ images are continuum-subtracted using the $K$-band images obtained closest in time and with the best agreement in seeing. The background-subtracted $K$-band image is divided by the average ratio of counts [$K$/H$_2$] obtained for a few isolated point sources in the $K$ and H$_2$ images.  This image is then subtracted from the background-subtracted H$_2$ image to obtain the continuum-subtracted emission line image. Since clouds were present during some of those observations, the continuum subtracted images from the different epochs were weighted according to the extinction from clouds and averaged. 

The near-IR emission from HOPS 373 consists of a compact source with a tail.  The magnitudes for these components, presented in Table~\ref{tab:nearirphot}, are estimated in two apertures of diameter 2\arcsec, centered on the compact source at $\alpha$=5:46:30.631 $\delta$=-00:02:35.43 (J2000) and on the tail closer to the YSO, $1\farcs74$ NE of the first position. 
Figure~\ref{fig:HOPS373_UKIRT_K} shows a 1\arcmin$\times$1\arcmin\ field surrounding HOPS 373 in $K$, generated from the average of the $K$-band images from the epochs with deeper integration. The contours derived from the averaged  continuum-subtracted H$_2$ image are overlaid on the $K$-band image.

We also obtained an acquisition image with Gemini North/GNIRS for our spectrum (Section~\ref{sec:obs.nirspec}).  Because HOPS 373 shows extended structure in near-infrared, seven K-band acquisition images were taken to locate the GNIRS slit position precisely. Each image consists of 12 coadded frames with the exposure time of 2 seconds. A blank sky image was generated by combining seven
 images after masking the emission region; it was then used for background subtraction and flat fielding.  The total exposure time of the integrated image is 168 s. 
The pixel size of 0\farcs15 and position angle of $157.49^\circ$ were used to put the imaging scale and orientation on the sky.

\subsection{Spectroscopy from 1.9--2.5 $\mu$m}
\label{sec:obs.nirspec}

We obtained a near-IR $JHK$ spectrum of HOPS 373 using GNIRS on Gemini North in Fast Turnaround Time, program GN-2020B-FT-110 (PI: Doug Johnstone) on 2020 October 3 (HJD=2459126.12).   The integration time was 2400 s, split into two ABBA sequences with individual exposures of 300 s.

The data were obtained in the cross-dispersed mode with the 32 mm grating, the short camera, and a 0\farcs3 slit to achieve $R=1700$ spectra from 1--2.5 $\mu$m.  Our focus here is on the $K$-band spectrum, since little emission is detected in $H$ and none in $J$.  The slit of 0\farcs3$\times$7\arcsec\ was aligned with the parallactic angle of $157.49^\circ$, almost perpendicular to the direction of the source extension, and centered on the compact source, which dominates the observed emission.  The spectra were obtained at an average airmass 1.07.  The data were reduced and extracted following standard techniques.  

The data were corrected for telluric absorption using B8 V star HD 39803 observed immediately after HOPS 373 at an airmass of 1.14.   The flux calibration of the HOPS 373 spectrum, with an estimated synthetic magnitude of $K=15.56$, is estimated by the relative brightness to the same telluric standard star, as measured in the acquisition images, and assuming that the standard star is constant in brightness across the band.

The wavelength solution is accurate to $\sim 0.5$ \AA, or 7 \kms.  However, each pixel covers 80--100 \kms, so any asymmetry in spatial extent of the source within the slit may cause additional shifts.  We corrected for the local standard of rest (LSR) velocity frame using the IRAF task {\it rvcorrect}.

\subsection{Radio observations with the VLA}
Observations toward HOPS 373 were conducted with the NSF's Karl G. Jansky Very Large Array (VLA)
located on the Plains of San Agustin in central New Mexico, USA. HOPS 373 was observed 
in multiple bands both before and after the outburst.

\subsubsection{C-band Observations at 5 cm}
Observations of HOPS 373 were conducted in C-band at a central frequency of 6~GHz on 2015 Oct 06 (2 executions) and on 2015 Oct 07, as part of archival program 15A-369.  Both observations were obtained  while the array configuration was transitioning from A to D configuration, with 21 useable antennas in the D-configuration. The 2015 Oct 07 data were  unusable due to a system issue that occurred during observations. We performed subsequent observations
on 2021 Mar 24 (program 21A-409) and 2021 May 15 (program 21A-423) in D-configuration, replicating the setup
of the archival observations. In 2015, J0541-0541 was used as the phase
calibrator and 3C147 was used as the bandpass and flux calibrator. For 2021,
we used J0552+0313 as the complex gain calibrator and 3C147 as the flux and bandpass calibrator.

We reduced and imaged the data using the 
VLA calibration pipeline in Common Astronomy Software Application \citep[CASA,][]{mcmullin2007}
version 6.1.2. The correlator
was configured for 3-bit mode where the entire 4 to 8~GHz band is covered with a single setting. We 
performed additional flagging for system issues (amplitude jumps and spws total swamped with 
radio frequency interference [RFI]) and re-ran the calibration pipeline. We imaged each observation
using the CASA task \textit{tclean} using \textit{robust=0.5} weighting. We also cut out the inner
1.4~k$\lambda$ baselines using the \textit{uvrange} parameter to remove contributions from large-scale
emission associated with the nearby H II region. The 2015 data were self-calibrated
to remove dynamic range artifacts associated with a bright source in the field of view. The same source was substantially fainter during the 2021 observations and therefore self-calibration was not required.

For the 2021 observations, we
restored the image using the same beam as the archival observations (14\farcs77$\times$11\farcs40) to perform data analysis 
with beam-matched data sets. The noise in each observation is 6.28, 5.35, 4.88 $\mu$Jy~beam$^{-1}$ in respective 
chronological order.

\subsubsection{K-band Observations at 1.3 cm}
D-configuration observations were conducted in K-band (22~GHz) on 2015 Oct 17  and 2015 Nov 21 as part of program 15B-229. On 2021 Apr 05 we acquired additional D-configuration observations (program 21A-409). The 2015 observations
used 8-bit samplers and observed spectral lines and continuum within
two 1-GHz basebands. The main lines targeted were NH$_3$ (1,1), (2,2), (3,3) and the H$_2$O 
(water) maser
transition ($J=6_{1,6}\rightarrow 5_{2,3})$) at 22.23507980 GHz.  For the follow-up observations
in 2021, we used 3-bit samplers to obtain the maximum continuum bandwidth, but also
observed the same lines as observed in 2015. We only discuss the water maser
emission and continuum data in this paper.

The data were all calibrated using the same methodology as described for C-band. Imaging
was performed with the CASA task \textit{tclean} using \textit{robust=0.5} for the continuum data,
and \textit{robust=2} for the line data with 0.5~\kms\ channels. The noise in the continuum images 
is 7.2 and 10.7~$\mu$Jy~beam$^{-1}$ for the 2015 and 2021 data, respectively, while the noise in the 
water maser data cubes is 1.23 and 5.55~$\mu$Jy~beam$^{-1}$ for the 2015 and 2021 data, respectively.

\subsubsection{Ka-band Observations at 9 mm}
Observations were conducted in C-configuration at Ka-band (33~GHz) on 2016 Jan 30 as part of program 16A-197 and again on 2021 Apr 07 in D-configuration (program 21A-409). Both observations 
utilized the 3-bit samplers to cover a total bandwidth of 8~GHz, but in 2021 we centered the two 4~GHz
basebands at $\sim$29 and 37~GHz to sample a wider fractional bandwidth.

The data were all calibrated using the same methodology as described for C-band. Imaging
was performed with the CASA task \textit{tclean} using \textit{robust=0.5} for
the D-configuration data observed in 2021. Then for the 
C-configuration data, we used \textit{robust=2.0} and \textit{uvtaper=70 k$\lambda$} to better match the D-configuration beam. 

\subsubsection{Flux Density Measurements with the VLA}
\label{sec:obs-vla-analysis}
To measure the flux densities of the continuum sources in C, K, and Ka-bands, we used the CASA
imfit task to fit the source with two Gaussian profiles in K and Ka-bands, where
the sources are resolved, and a single Gaussian profile in C-band, where the two sources are unresolved.
We provided imfit with initial estimates for the position, flux density, and source size and allowed
the fitting to converge on its own. At C-band, we fixed the source size to be a point source.

For the C-band data, we also measured the flux densities of all sources that appeared in both the 
2015 and 2021 data since numerous YSOs and background sources lie within the field of view. This
enables us to characterize any systematic offsets in flux density calibration.  Twenty-five sources detected in both the 2015 and 2021 observations show a median variation of +15~$\mu$Jy, with a corresponding median flux density ratio of 1.22.
Similarly, for the two 2021 epochs, 28 matched sources show a median variation of +1~$\mu$Jy, with a corresponding median flux density ratio of 1.01. 

The median flux density ratio of 1.22 for the 2015 to 2021 epoch is larger than the expected flux density
uncertainty of 5-10\% in C-band. Furthermore, the same absolute flux calibrator (3C147) was used for both the
2015 and 2021 observations and the calibrator has had consistent flux density within $\sim$1\% between
2016 and 2019. We therefore expect that the large difference of measured flux densities between epochs is due to
real variability in the sources and not flux density calibration error.

\subsection{ALMA Observations}
The Atacama Large Millimeter/submillimeter Array (ALMA), 
located in northern Chile on the Chajnantor plateau at an elevation
of $\sim$5000~m, consists of 50  12~m antennas that constitute the main
12-m array, ten 7~m antennas that form the ALMA Compact Array (also called the Morita Array),
and an additional four 12-m antennas that are used for total power observations. 
The analysis of HOPS 373 makes use of data from standalone observations with the
ACA and the 12m array.  No data combination is performed.

\subsubsection{ACA Observations at 1.33 mm}
The Atacama Compact Array (ACA) and Total Power (TP) antennas conducted the Band 6 observation toward HOPS 373 on 2019 March 21 as a part of program (2018.1.01565.S; PI: Tom Megeath). The beam size of the continuum is $4\farcs4\times6\farcs8$ and the reference frequency is 225.69 GHz (1.33 mm). The $^{12}$CO J=2-1 line (230.538 GHz) analyzed in this work is obtained from the spectral window having the reference frequency of 230.59 GHz and bandwidth of 250.00 MHz. The spectral resolution is 0.317 km s$^{-1}$. 

\begin{figure*}[t]
\includegraphics[width=0.95\textwidth,trim=10mm 127mm 10mm 23mm]{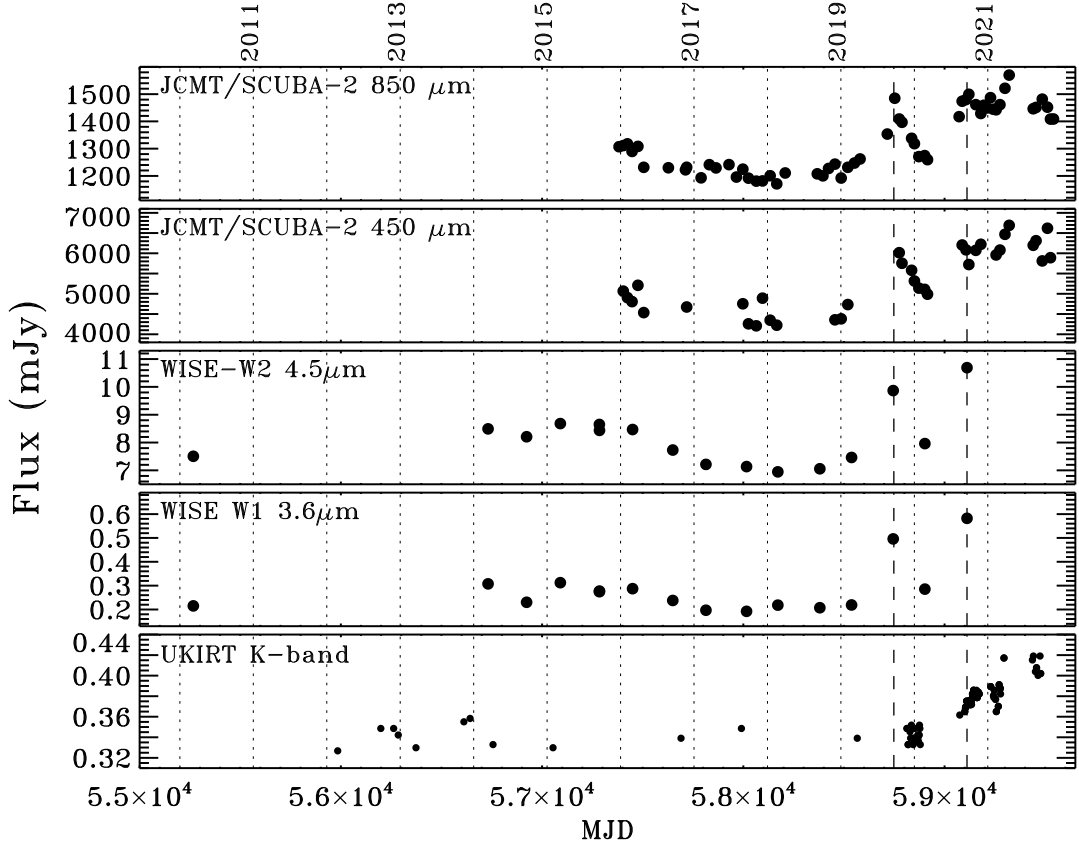}
\caption{Lightcurves of HOPS 373 at 850 and 450 $\mu$m from JCMT/SCUBA-2, at 4.5 and 3.6 $\mu$m from WISE and NEOWISE, and in the $K$-band with UKIRT.  The dotted vertical lines mark January 1 for each year.  The dashed vertical lines indicate the dates that are used for Table~\ref{tab:burstlist}.}
\label{fig:lightcurves}
\end{figure*}

\subsubsection{ALMA 12m Array and ACA Observations at 0.89 mm}
The observations of HOPS 373 with the 12m array were conducted as part of program 
2015.1.00041.S (PI: John J. Tobin) and were carried out in Band 7 on 2016 September 03, 04, and 2017 July 19.
The time on HOPS 373 during each execution was $\sim$20~seconds for a total
integration time of $\sim$1~minute. The maximum baseline length for the
2016 and 2017 observations was $\sim$2500~m and $\sim$3700~m, respectively.
The correlator was configured to have two basebands observed in low-resolution
continuum mode, each baseband having a bandwidth of 1.875~GHz divided into 128 channels
that are 31.25~MHz wide. The other two basebands were configured for spectral line
observations. The first was centered on $^{12}$CO ($J=3\rightarrow2$) at 345.79599~GHz, having
a total bandwidth of 937.5~MHz and 0.489~\kms\ channels. The second spectral line baseband
was centered on $^{13}$CO ($J=3\rightarrow2$) at 330.58797~GHz, having
a total bandwidth of 234.375~MHz and 0.128~\kms\ channels. The line-free regions of the
spectral line basebands were also used in the continuum imaging to result in an effective
bandwidth of $\sim$4.75~GHz at 0.89 mm. Reduction of the raw data and subsequent imaging was performed using CASA version 4.7.2. 
Self-calibration was also performed on the continuum data to increase
the S/N, and the self-calibration solutions were also applied to the 
spectral line data. The resultant 0.89 mm continuum image created using the 
CASA task \textit{clean} with \textit{robust=0.5} has a beam of $\sim$0\farcs11
and a noise of 0.27 mJy~beam$^{-1}$. The $^{12}$CO ($J=3\rightarrow2$) image was
created using natural weighting, but the visibilities were also tapered, starting at 500~k$\lambda$
to increase sensitivity to large scale structures, resulting in a $\sim$0\farcs25 beam with 
a noise level of 20~mJy~beam$^{-1}$ (see \citealt{tobin20} for further details on the 
observations and data analysis).

The ACA observation is conducted on 2018 October 2 as a part of program 2018.1.01284.S (PI: Tom Megeath). The continuum is observed with the beam size of 5\farcs0$\times$2\farcs9 at reference frequency of 338.239 GHz.

\subsection{Gaia Astrometry of the Parent Association}

HOPS 373 is likely associated with the NGC 2068 - 1 group described by \citet{kounkel18} from Gaia DR2 astrometry, located at 428 pc \citep{gaia18}.  The mean proper motion of this group is 0.254 mas/yr in right ascension and -0.573 mas/yr in declination.  This proper motion leads to spatial offsets of $\sim$ 0\farcs01 between the 2MASS epoch and later epochs from ALMA and NEOWISE.  This offset is negligible for our analysis and is not applied to our astrometry.  The velocity relative to the local standard of rest ($v_{lsr}$) is $\sim 10.3$ km s$^{-1}$ \citep[e.g.][]{mitchell01,kang15,nagy20}.

\section{Lightcurves of spatially unresolved emission}
\label{sec:lightcurves}

In this section, we describe the time variability of HOPS 373 and present a fiducial model to explain the sub-mm brightening as a response to increased accretion luminosity from a protostar deeply embedded in an envelope. 

\subsection{Qualitative and Quantitative Description of the Lightcurves} 

\begin{figure}[!t]
\includegraphics[width=0.48\textwidth, trim=10mm 130mm 10mm 20mm]{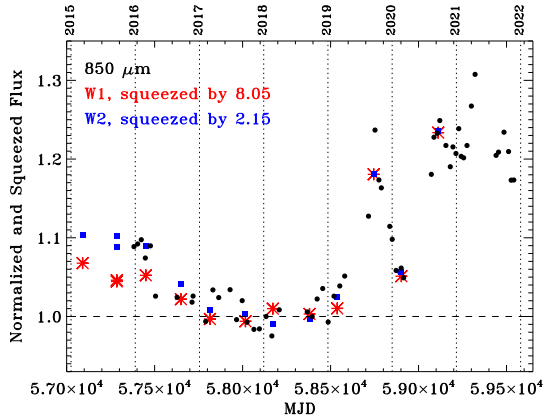}
\caption{The sub-mm and WISE lightcurves of HOPS 373.  The WISE W1 (red asterisks) and W2  (blue squares) photometry is squeezed so that the change between the quiescent level and the most recent burst match the contemporaneous sub-mm fluxes (black circles).  The dotted vertical lines indicate January 1 for each year.  \label{fig:squeezed}}
\end{figure}

The burst was initially detected in monitoring at 850 $\mu$m (Figure \ref{fig:scuba2}), triggering the follow-up investigation analyzed here.
Figure~\ref{fig:lightcurves} and Tables~\ref{tab:scuba2phot}--\ref{tab:nearirphot} present lightcurves for HOPS 373 in the sub-mm, mid-IR, and near-IR.  Figure~\ref{fig:squeezed} and Table~\ref{tab:burstlist} compare the size of the burst in each band.

Starting in December 2015, HOPS 373 had an initial flux density of $1.3$ Jy/beam at 850 $\mu$m for a few months, before decaying to a steady local minimum (quiescent) level of $1.2$ Jy/beam by April 2016.  An early point from the Gould Belt Survey in November 2014 \citep{Mairs2017GBS} is consistent with the initial level in December 2015. HOPS 373 then brightened by 25\% to 1.5 Jy/beam in September 2019 and decayed back to quiescence by 7 March 2020.   By the next data point, in August 2020, HOPS 373 was again bright and has stayed in this bright state in all epochs through the end of 2021.
The SCUBA-2 450 $\mu$m data are noisier and sparser but follow the same trend as the 850 $\mu$m measurements\footnote[2]{All uncertainties are $1\sigma$
}, with $F_{450} \propto F_{850}^{1.6\pm0.1}.$ 

The late 2019 outburst is narrower and more peaked in time than a Gaussian profile.  
The rise time is very uncertain and can be arbitrarily steep, since only one point is clearly in the increasing part of the lightcurve.  If the rise is exponential, the timescale near the peak must be shorter than 70 days (doubling time shorter than 50 days), based on the peak flux plus the one point in the rise.
The decay is better constrained, with  an e-folding timescale\footnote[3]{The decay is also consistent with a linear decay with a slope of 
$\sim (1.1\pm0.2)\times10^{-3}$ Jy/day, although the peak itself would be an outlier inconsistent with this fit.} of $105^{+35}_{-21}$ days (halving time of 73 days).
The brightening of 30\% from minima to maxima is the third largest change in flux seen in the JCMT Transient survey to date, surpassed by only by HOPS\,358 and Serpens Main EC 53 \citep{lee21}.

The NEOWISE mid-IR lightcurves also follow the same general pattern, with a lower cadence of one epoch every six months.  The 2015--2016 decay occurred from a previous stable period that was 25\% brighter in $W1$.  HOPS 373 then stayed faint until February 2019, with a bright epoch in September 2019, a faint epoch in March 2020, and then another bright epoch in September 2020.  The NEOWISE epoch in late 2019 occurred about 8 days before the first SCUBA-2 peak, including one exposure just 4 days earlier.  In 2020, the mid-IR returned to quiescence and then burst again, with an epoch four days after a SCUBA-2 observation.  The two mid-IR bright measurements are 2.46 and 2.88 times brighter than quiescence at $W1$ but only 1.39 and 1.51 times brighter than quiescence at $W2$.

The $K$-band monitoring is consistent with the sub-mm and mid-IR lightcurves, but with the scale of variability suppressed.
Prior to the first burst, the $K$-band emission showed stochastic fluctuations within a 0.1 mag range that are consistent with a constant brightness.  The $K$-band monitoring missed the first burst.  The most recent points are $\sim 0.15-0.25$ mag brighter than those in quiescence.

The sub-mm and mid-IR lightcurves follow each other closely, but with differences in the amplitude of change.  Averaging the two bursts and comparing against the mean of the preceding quiescent period (Table \ref{tab:burstlist}), the minimum to peak increase is a factor of 2.66 ($1.06$ mag) at W1, 1.45 (0.40 mag) at W2, and 1.24 (0.23 mag) at 850 $\mu$m.   Scaled to the 850 $\mu$m observations, both the $W1$ and $W2$ emission increased more during the 2020 peak than in the 2019 peak.   However, this difference may be the consequence of the offset in time between the mid-IR and sub-mm observations in the first burst.  The higher cadence of sub-mm observations and the long duration at a near-constant flux during the second burst makes the relative changes more reliable.

The previous {\it Spitzer} photometry  is consistent with the {\it WISE} photometry\footnote[4]{HOPS 373 is brighter in IRAC Band 1 than $W1$, likely because $W1$ has higher transmission than IRAC Band 1 at $<3$ $\mu$m.  A direct comparison is challenging for HOPS 373 because of the filter mismatch.}, with IRAC Band 1 of $13.86$ and IRAC Band 2 of 10.79 mag \citep{Stutz2013,getman17}.

\begin{table}[!t]
\begin{center}
\caption{Flux versus wavelength$^a$}
\begin{tabular}{lccccccc}
\multicolumn{1}{r}{MJD=} & 57750--58450 & \multicolumn{2}{c}{58748} &  \multicolumn{2}{c}{59111$^b$} \\
$\lambda$ ($\mu$m) & Quiescent & Burst 1 & Ratio & Burst 2 & Ratio\\
\hline
2.2 comp$^c$ & 3.42e-4 & -- & -- & 3.72e-4 & 1.087\\
2.2 tail$^c$   & 1.97e-4 & -- & -- & 2.30e-4 & 1.170\\
3.6 &  2.02e-4 & 4.96e-4 & 2.46 & 5.82e-4 & 2.88\\
4.5 & 7.10e-3 & 9.87e-3 &  1.39 & 10.7e-3 & 1.51\\
450 & 4.30  & -- & -- & 5.91 & 1.37 \\
850 & 1.20  & 1.485 & 1.24 & 1.490 & 1.24\\
\hline
\multicolumn{6}{l}{$^a$All fluxes in Jy; ratios are burst/quiescent flux.}\\
\multicolumn{6}{l}{$^b$Relative offsets from MJD=59111 may be more reliable,}\\
\multicolumn{6}{l}{~~~SCUBA-2 averaged from MJD 59107 and 59121.}\\
\multicolumn{6}{l}{$^c$Compact source and tail in $K$-band.}\\
\end{tabular}
\label{tab:burstlist}
\end{center}
\end{table}

\subsection{A Fiducial Model for the Sub-mm Lightcurve}
\label{sec:fiducial}

The variability in the sub-mm lightcurve is likely a response to a luminosity change within the deeply embedded protostar.  The variable emission at 850 $\mu$m, where the envelope is optically thin, is the consequence of a change in the dust temperature within the dense enshrouding envelope \citep[see for example,][]{johnstone13}.  
Infrared wavelengths are closer to the peak of the spectral energy distribution and should directly trace the emission from the protostar and inner disk, but with interpretations that are complicated by uncertain optical depth effects, including scattering. 
As we will show in Section~\ref{sec:morphology}, for this deeply embedded source, the outflow cavity is the surface of last emission for the energy that escapes in the mid-IR, although that energy should trace changes in the emission from the central protostar and its disk.

For sufficiently low dust temperatures, the 450 $\mu$m emission is somewhat shortward of the Planck function Rayleigh-Jeans limit, such that small temperature changes result in larger than linear brightness response. In contrast, the longer wavelength 850 $\mu$m emission is always closer to the Rayleigh-Jeans limit. Indeed, for a mean dust temperature in the envelope of 10 -- 15\,K at quiescence, the observed relationship between the sub-mm brightnesses, $F_{450} \propto F_{850}^{1.6}$, is well recovered, as the heating leads to a 
larger response at 450~$\mu$m versus 850~$\mu$m.

Along with the variation in the amplitude of brightness changes across wavelengths to underlying protostar luminosity changes, the \citet{johnstone13} model predicts a time delay for the sub-mm emission due to the finite light propagation that is required for the dust heating within the envelope. The crossing time of a 5000\,AU core is about 30 days.  Given that most of the core mass is on large scales, \citet{johnstone13} anticipated a smoothing of the light curve on timescales shorter than a month.

To roughly estimate the sub-mm lightcurve, we model the time-dependent sub-mm transport of energy through the envelope as a response to a jump in protostellar luminosity. Because the expected density and temperature profiles within the core decline as radial power-laws outward from the centre, the sub-mm time response is not uniform but rather highly peaked toward the first few days and with a long tail reaching out to months (heating the backside of the core introduces twice the delay from light travel time). Using this time dependent response to fit the observed sub-mm lightcurve in late 2019, we find that the limit on the exponential rise doubling time of the source shrinks to less than 30 days while the $\sim75$ day decay halving time stays the same. The timing of the underlying burst, however, is shifted earlier by about 25 days. This is not the entire story, however, as the sub-mm response is approximately a dust temperature response and thus the underlying protostar luminosity change should be much stronger \citep[i.e.\ expected to be more similar to the $W1$ lightcurve;][]{johnstone13, contreras20}. We therefore expect that the protostellar luminosity change during the burst event was significantly more peaked than that seen in the sub-mm, with a half-maximum width of only days for the rise and weeks for the decay.

\begin{figure*}
    \includegraphics[width=0.41\textwidth]{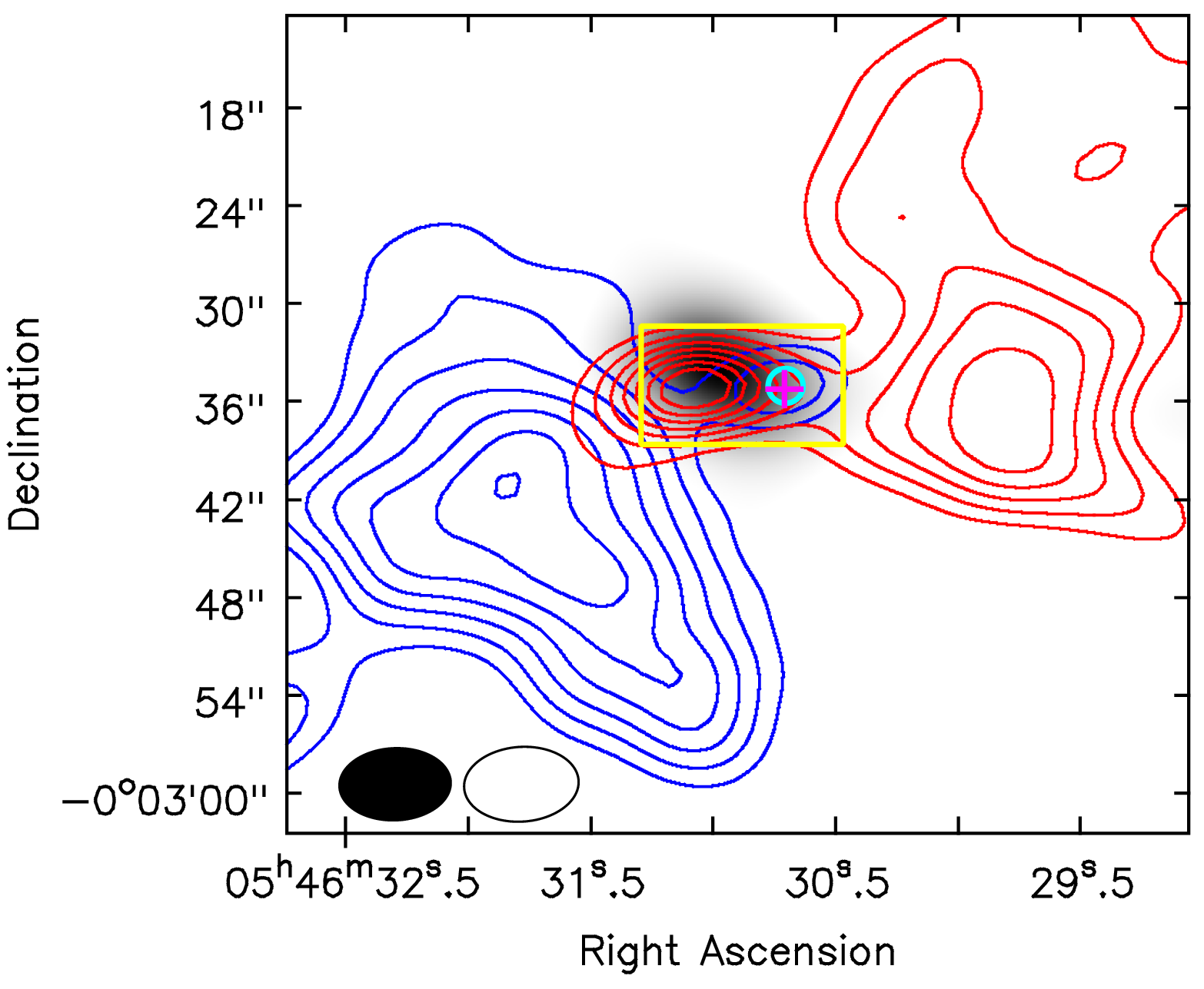}
    \includegraphics[width=0.57\textwidth]{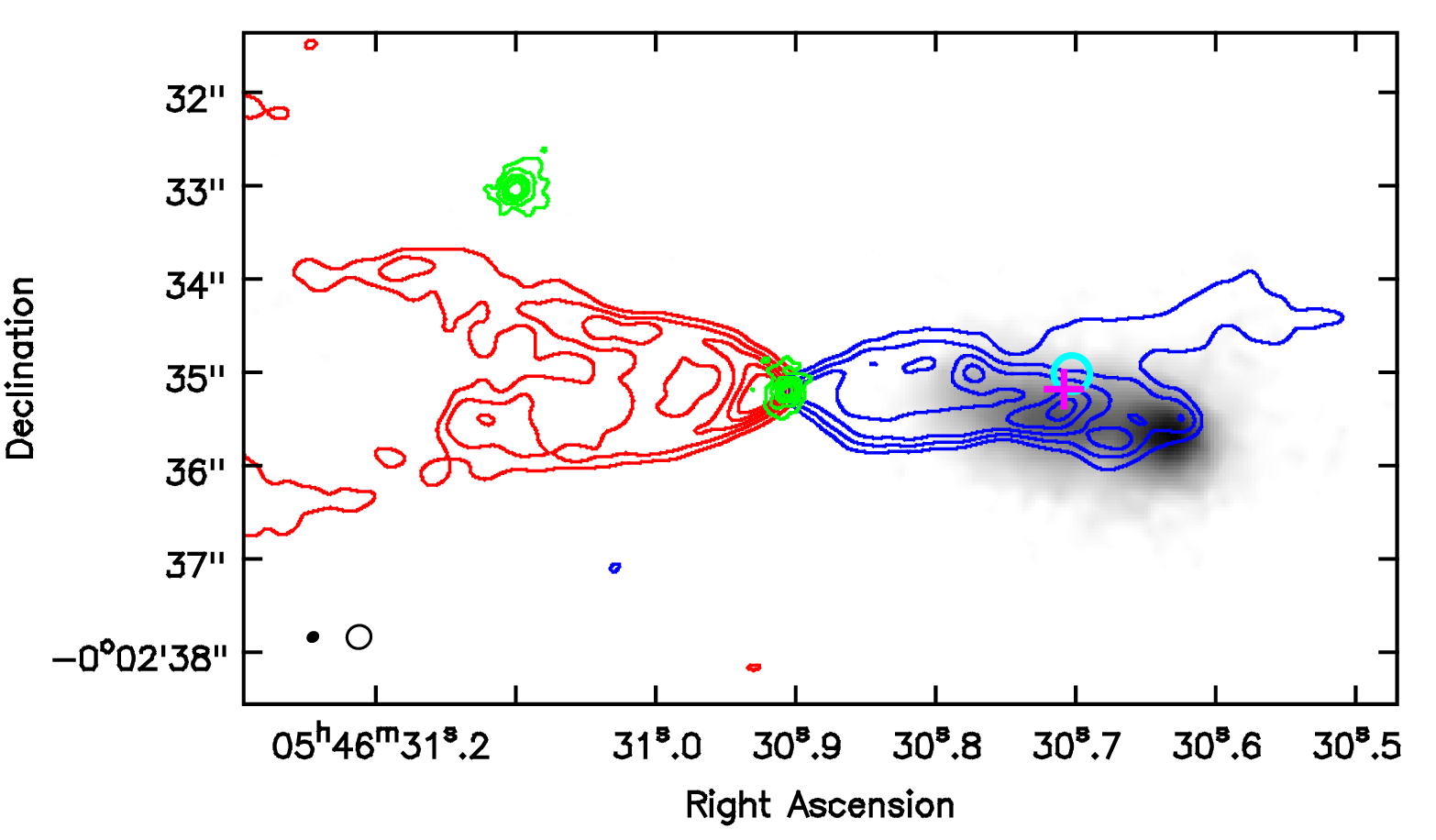}
    \caption{{\it Left}: $^{12}$CO (2-1) integrated intensity map (blue contours from $-6.2$ to $9.0$ \kms, red contours from 12.5 to 36.0 km s$^{-1}$) 
    overlaid on the 1.3 mm continuum image (grayscale). The contour levels are from 4$\sigma$ to 28$\sigma$ in steps of 4$\sigma$, with 1$\sigma$ = 2.4 Jy beam$^{-1}$ km s$^{-1}$. The cross (magenta) and circle (cyan) marks show the median centroid of the WISE observations at the quiescent and burst phases respectively. The yellow box is the field of view of the right panel. {\it Right:} $^{12}$CO (3-2) integrated intensity and the 0.89 mm continuum maps overlaid on the Gemini North/GNIRS K-band acquisition image. The blue and red contours show the CO integrated emission over the velocity ranges of from -16.0 to 9.0 km s$^{-1}$ and 13.0 to 37.0 km s$^{-1}$, with contour levels from 4$\sigma$ to 20$\sigma$ in the step of 4$\sigma$ with 1$\sigma$ = 0.2 Jy beam$^{-1}$ km s$^{-1}$. The green contours show 0.89 mm continuum emission associated with NE and SW sources with levels of 5$\sigma$ $\times$ (1, 2, 4, 6, 10, 20) with 1 $\sigma$ = 4.0$\times$10$^{-4}$ Jy beam$^{-1}$. The filled and open circles at bottom left present the beam size of the continuum and spectral line observations, respectively. \label{fig:almaoutflow}}
\end{figure*}

\section{Morphology of emission components}
\label{sec:morphology}
\begin{figure*}
\begin{center}
    \includegraphics[width=0.33\textwidth]{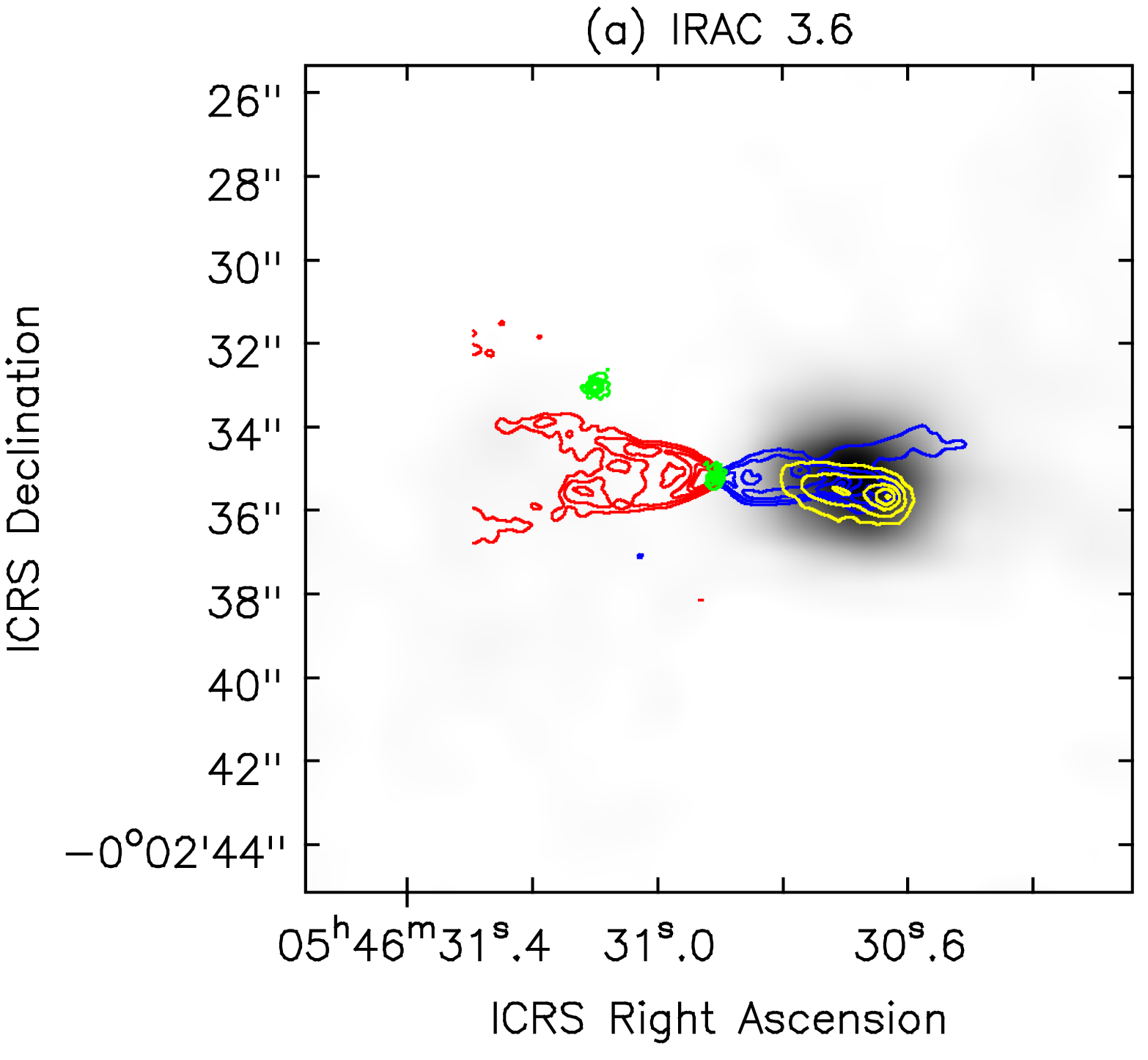}
    \includegraphics[width=0.33\textwidth]{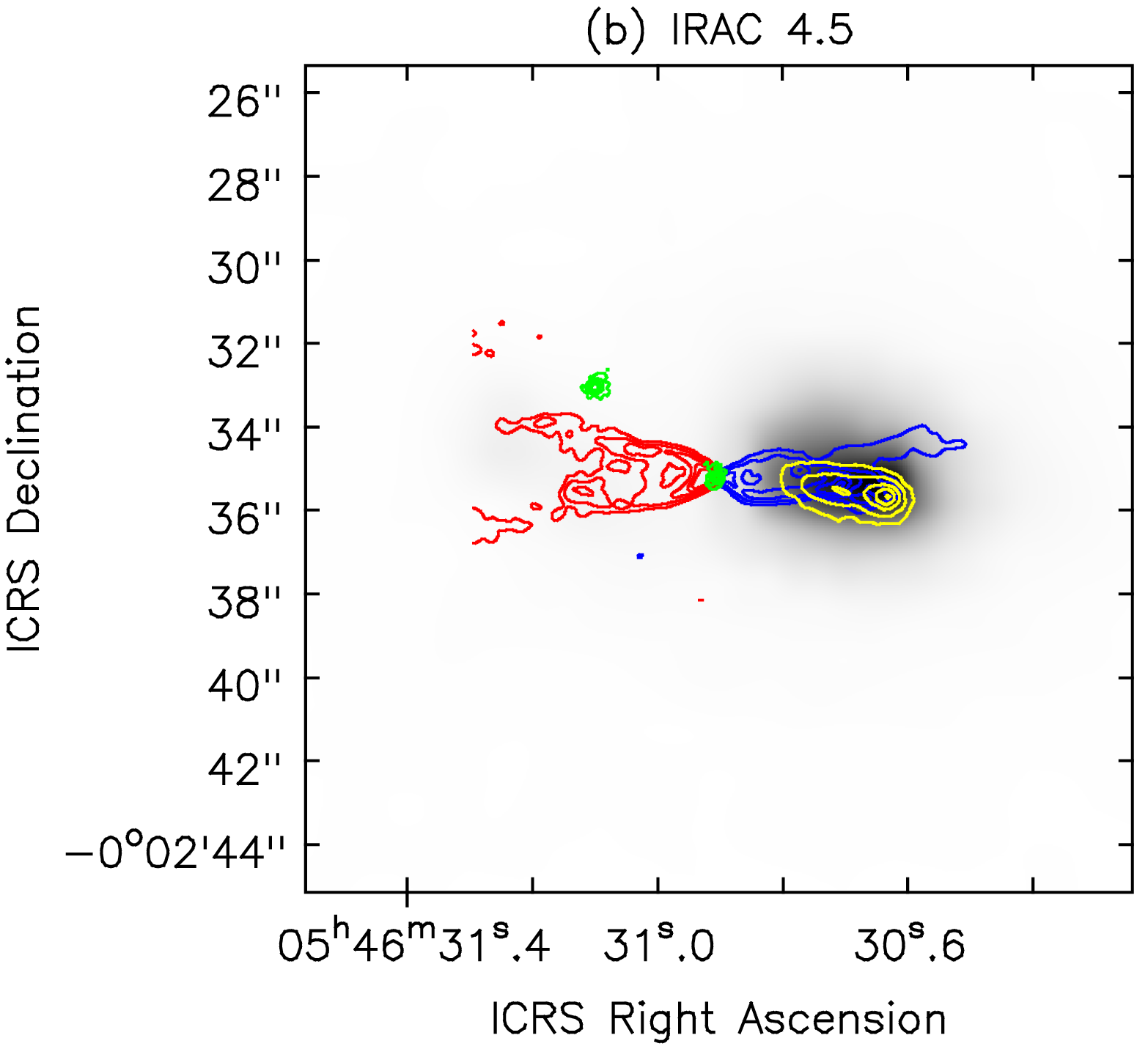} \\
    \includegraphics[width=0.33\textwidth]{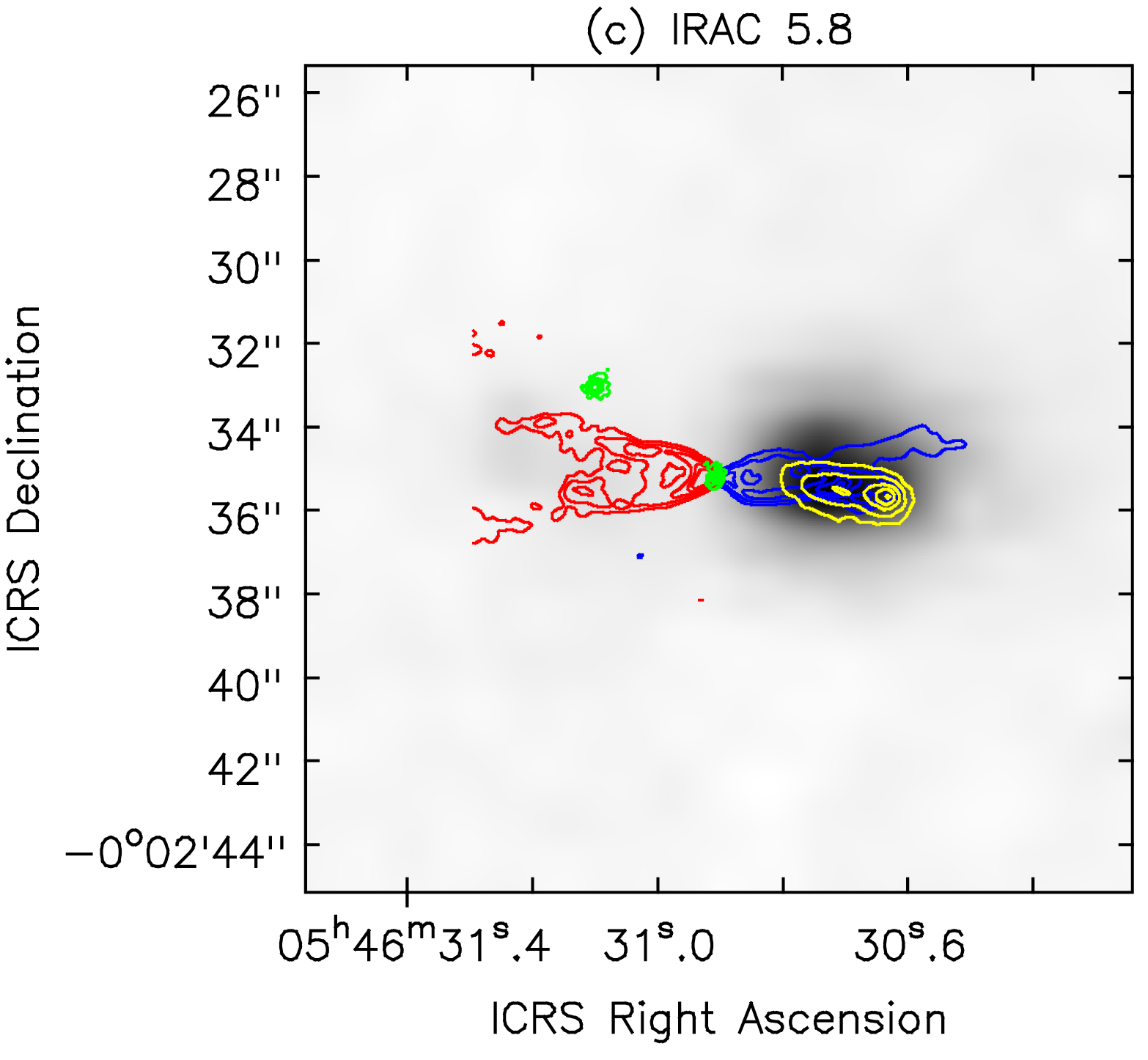}
    \includegraphics[width=0.33\textwidth]{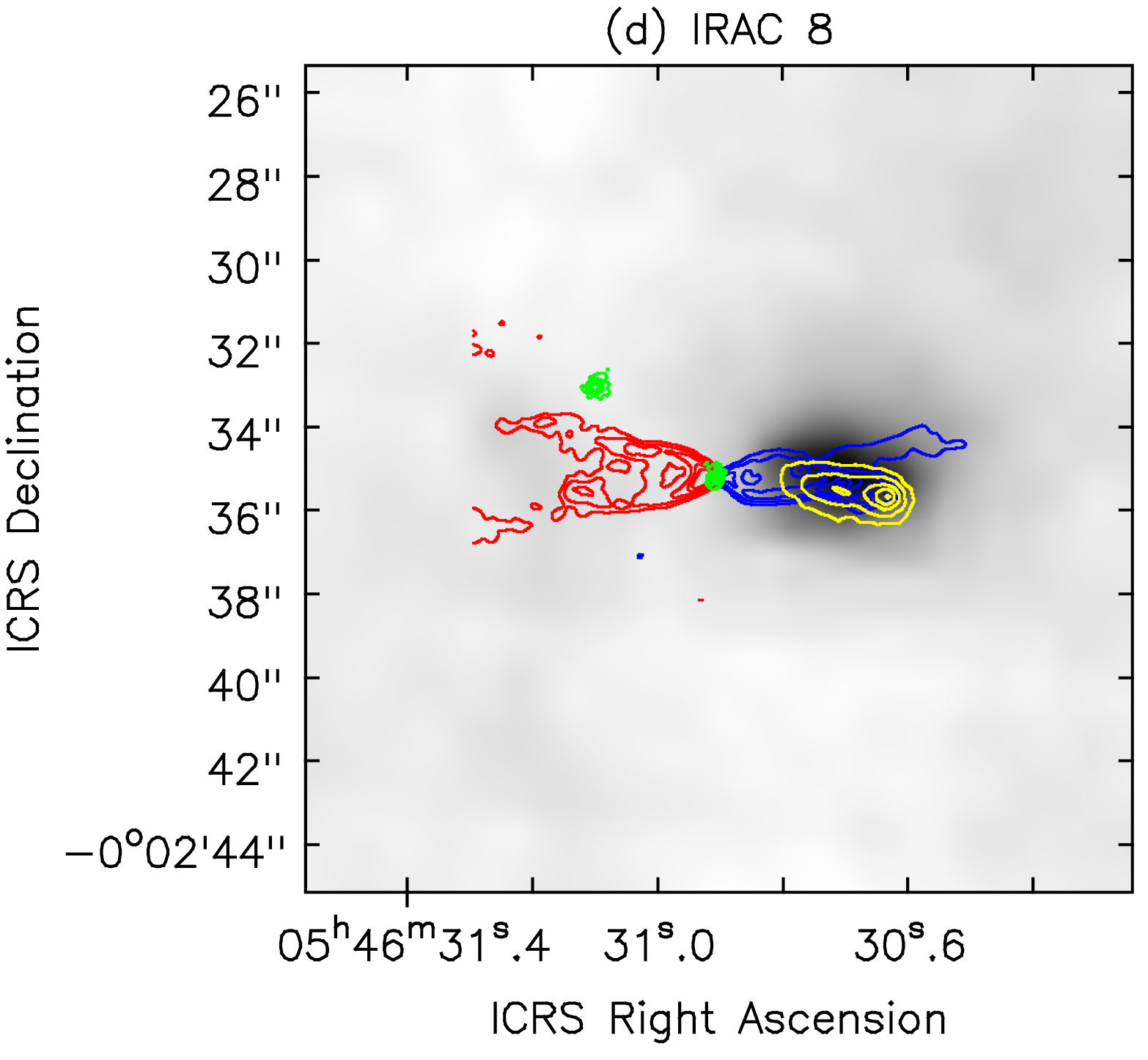} \\
    \includegraphics[width=0.33\textwidth]{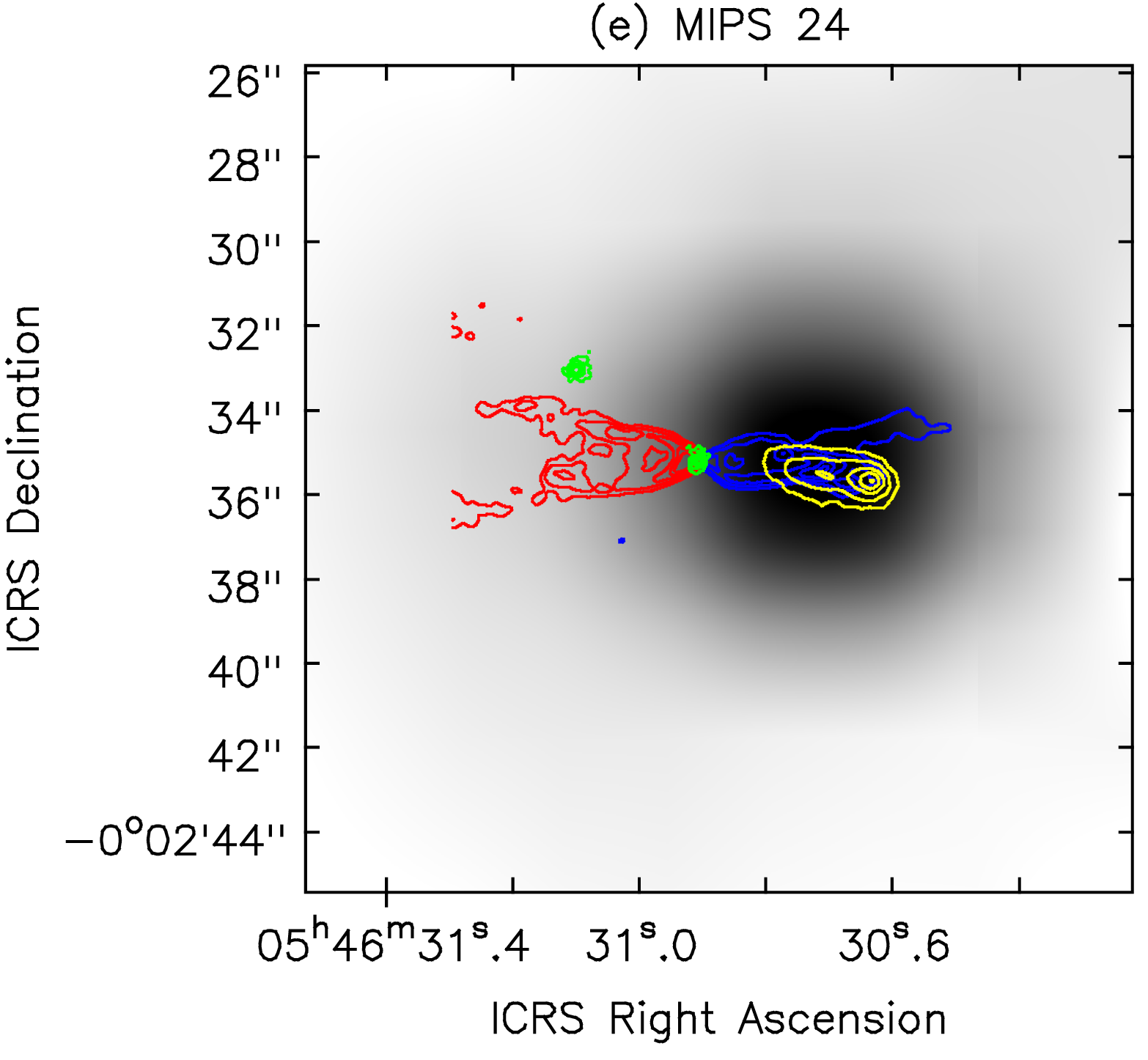}
    \includegraphics[width=0.33\textwidth]{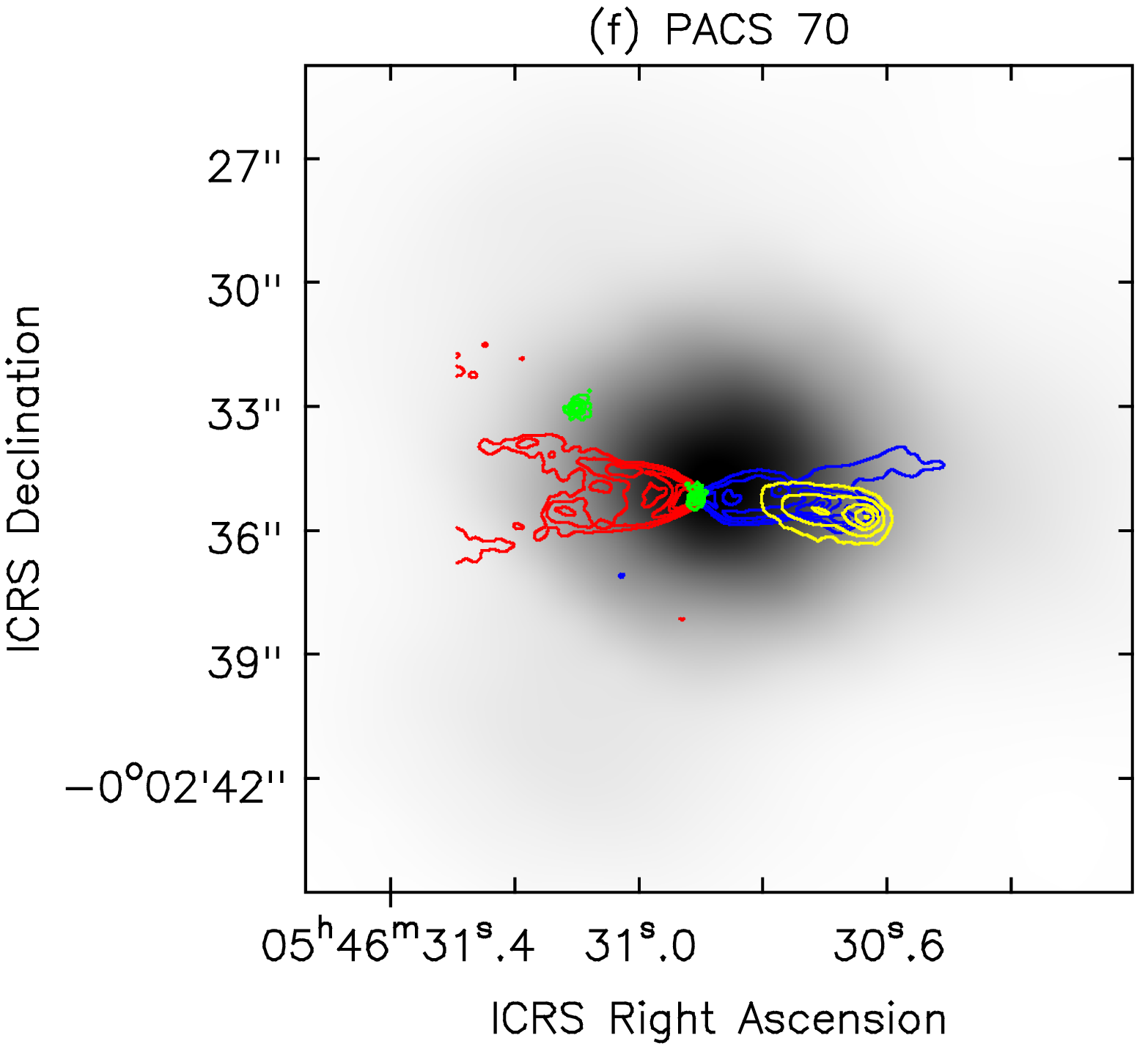} \\
    \includegraphics[width=0.33\textwidth]{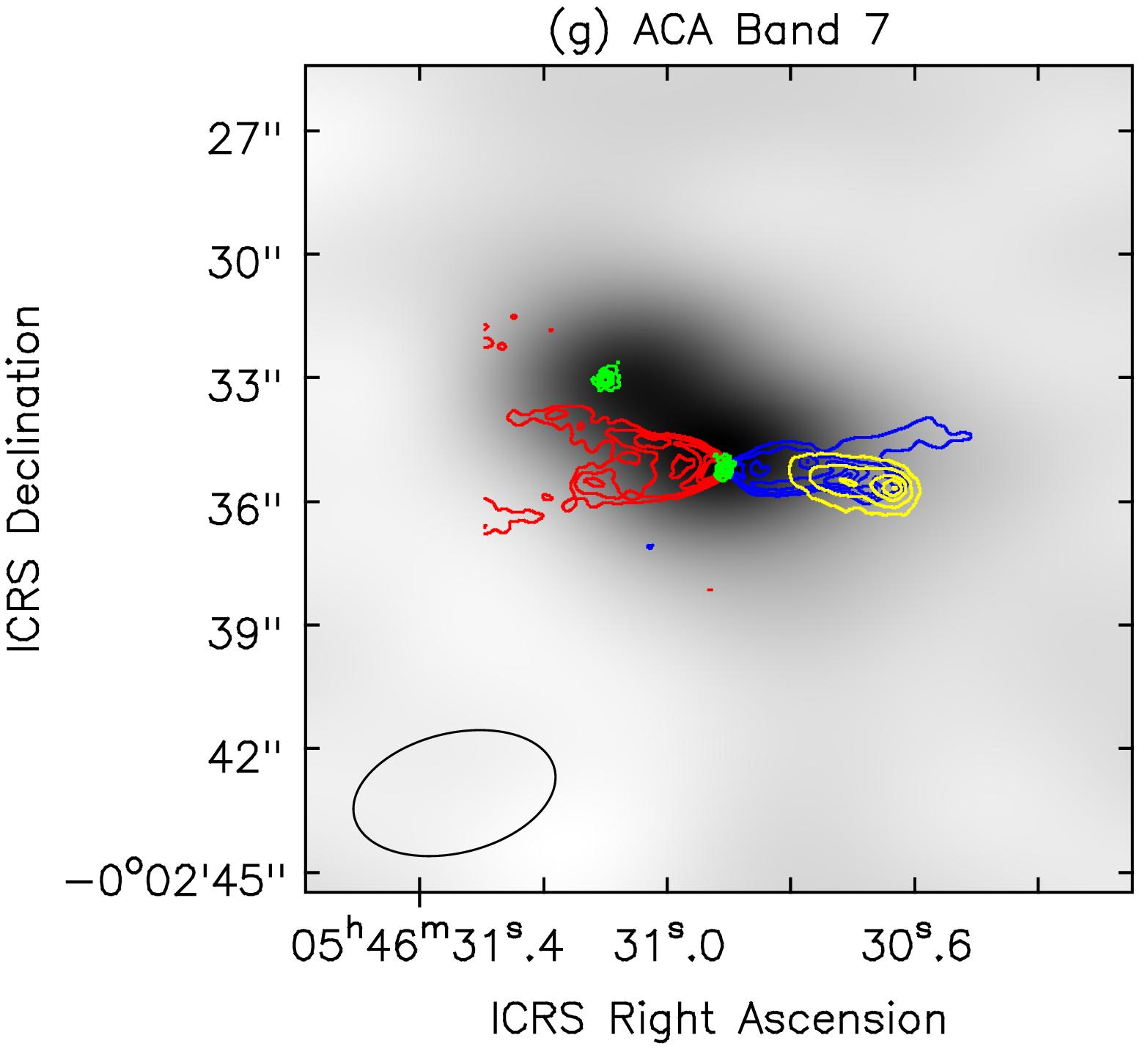}
    \includegraphics[width=0.33\textwidth]{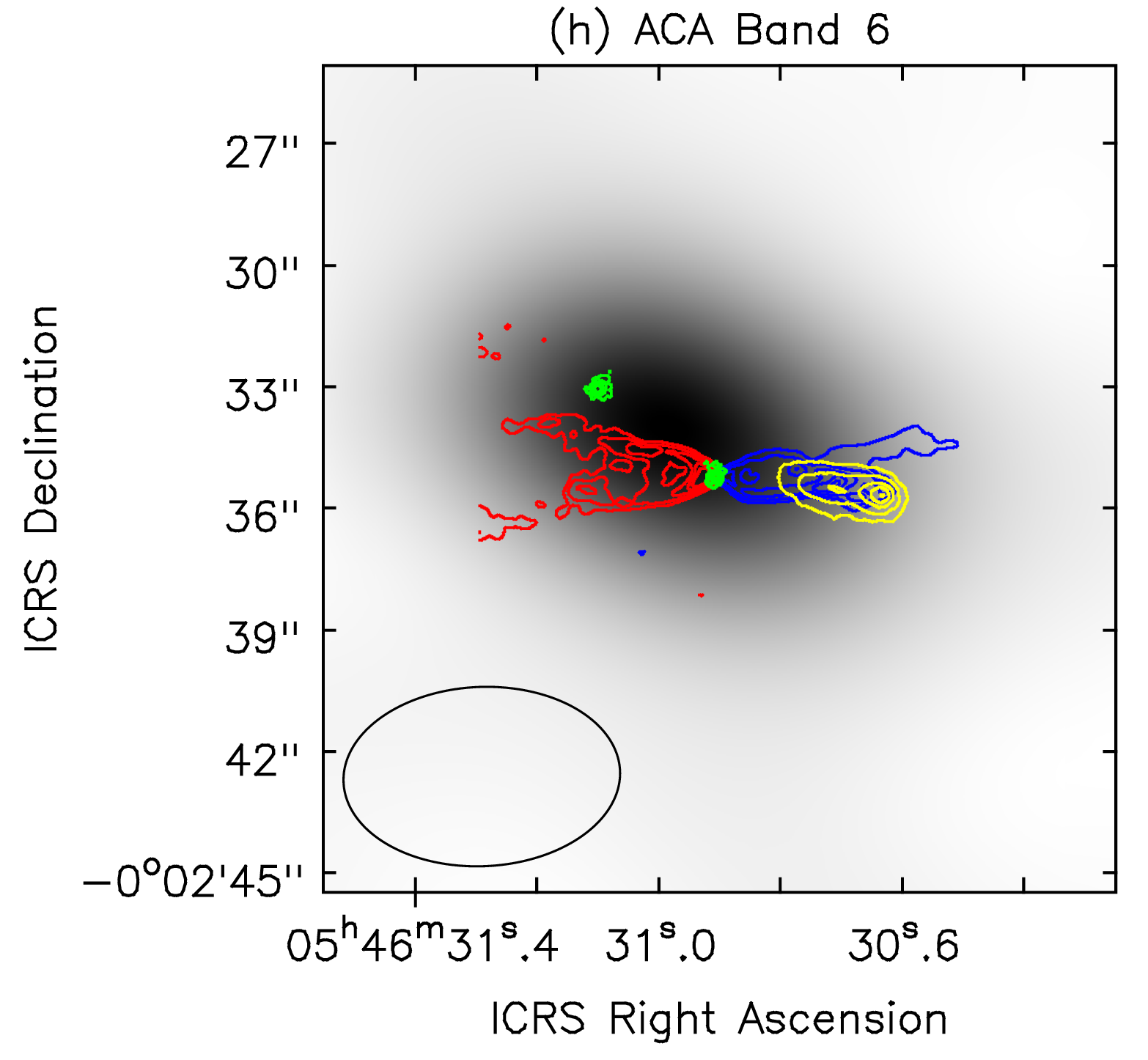} 
    \caption{The $^{12}$CO 3-2 (blue and red contours, following Figure~\ref{fig:almaoutflow}), 0.89 mm continuum emission (green) and 
     GNIRS Ks-band contours (yellow) map, superimposed on (a) IRAC 3.6~$\mu$m (Band 1) image; (b) IRAC 4.5~$\mu$m (Band 2) image; (c) IRAC 5.8~$\mu$m (Band 3) image; (d) IRAC 8~$\mu$m (Band 4) image; (e) MIPS 24~$\mu$m band image; (f) PACS 70~$\mu$m band image; (g) ACA Band 7 (0.89~mm) continuum image ; (h) and ACA Band 6 (1.3~mm) continuum image.  }
    \label{fig:almaoutflow_irimages}
    \end{center}
\end{figure*}

In the previous section, we demonstrate that HOPS 373 is variable in the mid-IR and sub-mm and interpret the sub-mm variability in terms of an accretion burst.  In this section, we dissect the source into distinct structures to understand the variability.  In the sub-mm, HOPS 373 has been resolved into two compact continuum sources with a small-scale CO outflow from the SW component (Figure~\ref{fig:almaoutflow}).  These structures together serve as benchmarks for interpreting the emission sources across wavelengths (Figure~\ref{fig:almaoutflow_irimages}).  In this section, we first describe the binary components and the outflow, and then describe the locations of emission in different bands.

For each instrument and image, the precise location of the emission related to HOPS 373 is determined by centroiding nearby compact sources (Table~\ref{tab:coordinates}).  The absolute images are then registered to the WISE astrometric frame. Appendix~\ref{appendix-a} describes the details of these positional shifts.

\begin{table}[!ht]
    \caption{Centroid positions}
    \label{tab:coordinates}
    \centering
    \begin{tabular}{llrr}
    Instrument & $\lambda$ &  RA$^a$ & Dec$^a$  \\
& & 05 46 00 +X$^b$ & -00 02 00 -Y$^b$\\
        \hline 
  2MASS & $1.2$--$2.5$ $\mu$m &      30.648  & 35.02  \\
 WISE & $3$--$25$ $\mu$m &       30.705  & 35.23  \\
  IRAC & 3.6 $\mu$m &      30.686  & 35.26  \\
 IRAC & 4.5 $\mu$m  &      30.692  & 35.26  \\
 IRAC & 5.8 $\mu$m   &     30.709  & 35.16  \\
 IRAC & 8.0 $\mu$m  &    30.698  & 35.07  \\
 MIPS & 24 $\mu$m  &     30.726  & 35.10  \\
 PACS & 70 $\mu$m &       30.859  & 35.31  \\
  PACS & 160 $\mu$m   &   30.855  & 34.87  \\
  SCUBA-2 &450 $\mu$m    &  30.902  & 34.17  \\
 SCUBA-2 &850 $\mu$m  &     30.913  & 34.43 \\
  ALMA ACA & 1.33 mm   &    30.973  & 34.11  \\
        \hline 
        \multicolumn{4}{l}{$^a$For uncertainty of centroid positions, see Appendix~\ref{appendix-a}}\\
        \multicolumn{4}{l}{$^b$As example, the first source is 05 46 30.648 -00 02 35.02}\\
    \end{tabular}
\end{table}

\begin{figure}
    \includegraphics[width=0.48\textwidth]{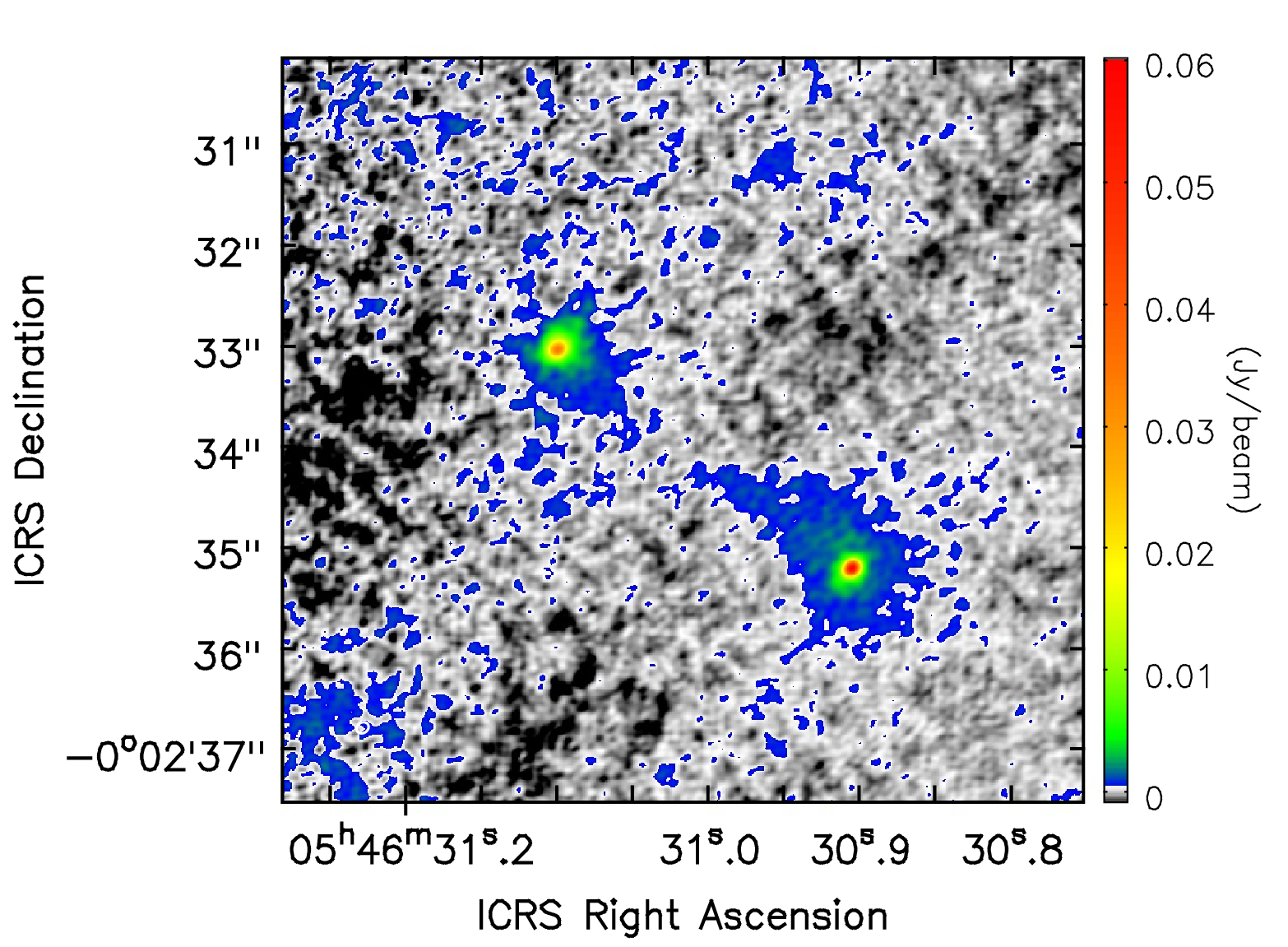}
    \caption{ALMA Band 7 (0.89 mm) continuum image of HOPS 373. The NE and SW sources show a weak extended continuum emission surrounding the compact emission.}
    \label{fig:weakcont}
\end{figure}

\subsection{Binarity in the sub-mm}
\label{sec:binary}

High-resolution mm imaging shows two distinct continuum sources with a separation of $3\farcs6$ and a position angle 232$^\circ$ \citep[Figure~\ref{fig:weakcont},][]{tobin15,tobin20}, using the NE component as the reference point. The continuum sources seem to be surrounded by a large envelope, which is resolved out in the high-resolution image. In Figure~\ref{fig:weakcont}, the diffuse emission surrounding the sources is associated with this unseen larger envelope. 

The binary components of the 0.89 mm emission are HOPS 373 NE centered at 05:46:31.100 -00:02:33.02 and HOPS 373 SW at 05:46:30.905 -00:02:35.20 \citep{tobin20}. The integrated flux of the NE and SW sources are 85.0$\pm$3.3 mJy and 81.8$\pm$1.7 mJy, respectively, as measured from the integrated flux in a 2D Gaussian fit of {\it imfit} in CASA.
 The total 0.89 mm emission\footnote[5]{The fluxes measured by \citet{tobin20} are $\sim 15\%$ higher from the same observation, due entirely to differences in subtracting the nearby emission.  The compact emission in the two components is located on top of diffuse emission on scales small enough that it does not resolve out.}  in the two continuum sources is $\sim0.17$ Jy, or about 9\% of the integrated flux in the 850 $\mu$m emission seen in SCUBA-2. Most of the emission is resolved out with the small beam.

The ACA 1.33 mm continuum emission, obtained with a 6\farcs8$\times$4\farcs4 beam, has a centroid of 05:46:30.971 -00:02:34.13, between the two compact components, is elongated in the position angle of those components, and has a flux of 362$\pm$9 mJy.  In the Rayleigh-Jeans limit, this flux at 1.33 mm implies a flux of 886 mJy at 850 $\mu$m, or 45\% of the total emission from the source.   In a simulation of the ACA observation, if only the two point sources are present, they would be marginally resolved.   However, the ACA continuum image shows an emission peak between the two sources, which requires some diffuse emission surrounds the two compact sources.

During the burst, the centroid of the residual (burst--quiescent) emission in the SCUBA-2 850 $\mu$m images is located $1\farcs1$ closer to the SW component than the quiescent emission, consistent with expectations if the SW component is the source of the variability.  The residual emission also has a compact profile, consistent with a FWHM\footnote[6]{The precise FWHM is uncertain since the profile width is much narrower than the $14\farcs1$ beam size of SCUBA-2 at 850 $\mu$m.} of $\sim 7\arcsec$, while the quiescent emission has a FWHM of 15\arcsec\ in the SCUBA-2 data.

\subsection{CO Outflows}

The ALMA images from the ACA and the 12 m array reveal large-scale outflows in CO 2--1 emission (see Figure~\ref{fig:almaoutflow} and also channel maps from the high-resolution 12m array observations of the CO outflow presented in Appendix A).  At large scales, the blueshifted outflow is located to the southeast of the source while the redshifted outflow is located to the northwest \citep[see also][]{mitchell01,nagy20}.  At small scales, the outflow direction is the opposite: the blueshifted outflow is launched to the west, while the redshifted outflow is launched to the east.  The small-scale outflow is driven by the southwestern component.

\begin{figure}
    \centering
    \includegraphics[width=0.48\textwidth]{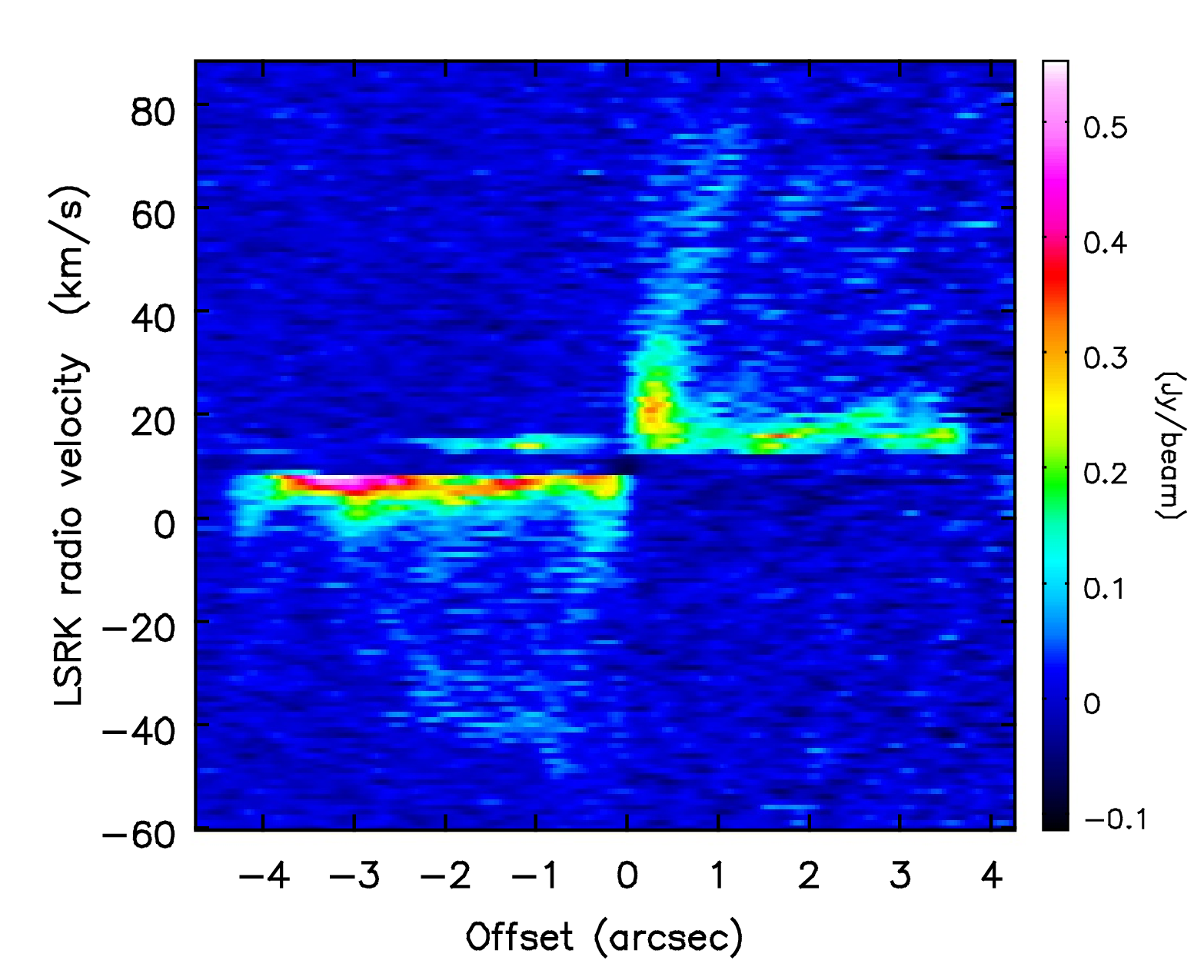} \\
    \includegraphics[width=0.45\textwidth]{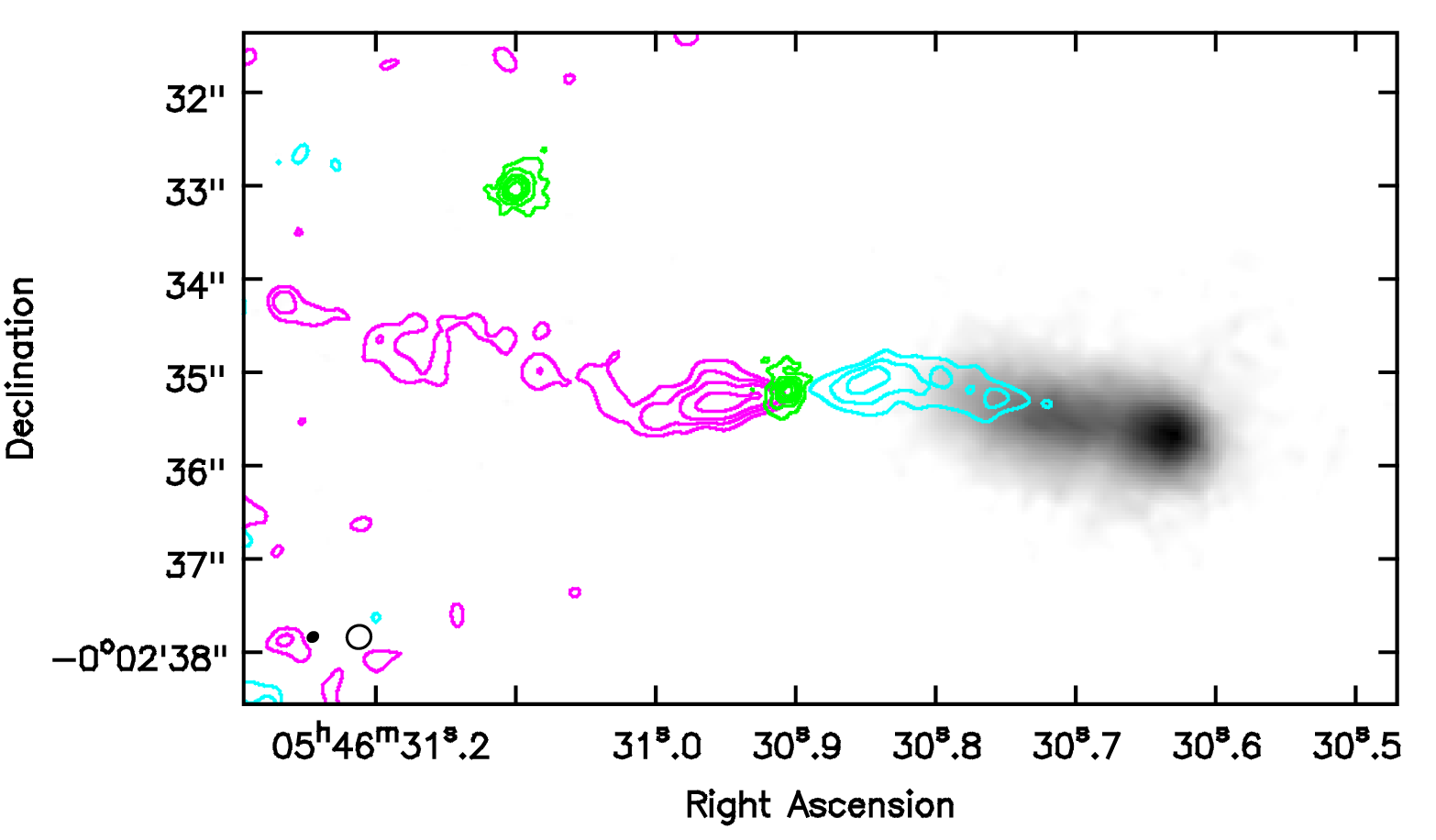}
    \caption{{\it Top:} Position-velocity diagram of CO 3-2, centered at  HOPS 373 SW and 
    aligned along the outflow axis, with a position angle of 90$^{\circ}$.  Extremely high velocity bullets are detected within $1\arcsec$ of the star in both the red and blueshifted jets.  The velocity shown here is not corrected for the source velocity of $\sim 10$ \kms.
     {\it Bottom:} $^{12}$CO (3-2) integrated intensity and 
    the 0.89 mm continuum maps (green contours) overlaid on 
    the GNIRS K-band acquisition image.  Cyan and 
    magenta contours show CO emission integrated over 
    the velocity ranges from -50.0 to -17.0 km s$^{-1}$ 
    and 38.0 to 75.0 km s$^{-1}$, respectively, with contour levels set to (3, 5, 7, and 10) $\times \sigma$, with 
    $\sigma = 0.19$ Jy beam$^{-1}$ km s$^{-1}$.}
    \label{fig:pvdiag}
\end{figure}

The position-velocity (PV) diagram (the upper panel in Figure~\ref{fig:pvdiag}) along the position angle of 90$^{\circ}$ centered on HOPS 373 SW shows that the blue and redshifted outflows extend to 4\arcsec\ from the source.  CO emission is also detected in extremely high-velocity components, or bullets, in a jet-like collimated morphology with velocities of $-60$ and 65 km s$^{-1}$ in each direction, relative to the source velocity of 10 \kms (the lower panel in Figure~\ref{fig:pvdiag}). 
 Such bullets are commonly detected in CO emission in jets from very young protostars \citep{tychoniec19}.
However, since the CO (3-2) line observation was carried in 2016, the jet-like feature is not directly related to the recent outburst.

The northeastern component is not associated with any detected small-scale outflow.  Any outflow from the northeastern component would have to either be low-velocity and absorbed by the cloud or large enough that it is resolved out.  The large-scale outflow is either a historical remnant of an outflow driven by the northeastern component, or the southwestern component has precessed such that the wind direction changed from red to blue.  A large change in outflow direction has been identified in another very young protostar, IRAS 15398-3359 \citep{okoda21}.

\begin{figure*}
\begin{center}
\includegraphics[width=0.45\textwidth]{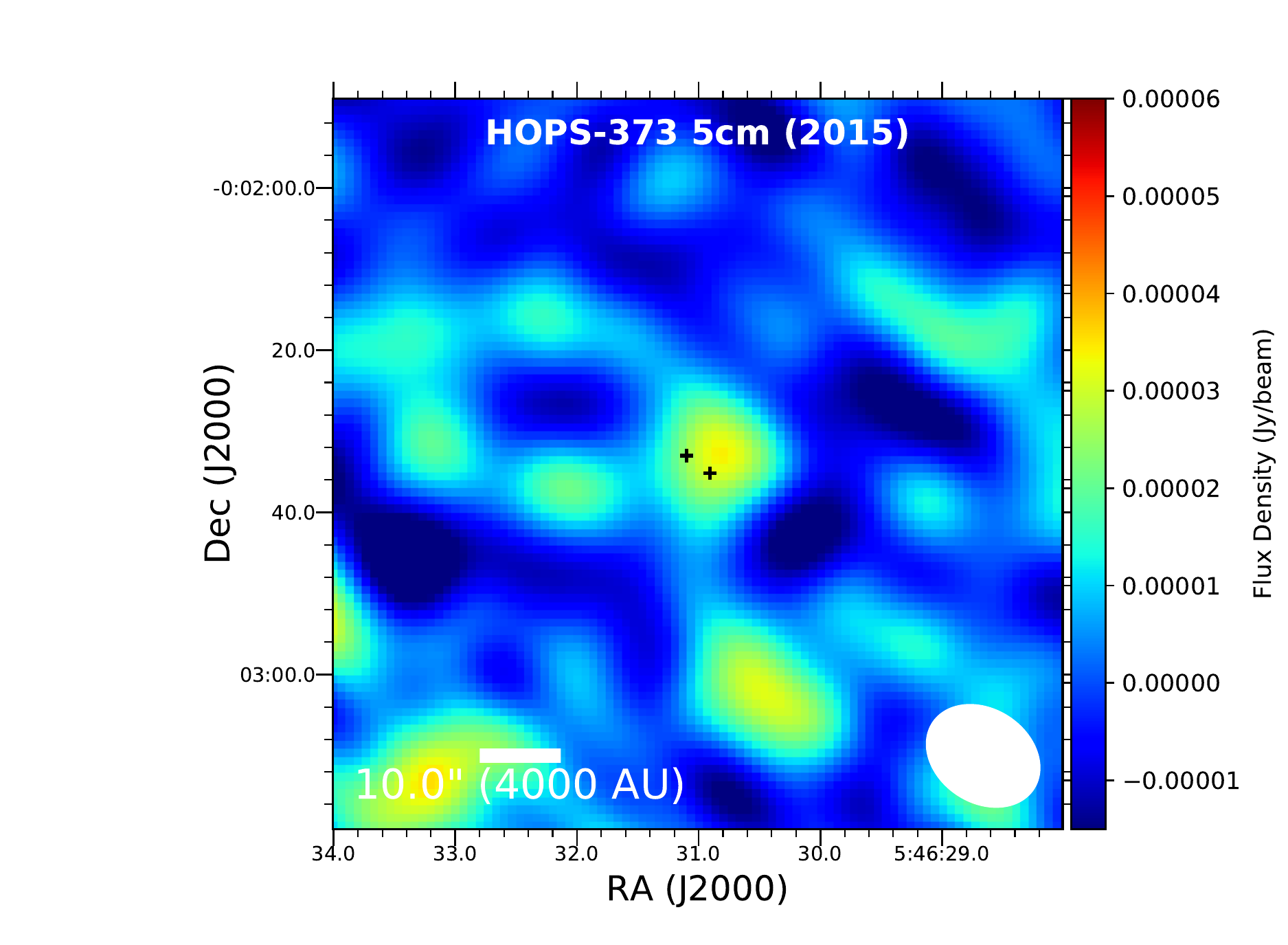}
\includegraphics[width=0.45\textwidth]{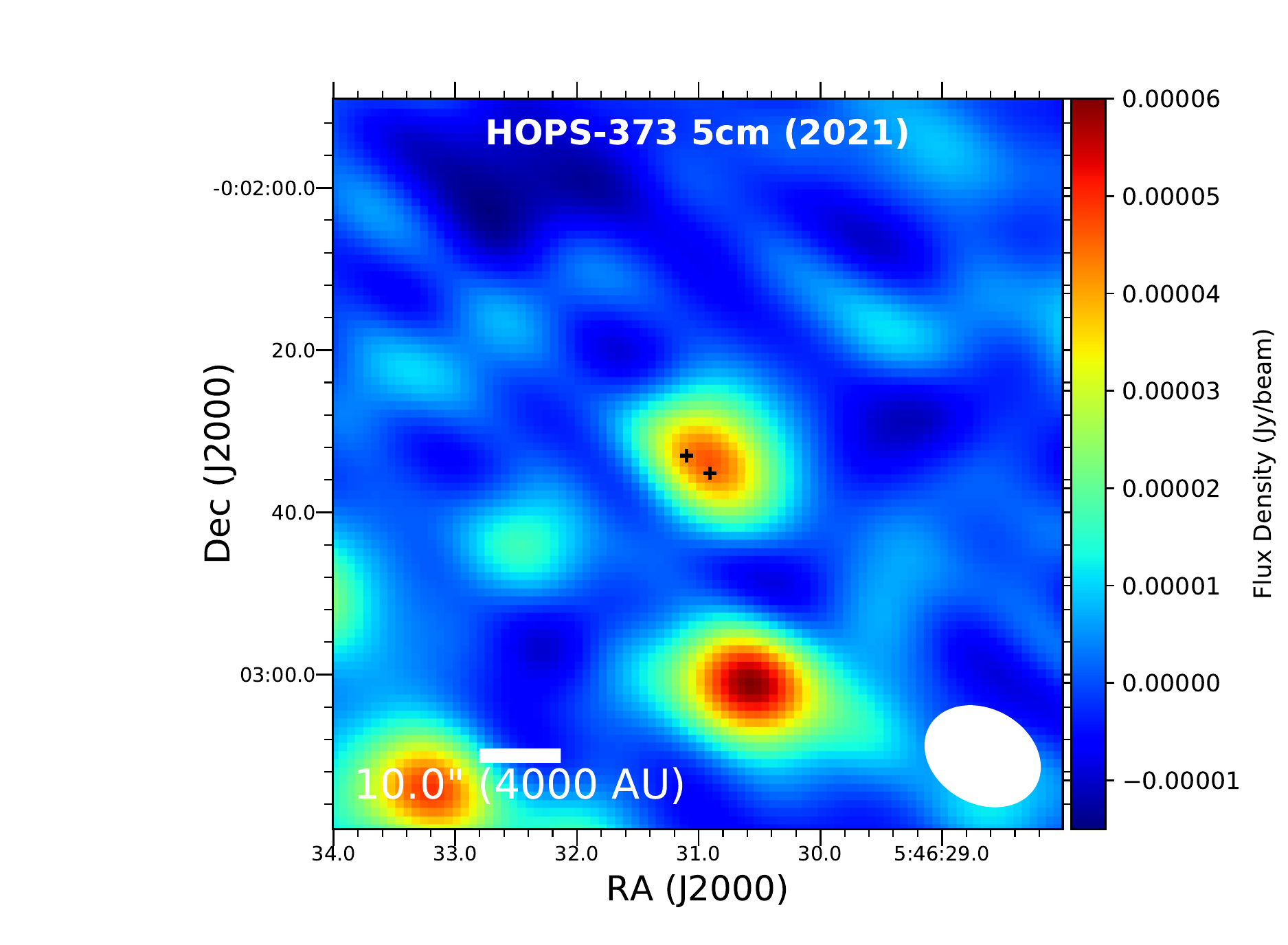}
\end{center}
\vspace{-5mm}
\caption{VLA images toward HOPS 373 at 5~cm; the 2015 data are on the left and 
the 2021 data are on the right.
The images clearly show that the 2021 data are brighter at 5~cm, potentially a calibration issue. The positions
of the SW and NE compact continuum components of HOPS 373 are marked by crosses in each panel.
The beam in each image is 14\farcs77$\times$11\farcs40.
}
\label{VLA-5cm}
\end{figure*}

\begin{figure*}
\begin{center}
\includegraphics[width=0.45\textwidth]{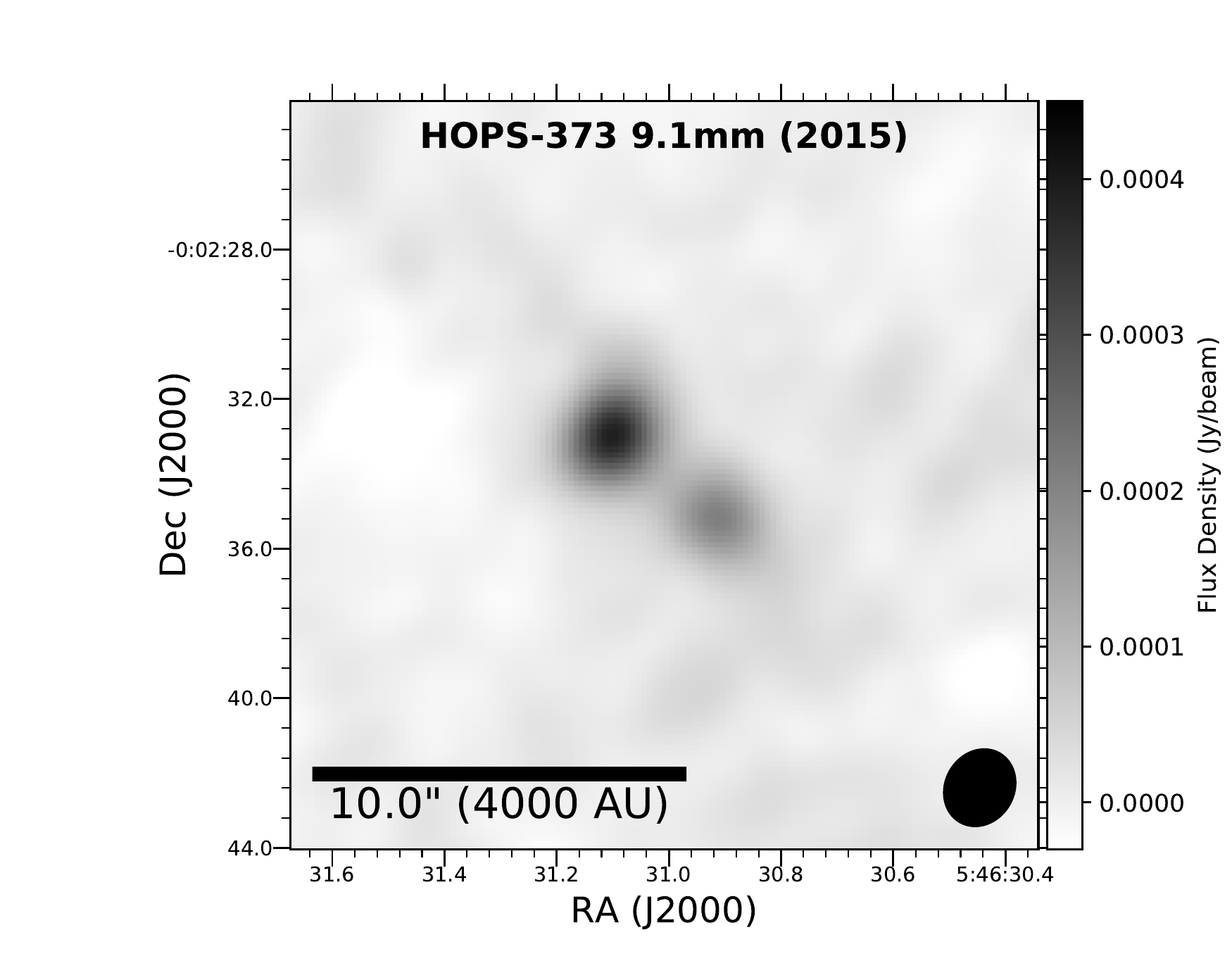}
\includegraphics[width=0.45\textwidth]{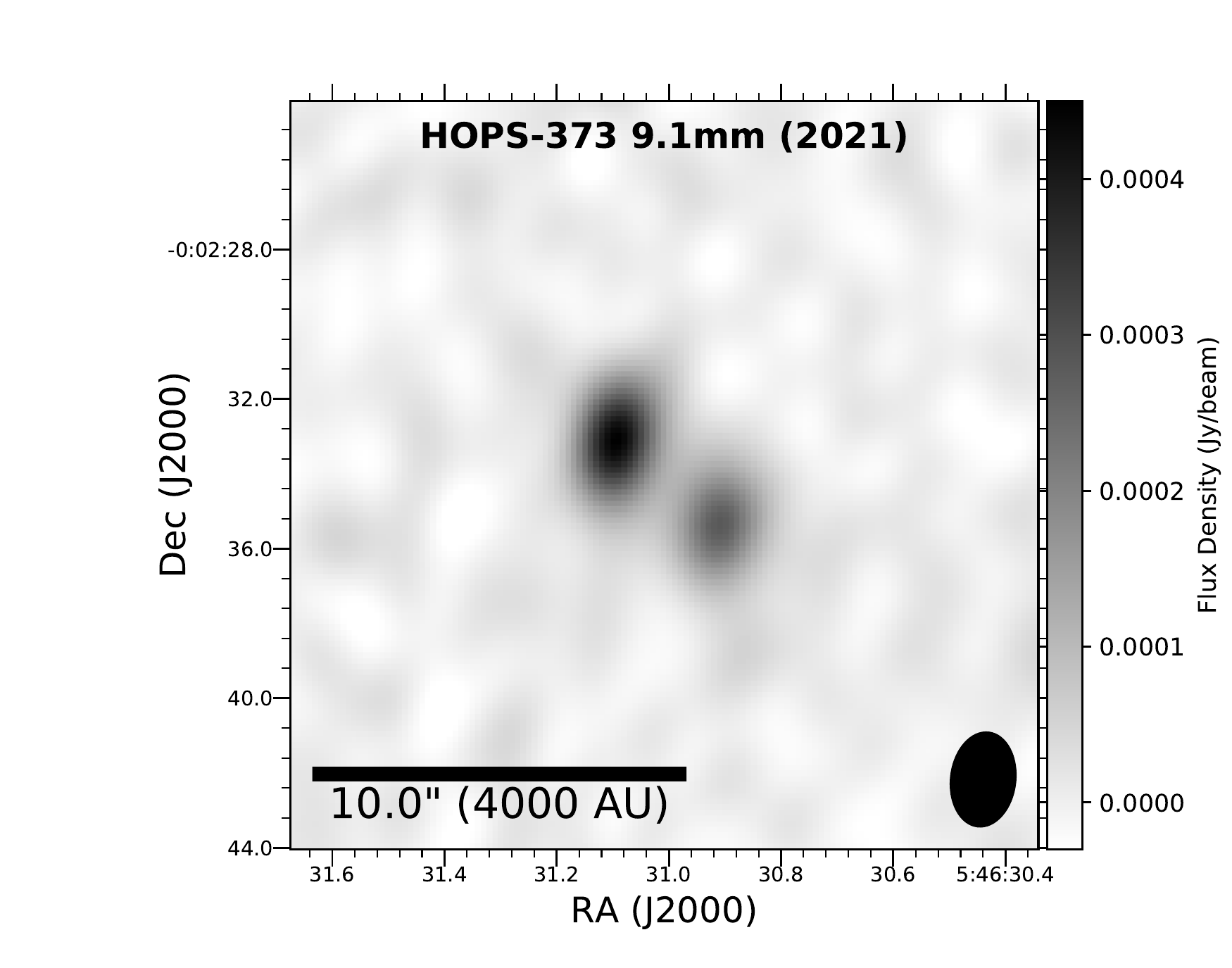}
\includegraphics[width=0.43\textwidth]{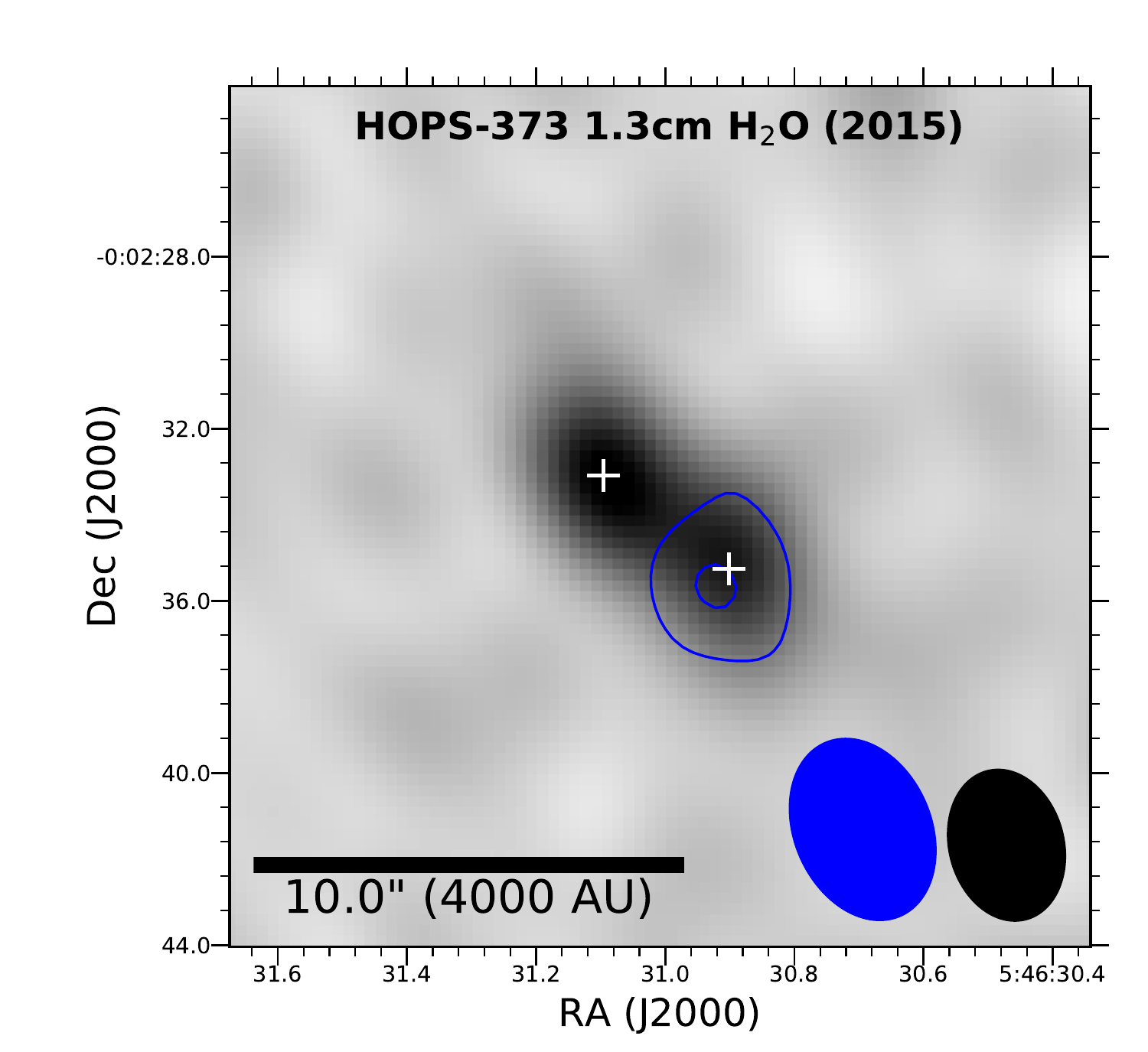}
\includegraphics[width=0.43\textwidth]{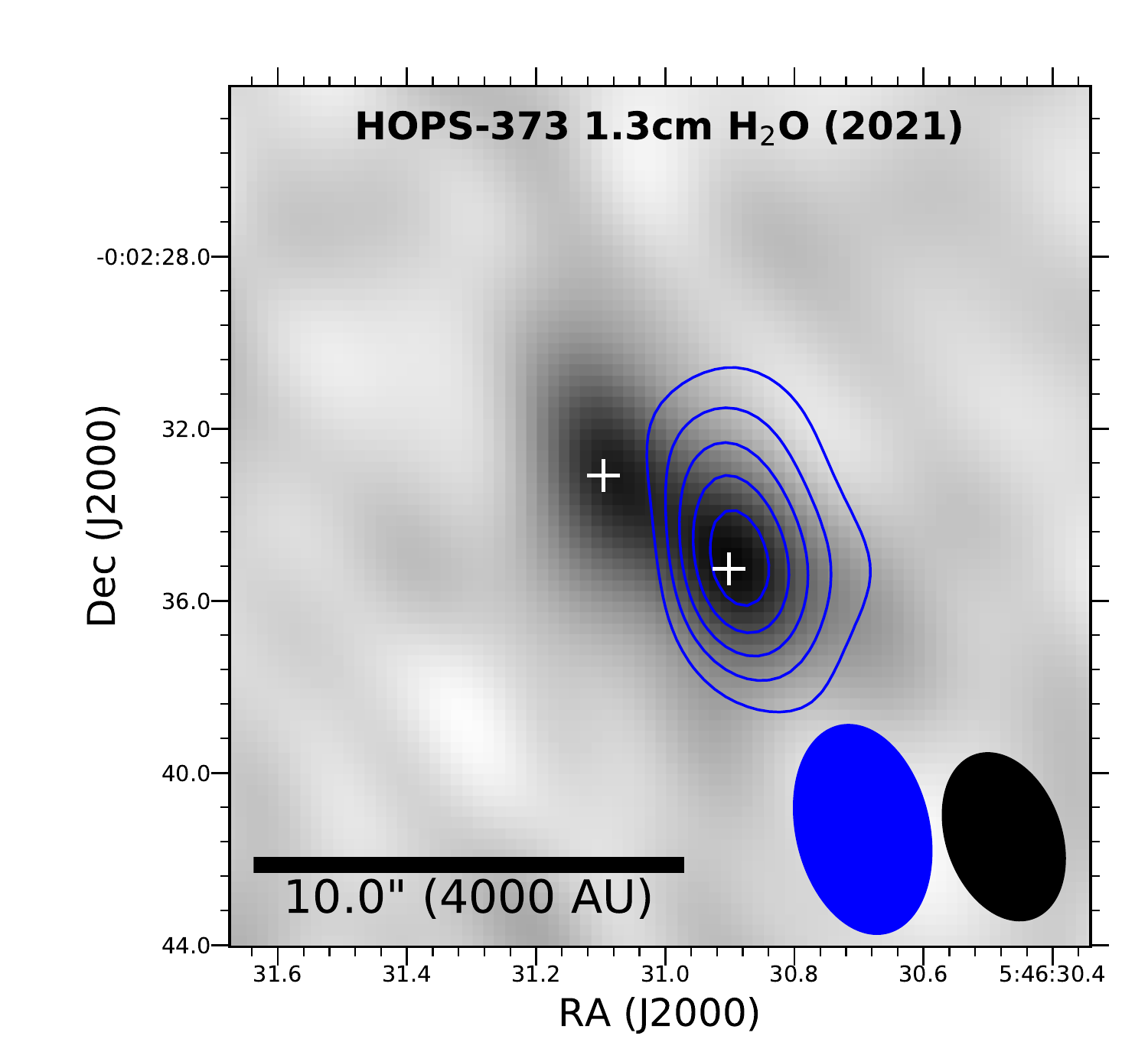}
\end{center}
\vspace{-3mm}
\caption{{\it Top:}  VLA images of HOPS 373 at 9.1~mm in 2015 (left) and 2021 (right).  Although both sources appear brighter in 2021, possibly because of absolute calibration error, the ratio of flux densities indicates that the SW source is brighter in 2021.  
{\it Bottom:} VLA images toward HOPS 373 at 1.3~cm with H$_2$O 
maser contours overlaid, in 2015 (left) and 2021 (right), both shown with the same color stretch. The positions
of the northeast and southwest components of HOPS 373 are marked by crosses.
The H$_2$O
maser emission is associated with the southwest source and brightened substantially in 2021
relative to 2015; the contours
start at 3$\sigma$ and increase on 2$\sigma$ increments where $\sigma_{2015}=1$~(mJy~\kms)/beam and $\sigma_{2021}=4$~(mJy~\kms)/beam demonstrating a 7$\times$ increase
in maser emission between 2015 and 2021.
The 2015 maser emission is integrated from -15.5 to -14.5~\kms, and
the 2021 maser emission is integrated from 9.5 to 10.5~\kms.
}
\label{VLA-water-maser}
\end{figure*}

\subsection{Radio emission from 0.9--5 cm}

The archival observations from the VLA were obtained in 2015 October, just prior to the start of JCMT monitoring.  At that time, the protostar was likely in a quiescent state, though brighter than the minimum.  Our follow-up C-band observations in 2021 occurred $\sim$6 months after the current and on-going burst began.

{\it Emission at 5 cm:} The large VLA beam encompasses both of the distinct sub-mm continuum sources.  The flux density of HOPS 373 was 32$\pm$6~$\mu$Jy in 2015 and 47$\pm$6~$\mu$Jy in 2021 (see Table~\ref{tab:vla-fluxes}), apparently brighter during the burst but consistent with no change to within $2\sigma$.
  As detailed in Section~\ref{sec:obs-vla-analysis}, 
there is a systematic offset in the distribution of flux density differences and ratios measured
for other sources in the field between 2015 and 2021. Therefore we cannot be confident
that the difference in flux density at C-band represents a true variation.
These uncertainties also do not include
the 5-10\% absolute flux density calibration uncertainty.

\begin{table}[!b]
\begin{center}
\caption{VLA Flux Densities$^a$}
\begin{tabular}{lccc}
           & Flux Density & Flux Density & Flux Density \\
$\lambda$  & 2015/2016 & 2021 Mar/Apr & 2021 May\\
(cm)      & ($\mu$Jy)     &  ($\mu$Jy) & ($\mu$Jy) \\
\hline
0.9 & 761$\pm$24 & 960$\pm$30 & \nodata \\
1.3 & 323$\pm$10 & 337$\pm$38 & \nodata \\
5.0 & 32$\pm$6   & 47$\pm$6   & 45$\pm$6\\
\hline
NE\\
\hline
0.9 & 486$\pm$17 & 569$\pm$21 & \nodata \\
1.3 & 170$\pm$7 & 175$\pm$27 & \nodata \\
\hline
SW\\
\hline
0.9 & 275$\pm$17 & 391$\pm$21 & \nodata \\
1.3 & 153$\pm$7 & 162$\pm$27 & \nodata \\
\hline
H$_2$O Maser
   & (mJy~\kms)     &  (mJy~\kms) & \\

\hline
1.349 & 5.5 $\pm$0.8 & 56.1$\pm$7 & \nodata\\
\hline
\multicolumn{4}{l}{$^a$Measured from fits with Gaussian profiles}
\end{tabular}
\label{tab:vla-fluxes}
\end{center}
\end{table}

{\it Emission at 1.3 cm:} At 1.3~cm the two continuum sources are marginally resolved when imaged at the same
resolution in 2015 and 2021 (Table~\ref{tab:vla-fluxes}). The NE component appears brighter than 
the SW at 1.3~cm in 2015,
while in 2021 the SW source appears to be the brighter (Figure \ref{VLA-water-maser}), though with flux densities consistent within the uncertainties.

{\it Emission at 0.9 cm:} At 0.9 cm, the NE source is brighter than the SW
source in both 2015 and 2021, but the sources are both overall brighter in 2021 relative
to 2015 (Figure \ref{VLA-water-maser}). Using the well-resolved and high S/N detections of each source at 9~mm, we 
find a NE/SW flux density ratio of 1.77$\pm$0.13 in 2015 and 1.45$\pm$0.12 in 2021, suggesting that the SW source brightened; again a constant ratio cannot be ruled out at the 2$\sigma$ level\footnote[7]{These ratios ignore the absolute flux calibration, since that uncertainty is applied in the same way to both targets.}. Most of the 9-mm emission is produced by thermal dust, so the brighter emission from the SW source indicates an increase in the dust temperature, in agreement with the JCMT sub-mm observations.

{\it Emission in water masers:} 
 The
water maser emission is only associated with the SW source, with a spatial position that did not change between 2015 and 2021 (Figure \ref{VLA-water-maser}). The
flux density increased by a factor of $\sim$10 from 2015 to 2021 (Table \ref{tab:vla-fluxes}) and changed in velocity from -15~\kms\ in 2015 to $\sim$10~\kms\ in 2021 (relative to the local standard of rest and not corrected for the source velocity). The maser lines are  narrow, with FWHM of $<$1~\kms\ in both epochs. The previously published single-dish maser observations toward the region detected maser activity at substantially 
higher flux densities and at velocities of $\sim$20~\kms\ \citep{haschick83}.

\subsection{Near- and Mid-IR emission from the outflow}
\label{sec:irloc}

The {\it WISE} $W1$ and $W2$ emission is centered $3\farcs1$ to the west of the ALMA-observed southwestern sub-mm continuum source and is spatially unresolved due to the $\sim 6^{\prime\prime}$ point spread functions.  The two epochs when HOPS 373 was bright occurred  in Sept.~2019 and Sept.~2020.  Compared to the centroid position from previous September epochs, in 2019 the centroid position in $W1$ is $0\farcs176$ W and $0\farcs093$ N of previous Sept. epochs; in 2020 the centroid is $0\farcs122$ W and $0\farcs077$ N.  In $W2$, the offsets are $0\farcs132$ west and $0.068$ north in 2019 and $0\farcs133$ W and $0.079$ N in 2020.  Based on previous epochs, each position has a $1\sigma$ uncertainty of $0\farcs025$.  These centroid positions include the quiescent emission and the emission added from the burst.  After subtracting the quiescent emission, the position of the burst would be even further away from the quiescent centroid.

The $K$-band images of HOPS 373, mostly H$_2$ emission (see Section~\ref{sec:molecular}), are dominated by a compact source 4.2$''$ west of the southwestern continuum component, with a tail of fainter emission extending back to the northeast toward the source (Figure~\ref{fig:almaoutflow}). The IRAC 4.5 $\mu$m emission also shows very faint emission to the east of this source, likely associated with the redshifted outflow (Figure~\ref{fig:HOPS373_UKIRT_K}).

The $K$-band tail follows the southern half of the blueshifted CO emission.  The $K$-band compact emission is located just beyond the extent of the CO emission, perhaps indicating a bow-shock at the end of the jet \citep{varricatt10}.

The $J$- and $H$-band emission seen with UKIRT is spatially consistent with the $K$-band emission and likely traces molecular and atomic line emission from the outflow.  
The total source brightness is $J=21.18\pm0.4$ and $H=18.78\pm0.09$, but the emission is too faint to divide into individual epochs or to separate into different components. \citet{spezzi15} reported $H=19.34$ from the VISTA Orion Survey, fainter than measured here, either because of variability or a smaller aperture over which the emission was measured.

\subsection{Far-IR emission from the SW component}

The {\it Herschel} PACS 70 and 160~$\mu$m emission is nearly centered on the HOPS 373 SW (Figure~\ref{fig:almaoutflow_irimages}). The binary is not resolved and the emission is not elongated in any direction. The total fluxes at 70 $\mu$m and 160 $\mu$m are 5.46 Jy and 36.3 Jy, respectively \citep{Furlan2016}. The far-IR emission seems to be mainly associated with HOPS 373 SW, attributed to the thermal emission from the HOPS 373 SW envelopes rather than the scattered light from the cavity wall or the shocked emission, seen in shorter wavelengths.

\subsection{Summary of source morphology}

The HOPS 373 protostar consists of two compact dust sources, HOPS 373 NE and SW, separated by $3\farcs6$, corresponding to a projected separation of 1500 AU at the distance of 428 pc.  Both a molecular outflow and maser emission are associated with the SW component.  No small-scale outflow is seen from HOPS 373 NE.

In Figure~\ref{fig:almaoutflow_irimages}, we identify the location where each emission component is detected.  The near- and mid-infrared emission centroids are all consistent with emission located 
along the blueshifted outflow from the central source, where the opacity should be reduced. 
At 70 $\mu$m, the emission is located nearly on the position of the SW source. In the ALMA ACA Band 6 imaging at 1.33 mm, the emission centroid is located between the sources, as is the emission at $\sim$850 $\mu$m from SCUBA-2.
The NE source does not show any emission feature over the mid-infrared images. However, as the wavelength gets longer through far-infrared to sub-mm, the thermal emission from each individual dust component becomes significant and the flux of the NE source increases.  

The variability in the infrared continuum and in the maser emission, along with the detected outflow, conclusively demonstrates that the SW component is the component that is actively accreting and variable.   The variability in maser emission is consistent with past associations between maser emission and accretion \citep[e.g.][]{burns15,burns20,hirota21,stecklum21}.  Assuming that the maser variability is related to the accretion event, then the water masers are responding to the increase in radiation field.  The masers may also brighten due to an increase in outflow activity, any such change would occur over longer timescale than observed for HOPS 373 and would mean that the correlation between maser emission and sub-mm emission is only a coincidence.

The NE source is only bright in the sub-mm and is not detected in the mid-IR.  The luminosity must be very low.  The compact object is the size of a disk, but the central source must be low-mass and is not actively driving any outflow.  The large-scale outflow may be a remnant outflow from the NE source or may be an outflow from the SW source that changed direction.

\section{Molecular Emission in the Near-Infrared}
\label{sec:molecular}

\begin{figure*}[!th]
    \includegraphics[width=0.95\textwidth]{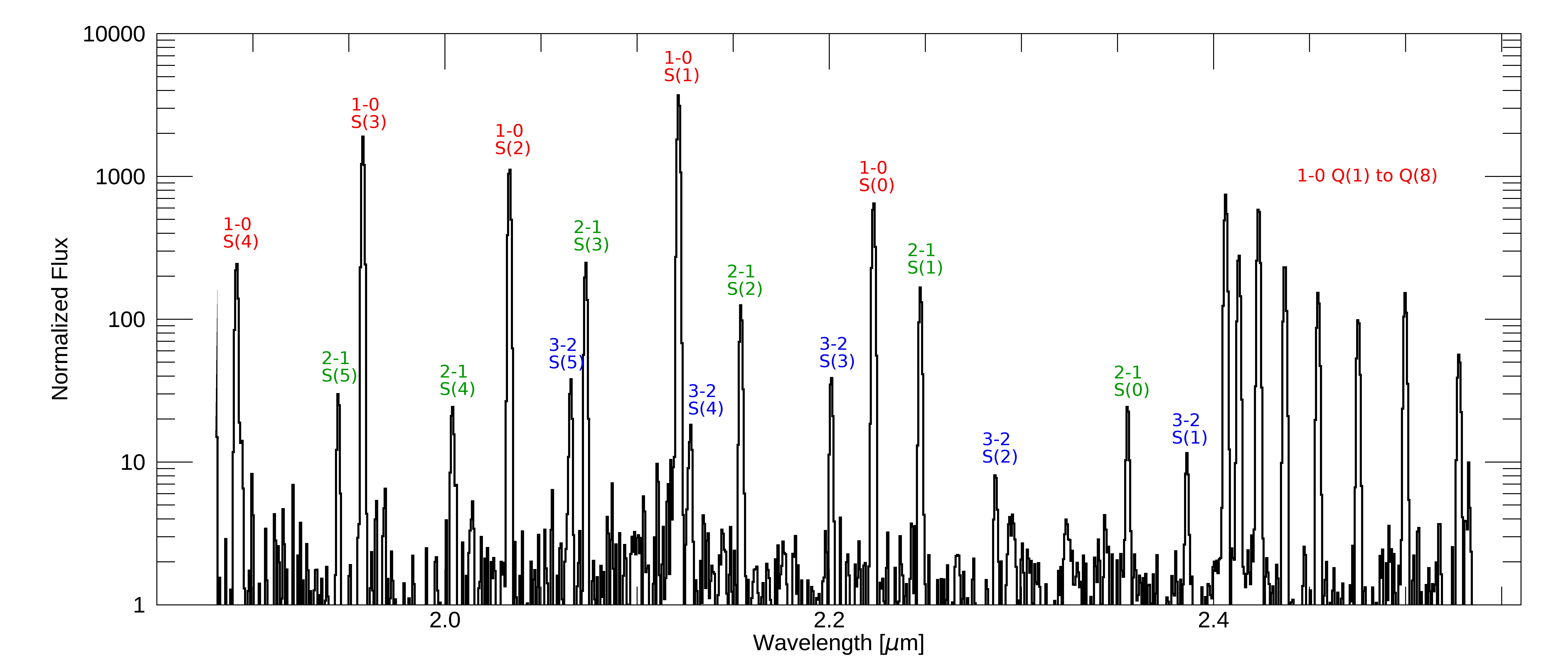}
    \caption{The normalized K-band spectrum of HOPS 373. The rovibrational transitions of H$_2$ dominate the K-band emission. The rovibrational transitions of $v$=1-0, 2-1, and 3-2 are denoted with red, green, and blue labels above the lines.}
    \label{fig:gnirs_kband}
\end{figure*}

The compact near-IR emission is located $\sim$4.3\arcsec\ away from HOPS 373 SW and has $\sim$1\arcsec\ in diameter. The elongated emission feature extends $\sim$3\arcsec\ to the east, towards the driving source.   We obtained the near-infrared spectrum from the compact source by placing the 0.3\arcsec$\times$7\arcsec\ slit nearly perpendicular to the extended near-IR emission feature (see Figure~\ref{fig:almaoutflow}).
The near-IR spectrum of the compact $K$-band source is dominated by rovibrational H$_2$ emission, detected in vibrationally excited lines up to $v=3$ (Figure~\ref{fig:gnirs_kband}).  Table~\ref{tab:h2list} provides the intensities of the H$_2$ lines, measured from fitting Gaussian profiles to the lines.  The flux error is estimated from the standard deviation in the continuum near each line. The H$_2$ 1-0 S(1) line center is shifted from the central wavelength by about -22 km s$^{-1}$ in LSR velocity, or $\sim -32$ \kms\ relative to the source velocity.

For an optically thin line, the intensity is proportional to the column density $N_{vj}$ in a given rovibrational level as follows 
\begin{equation}
    \frac{I_{\lambda}(v, J)}{\Omega} = \frac{Ahc}{4{\pi}{\lambda}}N(v, J)\cdot10^{-0.4A_{\lambda}}
\end{equation}
where $I$ is the line flux given in the unit of W cm$^{-2}$, $A$ is the Einstein coefficient, $h$ is the Planck constant, $c$ is the speed of light, $N$ is the column density at a given rovibrational level, and $\Omega$ is the area that the emission comes from, with 0\farcs3$\times$1\arcsec adopted here for simplicity.
The near-infrared extinction to the H$_2$ emission is estimated to be 8.4$\pm$0.1 at 1 $\mu$m from the H$_2$ line ratio of 1-0 S(1) to 1-0 Q(3) with the extinction law of A$_{\lambda}$/A$_{1}$ = $\lambda^{-2.27}$ where A$_{1}$ is the extinction at 1 $\mu$m \citep{maiz-apellaniz20}. Other H$_2$ line ratios yield $\rm{A}_1=4-10$ mag but are less reliable because they rely upon lines at wavelengths that are progressively longward of 2.4 $\mu$m, where our telluric correction is more uncertain.  For comparison, if we adopt the extinction law of \citet{wang19}, with a near-IR power-law index of 2.07, the H$_2$ 1-0 S(1) to 1-0 Q(3) line flux ratio would lead to extinctions of $A_K=1.5$ mag and $A_1=7.8$ mag.

\begin{figure}[!t]
    \includegraphics[width=0.48\textwidth]{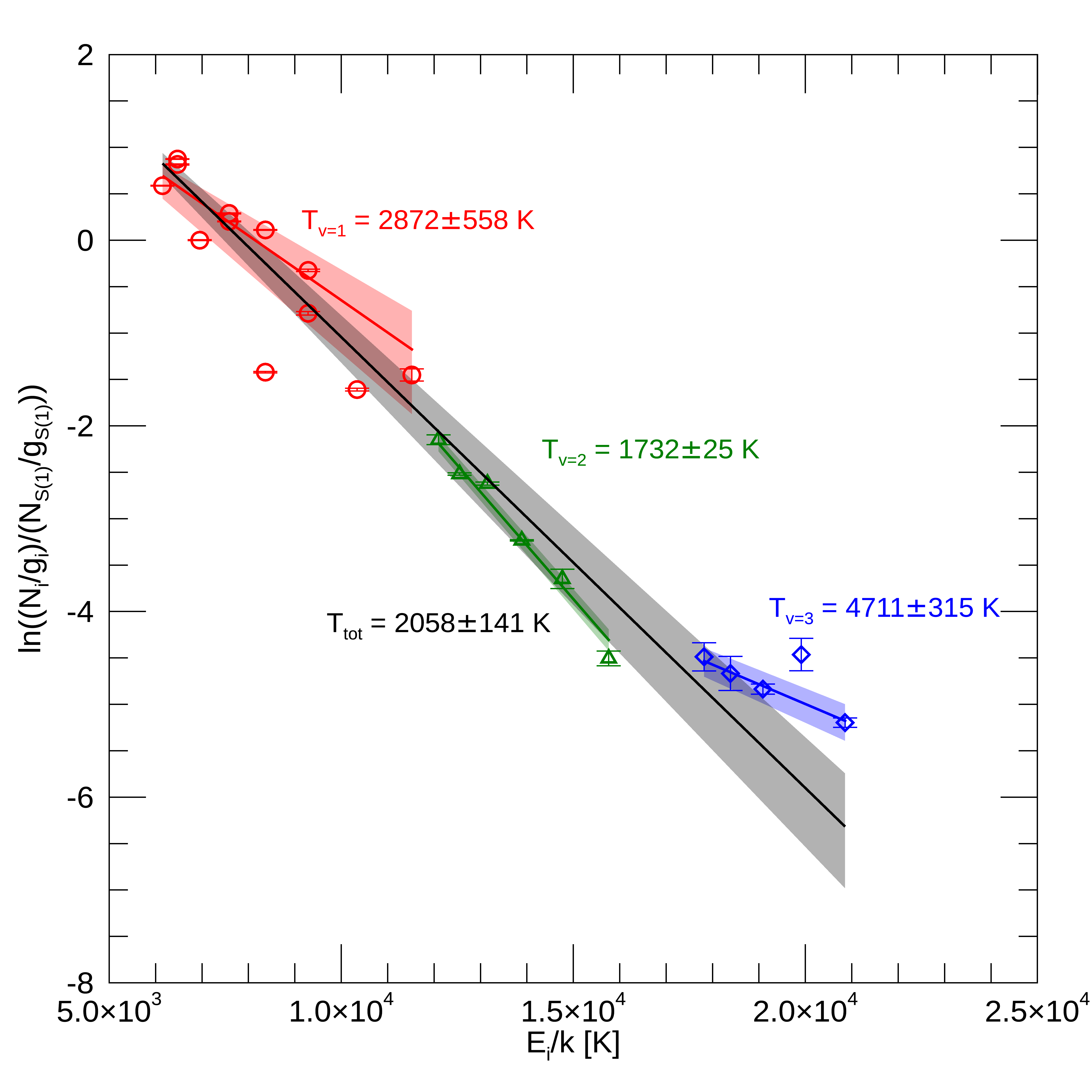}
    \caption{The excitation diagram of H$_2$ transitions for HOPS 373. The column densities per statistical weights of H$_2$ transitions are normalized by that of H$_2$ 1-0 S(1) line. The vibrational transitions of $v$=1-0, 2-1, and 3-2 are fitted by red, green, and blue straight lines, respectively. The H$_2$ 1-0 Q(5) line is excluded from the fit. The total H$_2$ transitions are fitted by the black solid line. The shaded regions show the 3 sigma error of the fitted lines.}
    \label{fig:ext.diagram}
\end{figure}

After correcting for the near-infrared extinction, the H$_2$ $v$=1-0, 2-1, and 3-2 rovibrational transitions are fitted with the excitation temperatures of about 2900, 1700, and 4700 K, respectively (Figure~\ref{fig:ext.diagram}). A combined fit to all
lines\footnote[8]{One outlier from the fit, $v$=1-0 Q(5) line at 2.4548 $\mu$m, is a factor of 2.2 weaker than expected, likely because the flux overlaps exactly with a telluric
absorption line.  The telluric absorption line is barely seen at low resolution but would be strong if resolved.  This line is ignored in our fits.}
leads to an excitation temperature of about 2100 K and a total column density of
4.4$\times$10$^{19}$ cm$^{-2}$ calculated assuming a 0\farcs3$\times$1\arcsec\  emitting
area on the sky.  These temperatures are roughly consistent with H$_2$ excitation
temperatures from other protostellar jets \citep[e.g.][]{giannini02,takami06,beck08,oh18} and, together with the 1-0 to 2-1 S(1) line ratio of $\sim$8 \citep{smith95}, indicate
thermal excitation from shocks.  The lines from the $v^\prime=3$ are somewhat stronger than thermal excitation, 
suggesting a possibility that populations in high vibrational levels are enhanced by UV
irradiation \citep[e.g.][]{black87,nomura07}.

The total flux of H$_2$ emission in the $K$-band between wavelengths from 2.03 to 2.37 $\mu$m is 4.0$\times$10$^{-21}$ W cm$^{-2}$, based on the best-fit to all lines.  This flux is 12.0 times brighter than the total continuum flux in the K-band, as measured from the spectrum.  Extrapolating from an H$_2$ excitation temperature of 2100 K and column density of 4.4$\times$10$^{19}$ cm$^{-2}$, extincted by $A_1=8.4$ mag, leads to H$_2$ fluxes of 3.21$\times$10$^{-20}$ W cm$^{-2}$ in the WISE W1 band and 2.97$\times$ 10$^{-20}$ W cm$^{-2}$ in the WISE W2 band (magnitudes of 16.8 and 15.8, respectively).  Even with some correction for emission outside of the slit, these magnitudes are much fainter than the observed $W1$ and $W2$ brightness. 

\begin{figure}[!t]
 \includegraphics[width=0.48\textwidth]{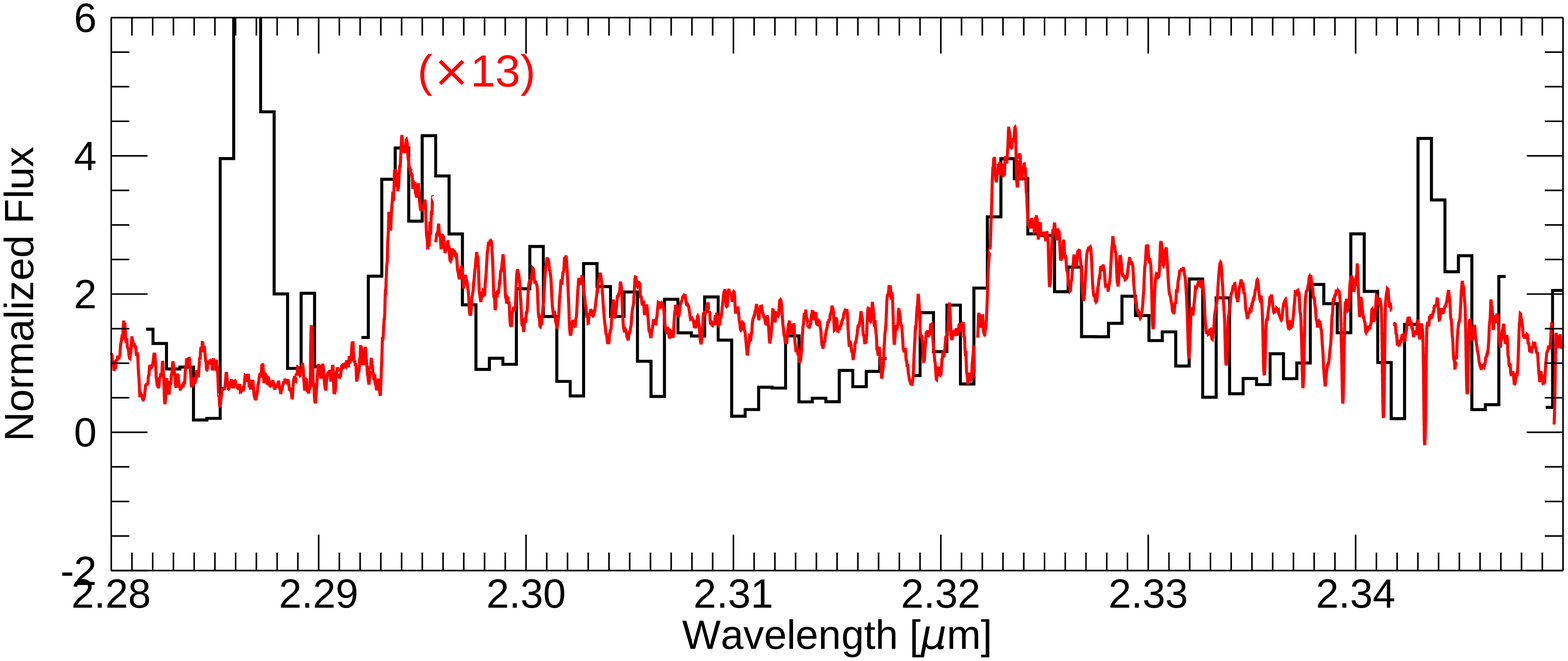}
\caption{GNIRS CO overtone band spectrum. The IGRINS spectrum of IRAS 03445+3242 \citep{lee16}, multiplied by 13, is overlaid with red. The spectrum of HOPS 373 is not continuum-subtracted. \label{fig:gnirs-Ks-CO}}
\end{figure}

The CO $v$=2-0 and 3-1 overtone bandheads are detected in emission with an integrated flux 50--90 times weaker than the summed H$_2$ line emission (Figure~\ref{fig:gnirs-Ks-CO}).  These lines typically trace emission at $\sim 3000$ K, hotter than either the sub-mm outflow emission seen in the outflow or the warmer far-IR emission \citep{tobin16}.  The critical density required to excite these levels, $\sim10^{12}-10^{13}$ cm$^{-3}$ \citep{najita96}, is associated with dense inner disks and not with outflows.

If the $W1$ and $W2$ variability are both caused by continuum emission that scales in the same way, then $\sim 73$\% of the quiescent emission in $W2$ would have to be produced by lines (assuming that the lines are non-variable). Although the H$_2$ component identified in the $K$-band cannot explain such line emission at $W1$, a cooler H$_2$ emission component may be present that could contribute flux at $W2$ but not in $W1$ or in the $K$-band \citep[see, e.g., excitation diagrams in][]{giannini06}.  This scenario would also be consistent with the warm ($\sim 300$ K) CO component detected in the far-IR \citep{tobin16}.  Alternately, strong shocks may also produce strong CO emission, as inferred in the 4--5 $\mu$m emission in photometry for the outflow shock HH 212 and the young protostar NGC 1333 IRAS 4B \citep{tappe12,herczeg12}.  Strong CO fundamental ($v=1-0$) band emission has been detected from the outflow of GSS 30 \citep{herczeg11}, an embedded protostar that also shows excited far-IR CO emission, like HOPS 373 \citep{green13,tobin16}.

\begin{table}[!bh]
\begin{center}
\caption{H$_2$ Linelist}
\begin{tabular}{ccccccc}
Line  & $\lambda$ & Flux  & Error\\
ID & ($\mu$m) & \multicolumn{2}{c}{($10^{-23}$ W cm$^{-2}$)}\\
\hline
\hline
1-0 S(4) & 1.8919 & 67.3 & 0.9 \\
2-1 S(5) & 1.9449 & 4.7 & 0.4 \\
1-0 S(3) & 1.9576 & 293.8 & 0.4 \\
2-1 S(4) & 2.0041 & 3.8 & 0.4 \\
1-0 S(2) & 2.0338 & 92.0 & 0.1 \\
3-2 S(5) & 2.0656 & 2.5 & 0.1 \\
2-1 S(3) & 2.0735 & 16.5 & 0.1 \\
1-0 S(1) & 2.1218 & 168.7 & 0.6 \\
3-2 S(4) & 2.1280 & 1.8 & 0.3 \\
2-1 S(2) & 2.1542 & 8.7 & 0.1 \\
3-2 S(3) & 2.2014 & 3.7 & 0.2 \\
1-0 S(0) & 2.2233 & 77.0 & 0.1 \\
2-1 S(1) & 2.2477 & 21.8 & 0.3 \\
3-2 S(2) & 2.2870 & 1.3 & 0.2 \\
2-1 S(0) & 2.3556 & 6.0 & 0.3 \\
3-2 S(1) & 2.3865 & 3.5 & 0.5 \\
1-0 Q(1) & 2.4066 & 201.1 & 0.5 \\
1-0 Q(2) & 2.4134 & 100.2 & 0.5 \\
1-0 Q(3) & 2.4237 & 173.2 & 0.6 \\
1-0 Q(4) & 2.4375 & 96.2 & 0.3 \\
1-0 Q(5) & 2.4548 & 62.7 & 0.4 \\
1-0 Q(6) & 2.4756 & 46.1 & 0.9 \\
1-0 Q(7) & 2.5000 & 69.5 & 1.0 \\
1-0 Q(8) & 2.5280 & 30.5 & 2.0 \\ 
\hline
\\
\\
\end{tabular}
\label{tab:h2list}
\end{center}
\end{table}

\section{The Dissection of the HOPS 373 Accretion Burst}
\label{sec:dissection}

The broadband wavelength coverage and $K$-band spectroscopy help us to dissect the response to an accretion burst of different structural components in the HOPS 373 protostar.  In this section, we step through the different wavelengths to describe the changes in the central source, as seen at long wavelengths, and how some of that emission escapes in the outflow cavity at short wavelengths.  We then describe the importance of these results for surveys that search for variable protostars.

\subsection{The sub-mm variability and change in luminosity}
\label{sec:submmvar}

The sub-mm continuum emission seen with JCMT/SCUBA-2 traces dust in the envelope, heated primarily by emission from accretion onto the central protostar (see Section \ref{sec:fiducial}).  Since the envelope acts as a bolometer, any change in the sub-mm emission should probe changes in the dust temperature profile caused by variable accretion luminosity \citep{johnstone13}.  The different scales for emission are important:  the SCUBA-2 imaging has an angular resolution of $14\farcs1$ ($\sim$6100~AU), so most of the envelope emission is detected in a single resolution element.  The ALMA 12-m Array observations have an angular resolution of $\sim 0.08^{\prime\prime}$, the typical scale of protoplanetary disks, and filter out most of the envelope emission, which occurs on scales larger than $\sim 1^{\prime\prime}$.

In the SCUBA-2 monitoring, HOPS 373 brightens by a factor of $\sim 1.25$ at 850 $\mu$m.   The single-dish SCUBA-2 and ACA sub-mm emission is centered between the two sources.  The variability is associated with the SW source, as inferred by the location of {\it WISE} mid-IR emission, ongoing outflow activity, and increase in H$_2$O maser emission.  
In the resolved ALMA observations of the continuum emission at 890 $\mu$m, obtained during a quiescent period of the NEOWISE lightcurve, the SW source is 88\% as bright as the NE source.  Nevertheless, while the NE component is bright in high-resolution sub-mm images, the images at shorter wavelengths indicate that this source is faint and contributes little to the heating and total luminosity of the envelope. The SW component dominates the emission in the far-IR, where the combined SED peaks \citep{Stutz2013}; the SW component is also not detected at shorter wavelengths.

The bolometric luminosity of $5.3$ L$_\odot$ \citep{kang15}, measured during a low luminosity epoch, corresponds to an accretion rate of approximately
\begin{equation}
 \dot{M}_{acc} =   1.4\times10^{-6} \frac{R*}{2 \times R_\odot} \frac{0.3 M_\odot}{M_*}~{\rm M}_\odot {\rm yr}^{-1}.
    \end{equation} 
    The SW component is expected to contribute most of the envelope heating.  With this assumption and scaling from radiative transfer models by \citet{baek20}, the 25\% brightness increase at 850 $\mu$m translates into an increase of the source luminosity by a factor $\sim 1.8$.  However, if both targets contribute equally to the source luminosity, and therefore the heating of the envelope (similar to the measured ratio on small scales with ALMA), then we would infer a brightness increase of 50\% at 850 $\mu$m from the southwest source.  This brightening would correspond to a luminosity increase of a factor of 3.3.

Our monitoring probes only changes during what we identify as the quiescent level of emission.   The far-IR CO emission is about 30 times stronger than that expected for its luminosity, based on correlations established for protostellar outflows \citep{manoj16}.  In addition, HOPS 373 is the only PACS Bright Red source detected in far-IR [OI] and OH emission in the \citet{tobin16} sample, indicating the presence of very strong shocks in the outflow. 

If the high far-IR CO luminosity is associated with a photo-dissociation region along the outflow cavity walls, then the current internal luminosity must be, higher than that from the SED fitting, given the cooling timescales.  Inspecting the SED model of HOPS 373 by \citet{Furlan2016}, the model flux peaks at shorter wavelengths than the observed SED. The envelope may be more massive than $\sim 0.3$ M$_\odot$ \citep{Furlan2016}, since no near- and mid-IR emission is detected from the protostar itself. We estimated the envelope mass from the 850 $\mu$m flux to be $\sim 3.6$ M$_\odot$ by using equation 1 in \citet[][]{johnstone01} with the dust temperature of 20 K and the dust opacity of 0.01 cm$^2$ g$^{-1}$ at 850 $\mu$m. The source luminosity may be underestimated, if some uncertain fraction of the energy escapes the system through the outflow cavities, although any underestimate would likely be much less than the factor of 30 needed to explain the CO emission.

On the other hand, if the far-IR CO emission is dominated by the shocked gas, then the internal luminosity and far-IR CO emission might not be necessary to be contemporaneous.
HOPS 373 is the only one of the PBRS with strong far-IR CO and H$_2$O lines, along with some OH and [\ion{O}{1}] emission \citep{manoj16,tobin16}.  The ALMA observation of CO emission shows a well-collimated jet and spot-like H$_2$ emission at along the outer boundary of CO outflows and at the termination of the outflow. These observational results indicate non-dissociative C-shocks, and a time difference of ~1000 yrs might be possible in the consideration of shock chemistry.  Other young sources, such as NGC 1333 IRAS 4B, HOPS 108, and HOPS 370, also have anomalously strong CO emission, perhaps because the emission is produced by non-dissociative C-shocks \citep[e.g.][]{herczeg12,manoj13,karska18}.

\subsection{Interpreting the brightness changes in the infrared}
\label{sec:brightnessinterp}

The warm dust and gas emission seen in the near- and mid-IR traces the outflow \citep[see schematic from][]{visser12}.  The H$_2$ emission is produced in shocks in the outflow and along the cavity walls. The continuum emission, which dominates the $W1$ imaging, is most likely produced by the warm inner disk, which irradiates the cavity walls, although we cannot rule out {\it in situ} emission from warm dust that could line the cavity walls. In this scenario, the mid-IR emission from the disk is detected in scattered light off of the cavity walls. The high extinction through the envelope to the central star absorbs all short-wavelength emission, with a dense envelope that causes the SED to peak at $100$~$\mu$m, while the line of sight extinction to the H$_2$ emission in the outflow is $A_1=8.4$ mag.  This extinction is caused by dust in the interstellar medium and in the circumstellar envelope. The disk origin for the infrared continuum emission is supported by the detection of CO overtone band emission and by the ratio of changes in the infrared compared with the sub-mm.

The CO overtone bands (${\Delta}v=2$) in our GNIRS spectrum are likely produced in the disk \citep[e.g.][]{brown13,ilee13}, since the critical density to excite the upper levels is higher than expected for outflow shocks. The detected CO emission would therefore be seen only because the outflow scatters emission that originates in the disk. This scenario strongly supports the idea that the $W1$ emission is scattered light.  For the high-mass protostar, IRAS 11101-5829, CO overtone emission is detected in scattered light by the outflow wall but is generated by the disk \citep{fedriani20}.  The high-mass  protostar S255IR NIRS 3 also seems to have a similar morphology, with variability in continuum and CO emission traced to light echoes \citep{caratti17}.  This specific scenario, with H$_2$ emission from extended winds and CO emission produced by the disk but seen only in scattered light, also has a direct analog with the post-main sequence, pre-planetary nebula IRAS 16342-3814 \citep{gledhill12}. 

The relation between the variability in W1 and that at 850 $\mu$m, $F_{\rm mid-IR} \propto F_{850}^{\eta}$, is $\eta \sim 4.6$ for HOPS 373, consistent with the empirical correlation found for mid-IR and sub-mm variability  for a subset of JCMT Transient Survey embedded protostars, $\eta = 5.53\pm 0.29$ \citep{contreras20}.  This correlation is also close to expectations from radiative transfer models that include mid-IR emission from disks scattering of light off of outflow cavities \citep{baek20}.

The variability is smaller in other IR bands, with $\eta \sim 1.8$ at $W2$ and $\eta\sim$0.5  at $K$-band (average of the bursts in Table~\ref{tab:burstlist}).  Since the $K$-band emission is dominated by H$_2$ lines, the observed variability must be produced by either large changes in the continuum emission or small changes in the H$_2$ emission.  The continuum variability at $W2$ is also likely veiled by stable molecular emission, either a cool H$_2$ component or CO emission from a strong shock in the outflow.\footnote[9]{The molecular emission from shocks should be constant on relevant timescales and change only on the longer (centuries) timescales for the outflow to travel $\sim 1000$ AU.}    

The raw $W2-W1$ color is an extreme outlier in protostar samples \citep[e.g.][]{gutermuth09,dunham15}.
If we correct the $W2$ photometry by assuming that the continuum variability at $W1$ and $W2$ are the same, then the $W2$ continuum brightness would be 1.4 mag fainter than measured, or $W2=12.3$, during quiescence; the remaining 75\% of the quiescent $W2$ emission is produced by either CO or H$_2$. The $W1-W2$ color is then $\sim 3.1$ mag, still an outlier among protostars.  Compared to the variable young protostar EC 53 ($W1-W2=2.3$ from \citealt{lee20} and $A_V\sim10$ from \citealt{dunham15}), HOPS 373 is 0.8 mag redder than EC 53, so if the emission sources are similar, then there should be $A_V=60$ mag more extinction to HOPS 373 than to EC 53 \citep[for the extinction curve of][]{wang19}, so $A_V\sim 70$ mag.  This excess extinction is the sum of extinction in our line of sight to the outflow emission (including any interstellar dust) and in the line between the outflow emission and the central source.  While this extinction estimate is highly uncertain, the very red color even after correcting for molecular line emission indicates that either the extinction to the mid-IR emission from the outflow is high, or the emission is from a very cool source.
The extinction of $A_V\sim 70$ mag is higher than that inferred from the H$_2$, but the H$_2$ emission in the slit is dominated by the compact source while the mid-IR emission is centered closer to the central object.  The extinction in our line of sight to the central source is probably even larger than this value.

The mid-IR (and any near-IR) continuum emission is either scattered light or produced {\it in situ} by dust along the cavity walls.  If this continuum is produced by scattered light, the quiescent brightness of $W1=15.0$ would correspond to a central source brightness of $W1\sim6$, after very roughly correcting for both the scattered light, as follows.  We assume that the scattering source intercepts 1\% of the stellar emission (5 mag reduction in brightness) and then re-radiates that emission over $4\pi$ steradians (2.75 mag reduction).  The albedo is 0.3--0.55 at $3.5$ $\mu$m \citep{weingartner01}, and the extinction (if the same as the H$_2$ extinction) causes a 0.3 mag reduction in brightness.  The source would still have a very red $W1-W2$ color, so the extinction to the mid-IR emission may be higher than that to the H$_2$ emission.
The $W2$ absolute brightness would then be comparable to the bright outbursting star FU Ori \citep[e.g.][]{zhu07}, in other words, very bright but still physically plausible.  Any infrared emission from the central star itself is entirely attenuated by the optically thick envelope and not directly detected. 
If the energy from the central source is beamed out of the cavity,  the bolometric luminosity of 5--6 L$_\odot$ may be somewhat underestimated because the fluxes in near- and mid-IR are not the total fluxes originated from the central protostar itself. This might be hinted at by the overluminous far-IR CO emission, as discussed in  Section 6.1.

An alternative to the scattering hypthesis for the infrared emission is that dust emission could be produced from the cavity walls themselves. Such emission could be explained by $\sim 200$ K dust emission along the observed emission area, leading to a very red spectrum with a flux that is roughly consistent with the observed brightness. In this scenario, excess luminosity from the protostar would heat the dust enough to increase the near-IR emission on light travel timescales ($<10$ days). However, the high density of the CO overtone emission makes it challenging to explain with {\it in situ} emission from the cavity walls. Additionally, the region where the 2000--3000 K CO emission is produced is hotter than the dust sublimation temperature, so dust emission would not be co-located exactly where the CO emission occurs. A time-dependent SED, including measurements at 5--15~$\mu$m where the albedo is much smaller than at shorter wavelengths, would break the degeneracy between scattered and {\it in situ} emission.

\subsection{Implications for Variability Surveys}
\label{sec:implication}

The morphology of the emission from HOPS 373 has important implications for variability surveys that include protostars.  The variability in the sub-mm indicates a modest burst in the source luminosity by about a factor of 2 (Section \ref{sec:submmvar}), presumably due to enhanced accretion.  The near- and mid-infrared emission from the central protostar is too extincted to be directly detected (Section \ref{sec:irloc}).  The observed emission escapes along the outflow cavities, where the infrared extinction is about 8 mag, as inferred from the H$_2$ lines.  The change in luminosity implied by the sub-mm variability is consistent with the level of variability seen at $W1$ (Section \ref{sec:brightnessinterp}).  

However, since the $W1$ emission emerges from the outflow cavity, any robust interpretation is empirical, depends on multiple lines of sight in a complicated geometry, and may be suspect.  
That the $W1$ emission is seen in scattered light does not necessarily affect the interpretation of the variability, unless there are optical depth changes anywhere along the lines of sight \citep[as seen for some large outbursts, including V2492 Cyg and V346 Nor,][]{hillenbrand13,kospal17}.

For HOPS 373, the near-IR and $W2$ emission are dominated by stable molecular emission, which reduces the detectability of any changes in the continuum emission.  
EGOs and the youngest (Class 0) protostars  also have $K$-band emission dominated by H$_2$ lines \citep{caratti15,laos21}, such that any continuum variability would be veiled and not large enough to trigger follow-up.  The variability in $W2$ is also less than expected, likely veiled by non-variable CO and/or H$_2$ emission.  The line emission is produced along the outflow cavity walls, detected here as H$_2$ and sub-mm CO emission, and at a strong bow shock that produces H$_2$ emission but no significant sub-mm CO emission \citep[see also far-IR CO emission,][]{tobin16}.  

Searches for variability in the $K$-band \citep[e.g., VVV Survey,][]{contreras17,lucas17} or in $W2$ \citep{park21} may miss variability in sources like HOPS 373.  Many spectra of outbursts identified in the VVV survey show strong H$_2$ emission \citep{Guo20}, although none are nearly as extreme as HOPS 373 because they would not have been identified as candidate outbursts.  The variability in $W1$ emission should be more reliable than $K$ or $W2$ of diagnosing changes in the warm dust emission from the disk.  Any IR color analysis would indicate molecular emission rather than any spectral index for the dust continuum emission.  For follow-up investigations of protostellar variables, sources that are found to have spatial offsets between the infrared and sub-mm emission are likely to yield similar $K$-band spectra as HOPS 373, with strong H$_2$ emission.

\subsection{Future experiments with high-cadence lightcurves}

The protostellar morphology is a confounding variable but also a potential source of leverage because of delays caused by light travel time and opacities.  
If the outflow is nearly in the plane of the sky, to maximize the light travel time for infrared light from the source scattered by the blueshifted outflow, then the emission at $\sim 3$ -- $4$ arcsec ($1400$ -- $1800$ AU) would by delayed by 7 -- 9 days.  Very high time resolution might be able to trace the outflow shape as the emission scatters off of dust at progressively larger distances from the central object.  Such reverberation mapping has been applied to variable protostars IRAS 18148-0440 \citep{connelley09}, LRLL 110 \citep{muzerolle13}, and L1527 IRS \citep{cook19}.  The infrared lightcurve from the central star would only be delayed and not appreciably smoothed out.

On the other hand, the sub-mm emission is expected to be delayed and smoothed out \citep[see Section~\ref{sec:fiducial} and][]{johnstone13}. The energy from the central star heats the envelope, which is optically thin at 850 $\mu$m.  The heating occurs in all directions away from the central star, including the far side of the envelope. Even though the envelope dust temperature equilibrates quickly, the associated light crossing timescale for the core is on the order of one month and thus any short timescale burst will be smoothed out. Competing with this relatively long smoothing function, the steep density and temperature radial gradients in the envelope make the initial sub-mm reaction to an instantaneous burst strong, with a long weaker tail to the response. The first burst of HOPS 373 has a sub-mm lightcurve with a steep rise ($<50$ day doubling time), which is expected to have been broadened by the envelope heating and light propagation times -- implying an underlying  rise that must have been even faster. Conversely, the broader observed decay, $\sim75$ day halving time, is not much influenced by the envelope response time.  Future simultaneous, well calibrated and high cadence (daily) monitoring of a protostellar burst in both the infrared and sub-mm could place much stronger constraints on the envelope structure.

\section{Conclusions}

In recent years, infrared and sub-mm variability surveys have been developed to search for large accretion outbursts on protostars.  In this paper, we evaluate multi-wavelength emission for the modest accretion burst of HOPS 373, a deeply embedded protostar in NGC 2068 in Orion, with the following results:

\begin{itemize}
    \item Variability in the sub-mm continuum emission provides an 
 indirect probe of variability of the central source luminosity, dominated by accretion. The source luminosity brightens by $\sim 1.8$ -- $3.3$, depending on the contribution to the quiescent luminosity from the northeast source.
    
      \item High-resolution mm imaging reveals two distinct compact sources, which complicates the conversion of sub-mm brightness variability to changes in accretion luminosity onto the varying source.  The southwest component is identified as the variable because it launches the small-scale outflow and is associated with maser emission, which is also much brighter in 2021 than in 2015.

      \item The observed near and mid-IR continuum emission is likely scattered light from the central protostar and disk scattered in the outflow cavity; similar to the spatially resolved variability in the scattered light nebulae of low-mass stars LRL 54361 \citep{muzerolle13} and L1527 IRS \citep{cook19}.  In the infrared and sub-mm, the variable emission from the protostar and its disk cannot be detected directly.
    
    \item For the youngest protostars, the $W1$-band is likely optimal for measuring continuum changes.
    The $K$- and $W2$-band emission are dominated by CO and H$_2$ emission lines produced by the outflow, along the cavity walls and at a bow shock.  The line emission is expected to be much more steady than the continuum emission, so these contributions will reduce any variability signal that might otherwise be measured from the continuum.

\end{itemize}

These results together indicate that photometric variability (or lack of variability) for protostars require spectroscopic and multi-wavelength investigations for physical interpretations.  
The $K$-band and $W2$ bands pose challenges for some subset of young protostars. Variability searches in $W1$ may be more reliable because of the lack of strong lines coincident with the filter transmission. With existing facilities, the sub-mm provides the most robust measurement of protostellar variability, but is limited by sensitivity and spatial resolution, and should be coupled with observations at shorter wavelengths and sub-mm observations with high resolution.

\section{Acknowledgements}

The authors thank the referee, Phil Lucas, for a helpful and careful report.  We also thank Xindi Tang, Miju Kang, Jenny Hatchell, Somnath Dutta, Ross Burns, and Jan Forbrich for helpful comments in the preparation of the manuscript.

SYY and JEL are supported by the National Research Foundation of Korea (NRF) grant funded by the Korea government (MSIT) (grant number 2021R1A2C1011718). G.J.H.\ is supported by general grants 12173003 and 11773002 awarded by the National Science Foundation of China.
D.J.\ is supported by NRC Canada and by an NSERC Discovery Grant. J.J.T. acknowledges support from  NSF AST-1814762.  M.O. acknowledges support from the Spanish MINECO/AEI
AYA2017-84390-C2-1-R (co-funded by FEDER) and
PID2020-114461GB-I00/AEI/10.13039/501100011033 grants, and from the
State Agency for Research of the Spanish MCIU through the ``Center of
Excellence Severo Ochoa'' award for the Instituto de Astrof{\'i}sica de
Andaluc{\'i}a (SEV-2017-0709).

The authors wish to recognize and acknowledge the very significant cultural role and reverence that the summit of Maunakea has always had within the indigenous Hawaiian community.  We are most fortunate to have the opportunity to conduct observations from this mountain.

The James Clerk Maxwell Telescope is operated by the East Asian Observatory on behalf of The National Astronomical Observatory of Japan; Academia Sinica Institute of Astronomy and Astrophysics; the Korea Astronomy and Space Science Institute; the Operation, Maintenance and Upgrading Fund for Astronomical Telescopes and Facility Instruments, budgeted from the Ministry of Finance (MOF) of China and administrated by the Chinese Academy of Sciences (CAS), as well as the National Key R\&D Program of China (No. 2017YFA0402700). Additional funding support is provided by the Science and Technology Facilities Council of the United Kingdom and participating universities in the United Kingdom and Canada. Additional funds for the construction of SCUBA-2 were provided by the Canada Foundation for Innovation. The James Clerk Maxwell Telescope has historically been operated by the Joint Astronomy Centre on behalf of the Science and Technology Facilities Council of the United Kingdom, the National Research Council of Canada and the Netherlands Organisation for Scientific Research.

This paper makes use of the following ALMA data: ADS/JAO.ALMA\#2015.1.00041.S and ADS/JAO.ALMA\#2018.1.01565.S
ALMA is a partnership of ESO (representing its member states), NSF (USA) and 
NINS (Japan), together with NRC (Canada), NSC and ASIAA (Taiwan), and 
KASI (Republic of Korea), in cooperation with the Republic of Chile. 
The Joint ALMA Observatory is operated by ESO, AUI/NRAO and NAOJ.
The National Radio Astronomy 
Observatory is a facility of the National Science Foundation 
operated under cooperative agreement by Associated Universities, Inc.

UKIRT is owned by the University of Hawaii (UH) and operated by the UH Institute for Astronomy. When some of the data reported here were obtained, UKIRT was supported by NASA and operated under an agreement among the University of Hawaii, the University of Arizona, and Lockheed Martin Advanced Technology Center; operations were enabled through the cooperation of the East Asian Observatory. We thank the UKIRT staff for carrying out the queue observations, and the Cambridge Astronomy Survey Unit for carrying out the WFCAM data processing.

\bibliographystyle{aasjournal}
\bibliography{ms}

\begin{thebibliography}{}
\expandafter\ifx\csname natexlab\endcsname\relax\def\natexlab#1{#1}\fi
\providecommand{\url}[1]{\href{#1}{#1}}
\providecommand{\dodoi}[1]{doi:~\href{http://doi.org/#1}{\nolinkurl{#1}}}
\providecommand{\doeprint}[1]{\href{http://ascl.net/#1}{\nolinkurl{http://ascl.net/#1}}}
\providecommand{\doarXiv}[1]{\href{https://arxiv.org/abs/#1}{\nolinkurl{https://arxiv.org/abs/#1}}}

\bibitem[{{Antoniucci} {et~al.}(2014){Antoniucci}, {Giannini}, {Li Causi}, \&
  {Lorenzetti}}]{antoniucci14}
{Antoniucci}, S., {Giannini}, T., {Li Causi}, G., \& {Lorenzetti}, D. 2014,
  \apj, 782, 51, \dodoi{10.1088/0004-637X/782/1/51}

\bibitem[{{Audard} {et~al.}(2014){Audard}, {{\'A}brah{\'a}m}, {Dunham},
  {Green}, {Grosso}, {Hamaguchi}, {Kastner}, {K{\'o}sp{\'a}l}, {Lodato},
  {Romanova}, {Skinner}, {Vorobyov}, \& {Zhu}}]{audard14}
{Audard}, M., {{\'A}brah{\'a}m}, P., {Dunham}, M.~M., {et~al.} 2014, in
  Protostars and Planets VI, ed. H.~{Beuther}, R.~S. {Klessen}, C.~P.
  {Dullemond}, \& T.~{Henning}, 387,
  \dodoi{10.2458/azu\_uapress\_9780816531240-ch017}

\bibitem[{{Bae} {et~al.}(2014){Bae}, {Hartmann}, {Zhu}, \& {Nelson}}]{bae14}
{Bae}, J., {Hartmann}, L., {Zhu}, Z., \& {Nelson}, R.~P. 2014, \apj, 795, 61,
  \dodoi{10.1088/0004-637X/795/1/61}

\bibitem[{{Baek} {et~al.}(2020){Baek}, {MacFarlane}, {Lee}, {Stamatellos},
  {Herczeg}, {Johnstone}, {Pe{\~n}a}, {Varricatt}, {Hodapp}, {Chen}, \&
  {Kang}}]{baek20}
{Baek}, G., {MacFarlane}, B.~A., {Lee}, J.-E., {et~al.} 2020, \apj, 895, 27,
  \dodoi{10.3847/1538-4357/ab8ad4}

\bibitem[{{Balog} {et~al.}(2014){Balog}, {Muzerolle}, {Flaherty}, {Detre},
  {Bouwmann}, {Furlan}, {Gutermuth}, {Juhasz}, {Bally}, {Nielbock}, {Klaas},
  {Krause}, {Henning}, \& {Marton}}]{balog14}
{Balog}, Z., {Muzerolle}, J., {Flaherty}, K., {et~al.} 2014, \apjl, 789, L38,
  \dodoi{10.1088/2041-8205/789/2/L38}

\bibitem[{{Beck} {et~al.}(2008){Beck}, {McGregor}, {Takami}, \& {Pyo}}]{beck08}
{Beck}, T.~L., {McGregor}, P.~J., {Takami}, M., \& {Pyo}, T.-S. 2008, \apj,
  676, 472, \dodoi{10.1086/527528}

\bibitem[{{Black} \& {van Dishoeck}(1987)}]{black87}
{Black}, J.~H., \& {van Dishoeck}, E.~F. 1987, \apj, 322, 412,
  \dodoi{10.1086/165740}

\bibitem[{{Brown} {et~al.}(2013){Brown}, {Pontoppidan}, {van Dishoeck},
  {Herczeg}, {Blake}, \& {Smette}}]{brown13}
{Brown}, J.~M., {Pontoppidan}, K.~M., {van Dishoeck}, E.~F., {et~al.} 2013,
  \apj, 770, 94, \dodoi{10.1088/0004-637X/770/2/94}

\bibitem[{{Burns} {et~al.}(2015){Burns}, {Imai}, {Handa}, {Omodaka},
  {Nakagawa}, {Nagayama}, \& {Ueno}}]{burns15}
{Burns}, R.~A., {Imai}, H., {Handa}, T., {et~al.} 2015, \mnras, 453, 3163,
  \dodoi{10.1093/mnras/stv1836}

\bibitem[{{Burns} {et~al.}(2020){Burns}, {Sugiyama}, {Hirota}, {Kim},
  {Sobolev}, {Stecklum}, {MacLeod}, {Yonekura}, {Olech}, {Orosz}, {Ellingsen},
  {Hyland}, {Caratti o Garatti}, {Brogan}, {Hunter}, {Phillips}, {van den
  Heever}, {Eisl{\"o}ffel}, {Linz}, {Surcis}, {Chibueze}, {Baan}, \&
  {Kramer}}]{burns20}
{Burns}, R.~A., {Sugiyama}, K., {Hirota}, T., {et~al.} 2020, Nature Astronomy,
  4, 506, \dodoi{10.1038/s41550-019-0989-3}

\bibitem[{{Caratti o Garatti} {et~al.}(2015){Caratti o Garatti}, {Stecklum},
  {Linz}, {Garcia Lopez}, \& {Sanna}}]{caratti15}
{Caratti o Garatti}, A., {Stecklum}, B., {Linz}, H., {Garcia Lopez}, R., \&
  {Sanna}, A. 2015, \aap, 573, A82, \dodoi{10.1051/0004-6361/201423992}

\bibitem[{{Caratti o Garatti} {et~al.}(2017){Caratti o Garatti}, {Stecklum},
  {Garcia Lopez}, {Eisl{\"o}ffel}, {Ray}, {Sanna}, {Cesaroni}, {Walmsley},
  {Oudmaijer}, {de Wit}, {Moscadelli}, {Greiner}, {Krabbe}, {Fischer}, {Klein},
  \& {Iba{\~n}ez}}]{caratti17}
{Caratti o Garatti}, A., {Stecklum}, B., {Garcia Lopez}, R., {et~al.} 2017,
  Nature Physics, 13, 276, \dodoi{10.1038/nphys3942}

\bibitem[{{Casali} {et~al.}(2007){Casali}, {Adamson}, {Alves de Oliveira},
  {Almaini}, {Burch}, {Chuter}, {Elliot}, {Folger}, {Foucaud}, {Hambly},
  {Hastie}, {Henry}, {Hirst}, {Irwin}, {Ives}, {Lawrence}, {Laidlaw}, {Lee},
  {Lewis}, {Lunney}, {McLay}, {Montgomery}, {Pickup}, {Read}, {Rees}, {Robson},
  {Sekiguchi}, {Vick}, {Warren}, \& {Woodward}}]{casali07}
{Casali}, M., {Adamson}, A., {Alves de Oliveira}, C., {et~al.} 2007, \aap, 467,
  777, \dodoi{10.1051/0004-6361:20066514}

\bibitem[{{Connelley} {et~al.}(2009){Connelley}, {Hodapp}, \&
  {Fuller}}]{connelley09}
{Connelley}, M.~S., {Hodapp}, K.~W., \& {Fuller}, G.~A. 2009, \aj, 137, 3494,
  \dodoi{10.1088/0004-6256/137/3/3494}

\bibitem[{{Contreras Pe{\~n}a} {et~al.}(2020){Contreras Pe{\~n}a}, {Johnstone},
  {Baek}, {Herczeg}, {Mairs}, {Scholz}, {Lee}, \& {JCMT Transient
  Team}}]{contreras20}
{Contreras Pe{\~n}a}, C., {Johnstone}, D., {Baek}, G., {et~al.} 2020, \mnras,
  495, 3614, \dodoi{10.1093/mnras/staa1254}

\bibitem[{{Contreras Pe{\~n}a} {et~al.}(2017){Contreras Pe{\~n}a}, {Lucas},
  {Kurtev}, {Minniti}, {Caratti o Garatti}, {Marocco}, {Thompson}, {Froebrich},
  {Kumar}, \& {Stimson}}]{contreras17}
{Contreras Pe{\~n}a}, C., {Lucas}, P.~W., {Kurtev}, R., {et~al.} 2017, Monthly
  Notices of the Royal Astronomical Society, 465, 3039,
  \dodoi{10.1093/mnras/stw2802}

\bibitem[{{Cook} {et~al.}(2019){Cook}, {Tobin}, {Skrutskie}, \&
  {Nelson}}]{cook19}
{Cook}, B.~T., {Tobin}, J.~J., {Skrutskie}, M.~F., \& {Nelson}, M.~J. 2019,
  \aap, 626, A51, \dodoi{10.1051/0004-6361/201935419}

\bibitem[{{Cutri} {et~al.}(2021){Cutri}, {Wright}, {Conrow}, {Fowler},
  {Eisenhardt}, {Grillmair}, {Kirkpatrick}, {Masci}, {McCallon}, {Wheelock},
  {Fajardo-Acosta}, {Yan}, {Benford}, {Harbut}, {Jarrett}, {Lake}, {Leisawitz},
  {Ressler}, {Stanford}, {Tsai}, {Liu}, {Helou}, {Mainzer}, {Gettngs},
  {Gonzalez}, {Hoffman}, {Marsh}, {Padgett}, {Skrutskie}, {Beck}, {Papin}, \&
  {Wittman}}]{cutri14}
{Cutri}, R.~M., {Wright}, E.~L., {Conrow}, T., {et~al.} 2021, VizieR Online
  Data Catalog, II/328

\bibitem[{{Cyganowski} {et~al.}(2008){Cyganowski}, {Whitney}, {Holden},
  {Braden}, {Brogan}, {Churchwell}, {Indebetouw}, {Watson}, {Babler},
  {Benjamin}, {Gomez}, {Meade}, {Povich}, {Robitaille}, \&
  {Watson}}]{cyganowski08}
{Cyganowski}, C.~J., {Whitney}, B.~A., {Holden}, E., {et~al.} 2008, \aj, 136,
  2391, \dodoi{10.1088/0004-6256/136/6/2391}

\bibitem[{{Dempsey} {et~al.}(2013){Dempsey}, {Friberg}, {Jenness}, {Tilanus},
  {Thomas}, {Holland}, {Bintley}, {Berry}, {Chapin}, {Chrysostomou}, {Davis},
  {Gibb}, {Parsons}, \& {Robson}}]{dempsey13}
{Dempsey}, J.~T., {Friberg}, P., {Jenness}, T., {et~al.} 2013, \mnras, 430,
  2534, \dodoi{10.1093/mnras/stt090}

\bibitem[{{Dunham} {et~al.}(2015){Dunham}, {Allen}, {Evans},
  {Broekhoven-Fiene}, {Cieza}, {Di Francesco}, {Gutermuth}, {Harvey},
  {Hatchell}, {Heiderman}, {Huard}, {Johnstone}, {Kirk}, {Matthews}, {Miller},
  {Peterson}, \& {Young}}]{dunham15}
{Dunham}, M.~M., {Allen}, L.~E., {Evans}, II, N.~J., {et~al.} 2015, \apjs, 220,
  11, \dodoi{10.1088/0067-0049/220/1/11}

\bibitem[{{Fazio} {et~al.}(2004){Fazio}, {Hora}, {Allen}, {Ashby}, {Barmby},
  {Deutsch}, {Huang}, {Kleiner}, {Marengo}, {Megeath}, {Melnick}, {Pahre},
  {Patten}, {Polizotti}, {Smith}, {Taylor}, {Wang}, {Willner}, {Hoffmann},
  {Pipher}, {Forrest}, {McMurty}, {McCreight}, {McKelvey}, {McMurray}, {Koch},
  {Moseley}, {Arendt}, {Mentzell}, {Marx}, {Losch}, {Mayman}, {Eichhorn},
  {Krebs}, {Jhabvala}, {Gezari}, {Fixsen}, {Flores}, {Shakoorzadeh}, {Jungo},
  {Hakun}, {Workman}, {Karpati}, {Kichak}, {Whitley}, {Mann}, {Tollestrup},
  {Eisenhardt}, {Stern}, {Gorjian}, {Bhattacharya}, {Carey}, {Nelson},
  {Glaccum}, {Lacy}, {Lowrance}, {Laine}, {Reach}, {Stauffer}, {Surace},
  {Wilson}, {Wright}, {Hoffman}, {Domingo}, \& {Cohen}}]{fazio04}
{Fazio}, G.~G., {Hora}, J.~L., {Allen}, L.~E., {et~al.} 2004, \apjs, 154, 10,
  \dodoi{10.1086/422843}

\bibitem[{{Fedriani} {et~al.}(2020){Fedriani}, {Caratti o Garatti},
  {Koutoulaki}, {Garcia-Lopez}, {Natta}, {Cesaroni}, {Oudmaijer}, {Coffey},
  {Ray}, \& {Stecklum}}]{fedriani20}
{Fedriani}, R., {Caratti o Garatti}, A., {Koutoulaki}, M., {et~al.} 2020, \aap,
  633, A128, \dodoi{10.1051/0004-6361/201936748}

\bibitem[{{Fischer} {et~al.}(2019){Fischer}, {Safron}, \&
  {Megeath}}]{fischer19}
{Fischer}, W.~J., {Safron}, E., \& {Megeath}, S.~T. 2019, \apj, 872, 183,
  \dodoi{10.3847/1538-4357/ab01dc}

\bibitem[{{Fischer} {et~al.}(2020){Fischer}, {Megeath}, {Furlan}, {Stutz},
  {Stanke}, {Tobin}, {Osorio}, {Manoj}, {Di Francesco}, {Allen}, {Watson},
  {Wilson}, \& {Henning}}]{Fischer20}
{Fischer}, W.~J., {Megeath}, S.~T., {Furlan}, E., {et~al.} 2020, \apj, 905,
  119, \dodoi{10.3847/1538-4357/abc7cb}

\bibitem[{{Furlan} {et~al.}(2016){Furlan}, {Fischer}, {Ali}, {Stutz}, {Stanke},
  {Tobin}, {Megeath}, {Osorio}, {Hartmann}, \& {Calvet}}]{Furlan2016}
{Furlan}, E., {Fischer}, W.~J., {Ali}, B., {et~al.} 2016, The Astrophysical
  Journal Supplement Series, 224, 5, \dodoi{10.3847/0067-0049/224/1/5}

\bibitem[{{Gaia Collaboration} {et~al.}(2018){Gaia Collaboration}, {Brown},
  {Vallenari}, {Prusti}, {de Bruijne}, {Babusiaux}, {Bailer-Jones}, {Biermann},
  {Evans}, {Eyer}, {Jansen}, {Jordi}, {Klioner}, {Lammers}, {Lindegren},
  {Luri}, {Mignard}, {Panem}, {Pourbaix}, {Randich}, {Sartoretti}, {Siddiqui},
  {Soubiran}, {van Leeuwen}, {Walton}, {Arenou}, {Bastian}, {Cropper},
  {Drimmel}, {Katz}, {Lattanzi}, {Bakker}, {Cacciari}, {Casta{\~n}eda},
  {Chaoul}, {Cheek}, {De Angeli}, {Fabricius}, {Guerra}, {Holl}, {Masana},
  {Messineo}, {Mowlavi}, {Nienartowicz}, {Panuzzo}, {Portell}, {Riello},
  {Seabroke}, {Tanga}, {Th{\'e}venin}, {Gracia-Abril}, {Comoretto},
  {Garcia-Reinaldos}, {Teyssier}, {Altmann}, {Andrae}, {Audard},
  {Bellas-Velidis}, {Benson}, {Berthier}, {Blomme}, {Burgess}, {Busso},
  {Carry}, {Cellino}, {Clementini}, {Clotet}, {Creevey}, {Davidson}, {De
  Ridder}, {Delchambre}, {Dell'Oro}, {Ducourant},
  {Fern{\'a}ndez-Hern{\'a}ndez}, {Fouesneau}, {Fr{\'e}mat}, {Galluccio},
  {Garc{\'\i}a-Torres}, {Gonz{\'a}lez-N{\'u}{\~n}ez}, {Gonz{\'a}lez-Vidal},
  {Gosset}, {Guy}, {Halbwachs}, {Hambly}, {Harrison}, {Hern{\'a}ndez},
  {Hestroffer}, {Hodgkin}, {Hutton}, {Jasniewicz}, {Jean-Antoine-Piccolo},
  {Jordan}, {Korn}, {Krone-Martins}, {Lanzafame}, {Lebzelter}, {L{\"o}ffler},
  {Manteiga}, {Marrese}, {Mart{\'\i}n-Fleitas}, {Moitinho}, {Mora}, {Muinonen},
  {Osinde}, {Pancino}, {Pauwels}, {Petit}, {Recio-Blanco}, {Richards},
  {Rimoldini}, {Robin}, {Sarro}, {Siopis}, {Smith}, {Sozzetti}, {S{\"u}veges},
  {Torra}, {van Reeven}, {Abbas}, {Abreu Aramburu}, {Accart}, {Aerts},
  {Altavilla}, {{\'A}lvarez}, {Alvarez}, {Alves}, {Anderson}, {Andrei},
  {Anglada Varela}, {Antiche}, {Antoja}, {Arcay}, {Astraatmadja}, {Bach},
  {Baker}, {Balaguer-N{\'u}{\~n}ez}, {Balm}, {Barache}, {Barata}, {Barbato},
  {Barblan}, {Barklem}, {Barrado}, {Barros}, {Barstow}, {Bartholom{\'e}
  Mu{\~n}oz}, {Bassilana}, {Becciani}, {Bellazzini}, {Berihuete}, {Bertone},
  {Bianchi}, {Bienaym{\'e}}, {Blanco-Cuaresma}, {Boch}, {Boeche}, {Bombrun},
  {Borrachero}, {Bossini}, {Bouquillon}, {Bourda}, {Bragaglia}, {Bramante},
  {Breddels}, {Bressan}, {Brouillet}, {Br{\"u}semeister}, {Brugaletta},
  {Bucciarelli}, {Burlacu}, {Busonero}, {Butkevich}, {Buzzi}, {Caffau},
  {Cancelliere}, {Cannizzaro}, {Cantat-Gaudin}, {Carballo}, {Carlucci},
  {Carrasco}, {Casamiquela}, {Castellani}, {Castro-Ginard}, {Charlot},
  {Chemin}, {Chiavassa}, {Cocozza}, {Costigan}, {Cowell}, {Crifo}, {Crosta},
  {Crowley}, {Cuypers}, {Dafonte}, {Damerdji}, {Dapergolas}, {David}, {David},
  {de Laverny}, {De Luise}, {De March}, {de Martino}, {de Souza}, {de Torres},
  {Debosscher}, {del Pozo}, {Delbo}, {Delgado}, {Delgado}, {Di Matteo},
  {Diakite}, {Diener}, {Distefano}, {Dolding}, {Drazinos}, {Dur{\'a}n},
  {Edvardsson}, {Enke}, {Eriksson}, {Esquej}, {Eynard Bontemps}, {Fabre},
  {Fabrizio}, {Faigler}, {Falc{\~a}o}, {Farr{\`a}s Casas}, {Federici},
  {Fedorets}, {Fernique}, {Figueras}, {Filippi}, {Findeisen}, {Fonti},
  {Fraile}, {Fraser}, {Fr{\'e}zouls}, {Gai}, {Galleti}, {Garabato},
  {Garc{\'\i}a-Sedano}, {Garofalo}, {Garralda}, {Gavel}, {Gavras}, {Gerssen},
  {Geyer}, {Giacobbe}, {Gilmore}, {Girona}, {Giuffrida}, {Glass}, {Gomes},
  {Granvik}, {Gueguen}, {Guerrier}, {Guiraud}, {Guti{\'e}rrez-S{\'a}nchez},
  {Haigron}, {Hatzidimitriou}, {Hauser}, {Haywood}, {Heiter}, {Helmi}, {Heu},
  {Hilger}, {Hobbs}, {Hofmann}, {Holland}, {Huckle}, {Hypki}, {Icardi},
  {Jan{\ss}en}, {Jevardat de Fombelle}, {Jonker}, {Juh{\'a}sz}, {Julbe},
  {Karampelas}, {Kewley}, {Klar}, {Kochoska}, {Kohley}, {Kolenberg},
  {Kontizas}, {Kontizas}, {Koposov}, {Kordopatis}, {Kostrzewa-Rutkowska},
  {Koubsky}, {Lambert}, {Lanza}, {Lasne}, {Lavigne}, {Le Fustec}, {Le
  Poncin-Lafitte}, {Lebreton}, {Leccia}, {Leclerc}, {Lecoeur-Taibi},
  {Lenhardt}, {Leroux}, {Liao}, {Licata}, {Lindstr{\o}m}, {Lister}, {Livanou},
  {Lobel}, {L{\'o}pez}, {Managau}, {Mann}, {Mantelet}, {Marchal}, {Marchant},
  {Marconi}, {Marinoni}, {Marschalk{\'o}}, {Marshall}, {Martino}, {Marton},
  {Mary}, {Massari}, {Matijevi{\v{c}}}, {Mazeh}, {McMillan}, {Messina},
  {Michalik}, {Millar}, {Molina}, {Molinaro}, {Moln{\'a}r}, {Montegriffo},
  {Mor}, {Morbidelli}, {Morel}, {Morris}, {Mulone}, {Muraveva}, {Musella},
  {Nelemans}, {Nicastro}, {Noval}, {O'Mullane}, {Ord{\'e}novic},
  {Ord{\'o}{\~n}ez-Blanco}, {Osborne}, {Pagani}, {Pagano}, {Pailler},
  {Palacin}, {Palaversa}, {Panahi}, {Pawlak}, {Piersimoni}, {Pineau}, {Plachy},
  {Plum}, {Poggio}, {Poujoulet}, {Pr{\v{s}}a}, {Pulone}, {Racero}, {Ragaini},
  {Rambaux}, {Ramos-Lerate}, {Regibo}, {Reyl{\'e}}, {Riclet}, {Ripepi}, {Riva},
  {Rivard}, {Rixon}, {Roegiers}, {Roelens}, {Romero-G{\'o}mez}, {Rowell},
  {Royer}, {Ruiz-Dern}, {Sadowski}, {Sagrist{\`a} Sell{\'e}s}, {Sahlmann},
  {Salgado}, {Salguero}, {Sanna}, {Santana-Ros}, {Sarasso}, {Savietto},
  {Schultheis}, {Sciacca}, {Segol}, {Segovia}, {S{\'e}gransan}, {Shih},
  {Siltala}, {Silva}, {Smart}, {Smith}, {Solano}, {Solitro}, {Sordo}, {Soria
  Nieto}, {Souchay}, {Spagna}, {Spoto}, {Stampa}, {Steele},
  {Steidelm{\"u}ller}, {Stephenson}, {Stoev}, {Suess}, {Surdej}, {Szabados},
  {Szegedi-Elek}, {Tapiador}, {Taris}, {Tauran}, {Taylor}, {Teixeira},
  {Terrett}, {Teyssandier}, {Thuillot}, {Titarenko}, {Torra Clotet}, {Turon},
  {Ulla}, {Utrilla}, {Uzzi}, {Vaillant}, {Valentini}, {Valette}, {van Elteren},
  {Van Hemelryck}, {van Leeuwen}, {Vaschetto}, {Vecchiato}, {Veljanoski},
  {Viala}, {Vicente}, {Vogt}, {von Essen}, {Voss}, {Votruba}, {Voutsinas},
  {Walmsley}, {Weiler}, {Wertz}, {Wevers}, {Wyrzykowski}, {Yoldas},
  {{\v{Z}}erjal}, {Ziaeepour}, {Zorec}, {Zschocke}, {Zucker}, {Zurbach}, \&
  {Zwitter}}]{gaia18}
{Gaia Collaboration}, {Brown}, A.~G.~A., {Vallenari}, A., {et~al.} 2018, \aap,
  616, A1, \dodoi{10.1051/0004-6361/201833051}

\bibitem[{{Getman} {et~al.}(2017){Getman}, {Broos}, {Kuhn}, {Feigelson},
  {Richert}, {Ota}, {Bate}, \& {Garmire}}]{getman17}
{Getman}, K.~V., {Broos}, P.~S., {Kuhn}, M.~A., {et~al.} 2017, \apjs, 229, 28,
  \dodoi{10.3847/1538-4365/229/2/28}

\bibitem[{{Giannini} {et~al.}(2006){Giannini}, {McCoey}, {Nisini}, {Cabrit},
  {Caratti o Garatti}, {Calzoletti}, \& {Flower}}]{giannini06}
{Giannini}, T., {McCoey}, C., {Nisini}, B., {et~al.} 2006, \aap, 459, 821,
  \dodoi{10.1051/0004-6361:20065127}

\bibitem[{{Giannini} {et~al.}(2002){Giannini}, {Nisini}, {Caratti o Garatti},
  \& {Lorenzetti}}]{giannini02}
{Giannini}, T., {Nisini}, B., {Caratti o Garatti}, A., \& {Lorenzetti}, D.
  2002, \apjl, 570, L33, \dodoi{10.1086/340883}

\bibitem[{{Gibb} \& {Little}(2000)}]{gibb00}
{Gibb}, A.~G., \& {Little}, L.~T. 2000, \mnras, 313, 663,
  \dodoi{10.1046/j.1365-8711.2000.03235.x}

\bibitem[{{Gledhill} \& {Forde}(2012)}]{gledhill12}
{Gledhill}, T.~M., \& {Forde}, K.~P. 2012, \mnras, 421, 346,
  \dodoi{10.1111/j.1365-2966.2011.20309.x}

\bibitem[{{Green} {et~al.}(2013){Green}, {Evans}, {K{\'o}sp{\'a}l}, {Herczeg},
  {Quanz}, {Henning}, {van Kempen}, {Lee}, {Dunham}, {Meeus}, {Bouwman},
  {Chen}, {G{\"u}del}, {Skinner}, {Liebhart}, \& {Merello}}]{green13}
{Green}, J.~D., {Evans}, II, N.~J., {K{\'o}sp{\'a}l}, {\'A}., {et~al.} 2013,
  \apj, 772, 117, \dodoi{10.1088/0004-637X/772/2/117}

\bibitem[{{Guo} {et~al.}(2020){Guo}, {Lucas}, {Contreras Pe{\~n}a}, {Kurtev},
  {Smith}, {Borissova}, {Alonso-Garc{\'\i}a}, {Minniti}, {Caratti o Garatti},
  \& {Froebrich}}]{Guo20}
{Guo}, Z., {Lucas}, P.~W., {Contreras Pe{\~n}a}, C., {et~al.} 2020, \mnras,
  492, 294, \dodoi{10.1093/mnras/stz3374}

\bibitem[{{Gutermuth} {et~al.}(2009){Gutermuth}, {Megeath}, {Myers}, {Allen},
  {Pipher}, \& {Fazio}}]{gutermuth09}
{Gutermuth}, R.~A., {Megeath}, S.~T., {Myers}, P.~C., {et~al.} 2009, \apjs,
  184, 18, \dodoi{10.1088/0067-0049/184/1/18}

\bibitem[{{Hartmann} \& {Kenyon}(1996)}]{hartmann96}
{Hartmann}, L., \& {Kenyon}, S.~J. 1996, \araa, 34, 207,
  \dodoi{10.1146/annurev.astro.34.1.207}

\bibitem[{{Haschick} {et~al.}(1983){Haschick}, {Moran}, {Rodriguez}, \&
  {Ho}}]{haschick83}
{Haschick}, A.~D., {Moran}, J.~M., {Rodriguez}, L.~F., \& {Ho}, P.~T.~P. 1983,
  \apj, 265, 281, \dodoi{10.1086/160673}

\bibitem[{{Herczeg} {et~al.}(2011){Herczeg}, {Brown}, {van Dishoeck}, \&
  {Pontoppidan}}]{herczeg11}
{Herczeg}, G.~J., {Brown}, J.~M., {van Dishoeck}, E.~F., \& {Pontoppidan},
  K.~M. 2011, \aap, 533, A112, \dodoi{10.1051/0004-6361/201016246}

\bibitem[{{Herczeg} {et~al.}(2012){Herczeg}, {Karska}, {Bruderer},
  {Kristensen}, {van Dishoeck}, {J{\o}rgensen}, {Visser}, {Wampfler}, {Bergin},
  {Y{\i}ld{\i}z}, {Pontoppidan}, \& {Gracia-Carpio}}]{herczeg12}
{Herczeg}, G.~J., {Karska}, A., {Bruderer}, S., {et~al.} 2012, \aap, 540, A84,
  \dodoi{10.1051/0004-6361/201117914}

\bibitem[{{Herczeg} {et~al.}(2017){Herczeg}, {Johnstone}, {Mairs}, {Hatchell},
  {Lee}, {Bower}, {Chen}, {Aikawa}, {Yoo}, \& {Kang}}]{herczeg2017}
{Herczeg}, G.~J., {Johnstone}, D., {Mairs}, S., {et~al.} 2017, The
  Astrophysical Journal, 849, 43, \dodoi{10.3847/1538-4357/aa8b62}

\bibitem[{{Hillenbrand} {et~al.}(2013){Hillenbrand}, {Miller}, {Covey},
  {Carpenter}, {Cenko}, {Silverman}, {Muirhead}, {Fischer}, {Crepp}, {Bloom},
  \& {Filippenko}}]{hillenbrand13}
{Hillenbrand}, L.~A., {Miller}, A.~A., {Covey}, K.~R., {et~al.} 2013, \aj, 145,
  59, \dodoi{10.1088/0004-6256/145/3/59}

\bibitem[{{Hillenbrand} {et~al.}(2018){Hillenbrand}, {Contreras Pe{\~n}a},
  {Morrell}, {Naylor}, {Kuhn}, {Cutri}, {Rebull}, {Hodgkin}, {Froebrich}, \&
  {Mainzer}}]{hillenbrand18}
{Hillenbrand}, L.~A., {Contreras Pe{\~n}a}, C., {Morrell}, S., {et~al.} 2018,
  \apj, 869, 146, \dodoi{10.3847/1538-4357/aaf414}

\bibitem[{{Hirota} {et~al.}(2021){Hirota}, {Cesaroni}, {Moscadelli},
  {Sugiyama}, {Burns}, {Kim}, {Sunada}, \& {Yonekura}}]{hirota21}
{Hirota}, T., {Cesaroni}, R., {Moscadelli}, L., {et~al.} 2021, \aap, 647, A23,
  \dodoi{10.1051/0004-6361/202039798}

\bibitem[{{Holland} {et~al.}(2013){Holland}, {Bintley}, {Chapin},
  {Chrysostomou}, {Davis}, {Dempsey}, {Duncan}, {Fich}, {Friberg}, {Halpern},
  {Irwin}, {Jenness}, {Kelly}, {MacIntosh}, {Robson}, {Scott}, {Ade},
  {Atad-Ettedgui}, {Berry}, {Craig}, {Gao}, {Gibb}, {Hilton}, {Hollister},
  {Kycia}, {Lunney}, {McGregor}, {Montgomery}, {Parkes}, {Tilanus}, {Ullom},
  {Walther}, {Walton}, {Woodcraft}, {Amiri}, {Atkinson}, {Burger}, {Chuter},
  {Coulson}, {Doriese}, {Dunare}, {Economou}, {Niemack}, {Parsons},
  {Reintsema}, {Sibthorpe}, {Smail}, {Sudiwala}, \& {Thomas}}]{Holland2013}
{Holland}, W.~S., {Bintley}, D., {Chapin}, E.~L., {et~al.} 2013, Monthly
  Notices of the Royal Astronomical Society, 430, 2513,
  \dodoi{10.1093/mnras/sts612}

\bibitem[{{Hsieh} {et~al.}(2019){Hsieh}, {Murillo}, {Belloche}, {Hirano},
  {Walsh}, {van Dishoeck}, {J{\o}rgensen}, \& {Lai}}]{hsieh19}
{Hsieh}, T.-H., {Murillo}, N.~M., {Belloche}, A., {et~al.} 2019, \apj, 884,
  149, \dodoi{10.3847/1538-4357/ab425a}

\bibitem[{{Ilee} {et~al.}(2013){Ilee}, {Wheelwright}, {Oudmaijer}, {de Wit},
  {Maud}, {Hoare}, {Lumsden}, {Moore}, {Urquhart}, \& {Mottram}}]{ilee13}
{Ilee}, J.~D., {Wheelwright}, H.~E., {Oudmaijer}, R.~D., {et~al.} 2013, \mnras,
  429, 2960, \dodoi{10.1093/mnras/sts537}

\bibitem[{{Johnstone} {et~al.}(2001){Johnstone}, {Fich}, {Mitchell}, \&
  {Moriarty-Schieven}}]{johnstone01}
{Johnstone}, D., {Fich}, M., {Mitchell}, G.~F., \& {Moriarty-Schieven}, G.
  2001, \apj, 559, 307, \dodoi{10.1086/322323}

\bibitem[{{Johnstone} {et~al.}(2013){Johnstone}, {Hendricks}, {Herczeg}, \&
  {Bruderer}}]{johnstone13}
{Johnstone}, D., {Hendricks}, B., {Herczeg}, G.~J., \& {Bruderer}, S. 2013,
  \apj, 765, 133, \dodoi{10.1088/0004-637X/765/2/133}

\bibitem[{{Johnstone} {et~al.}(2018){Johnstone}, {Herczeg}, {Mairs},
  {Hatchell}, {Bower}, {Kirk}, {Lane}, {Bell}, {Graves}, \&
  {Aikawa}}]{Johnstone2018}
{Johnstone}, D., {Herczeg}, G.~J., {Mairs}, S., {et~al.} 2018, The
  Astrophysical Journal, 854, 31, \dodoi{10.3847/1538-4357/aaa764}

\bibitem[{{J{\o}rgensen} {et~al.}(2015){J{\o}rgensen}, {Visser}, {Williams}, \&
  {Bergin}}]{jorgensen15}
{J{\o}rgensen}, J.~K., {Visser}, R., {Williams}, J.~P., \& {Bergin}, E.~A.
  2015, \aap, 579, A23, \dodoi{10.1051/0004-6361/201425317}

\bibitem[{{Kang} {et~al.}(2015){Kang}, {Choi}, {Stutz}, \&
  {Tatematsu}}]{kang15}
{Kang}, M., {Choi}, M., {Stutz}, A.~M., \& {Tatematsu}, K. 2015, \apj, 814, 31,
  \dodoi{10.1088/0004-637X/814/1/31}

\bibitem[{{Karska} {et~al.}(2018){Karska}, {Kaufman}, {Kristensen}, {van
  Dishoeck}, {Herczeg}, {Mottram}, {Tychoniec}, {Lindberg}, {Evans}, {Green},
  {Yang}, {Gusdorf}, {Itrich}, \& {Si{\'o}dmiak}}]{karska18}
{Karska}, A., {Kaufman}, M.~J., {Kristensen}, L.~E., {et~al.} 2018, \apjs, 235,
  30, \dodoi{10.3847/1538-4365/aaaec5}

\bibitem[{{Kirk} {et~al.}(2016){Kirk}, {Di Francesco}, {Johnstone},
  {Duarte-Cabral}, {Sadavoy}, {Hatchell}, {Mottram}, {Buckle}, {Berry},
  {Broekhoven-Fiene}, {Currie}, {Fich}, {Jenness}, {Nutter}, {Pattle},
  {Pineda}, {Quinn}, {Salji}, {Tisi}, {Hogerheijde}, {Ward-Thompson},
  {Bastien}, {Bresnahan}, {Butner}, {Chen}, {Chrysostomou}, {Coude}, {Davis},
  {Drabek-Maunder}, {Fiege}, {Friberg}, {Friesen}, {Fuller}, {Graves},
  {Greaves}, {Gregson}, {Holland}, {Joncas}, {Kirk}, {Knee}, {Mairs}, {Marsh},
  {Matthews}, {Moriarty-Schieven}, {Mowat}, {Rawlings}, {Richer}, {Robertson},
  {Rosolowsky}, {Rumble}, {Thomas}, {Tothill}, {Viti}, {White}, {Wouterloot},
  {Yates}, \& {Zhu}}]{kirk2016}
{Kirk}, H., {Di Francesco}, J., {Johnstone}, D., {et~al.} 2016, \apj, 817, 167,
  \dodoi{10.3847/0004-637X/817/2/167}

\bibitem[{{K{\'o}sp{\'a}l} {et~al.}(2007){K{\'o}sp{\'a}l}, {{\'A}brah{\'a}m},
  {Prusti}, {Acosta-Pulido}, {Hony}, {Mo{\'o}r}, \& {Siebenmorgen}}]{kospal07}
{K{\'o}sp{\'a}l}, {\'A}., {{\'A}brah{\'a}m}, P., {Prusti}, T., {et~al.} 2007,
  \aap, 470, 211, \dodoi{10.1051/0004-6361:20066108}

\bibitem[{{K{\'o}sp{\'a}l} {et~al.}(2017){K{\'o}sp{\'a}l}, {{\'A}brah{\'a}m},
  {Westhues}, \& {Haas}}]{kospal17}
{K{\'o}sp{\'a}l}, {\'A}., {{\'A}brah{\'a}m}, P., {Westhues}, C., \& {Haas}, M.
  2017, \aap, 597, L10, \dodoi{10.1051/0004-6361/201629447}

\bibitem[{{K{\'o}sp{\'a}l} {et~al.}(2020){K{\'o}sp{\'a}l}, {Szab{\'o}},
  {{\'A}brah{\'a}m}, {Kraus}, {Takami}, {Lucas}, {Contreras Pe{\~n}a}, \&
  {Udalski}}]{kospal20}
{K{\'o}sp{\'a}l}, {\'A}., {Szab{\'o}}, Z.~M., {{\'A}brah{\'a}m}, P., {et~al.}
  2020, \apj, 889, 148, \dodoi{10.3847/1538-4357/ab6174}

\bibitem[{{Kounkel} {et~al.}(2018){Kounkel}, {Covey}, {Su{\'a}rez},
  {Rom{\'a}n-Z{\'u}{\~n}iga}, {Hernandez}, {Stassun}, {Jaehnig}, {Feigelson},
  {Pe{\~n}a Ram{\'\i}rez}, {Roman-Lopes}, {Da Rio}, {Stringfellow}, {Kim},
  {Borissova}, {Fern{\'a}ndez-Trincado}, {Burgasser},
  {Garc{\'\i}a-Hern{\'a}ndez}, {Zamora}, {Pan}, \& {Nitschelm}}]{kounkel18}
{Kounkel}, M., {Covey}, K., {Su{\'a}rez}, G., {et~al.} 2018, \aj, 156, 84,
  \dodoi{10.3847/1538-3881/aad1f1}

\bibitem[{{Laos} {et~al.}(2021){Laos}, {Greene}, {Najita}, \&
  {Stassun}}]{laos21}
{Laos}, S., {Greene}, T.~P., {Najita}, J.~R., \& {Stassun}, K.~G. 2021, \apj,
  921, 110, \dodoi{10.3847/1538-4357/ac1f1b}

\bibitem[{{Lee}(2007)}]{lee07}
{Lee}, J.-E. 2007, Journal of Korean Astronomical Society, 40, 83,
  \dodoi{10.5303/JKAS.2007.40.4.083}

\bibitem[{{Lee} {et~al.}(2010){Lee}, {Lee}, {Shinn}, {Dunham}, {Kim}, {Kim},
  {Bourke}, {Evans}, \& {Choi}}]{lee10}
{Lee}, J.-E., {Lee}, H.-G., {Shinn}, J.-H., {et~al.} 2010, \apjl, 709, L74,
  \dodoi{10.1088/2041-8205/709/1/L74}

\bibitem[{{Lee} {et~al.}(2016){Lee}, {Lee}, {Park}, {Lee}, {Kidder}, {Mace}, \&
  {Jaffe}}]{lee16}
{Lee}, S., {Lee}, J.-E., {Park}, S., {et~al.} 2016, \apj, 826, 179,
  \dodoi{10.3847/0004-637X/826/2/179}

\bibitem[{{Lee} {et~al.}(2020){Lee}, {Johnstone}, {Lee}, {Herczeg}, {Mairs},
  {Varricatt}, {Hodapp}, {Naylor}, {Pe{\~n}a}, {Baek}, {Haas}, {Chini}, \&
  {JCMT Transient Team}}]{lee20}
{Lee}, Y.-H., {Johnstone}, D., {Lee}, J.-E., {et~al.} 2020, \apj, 903, 5,
  \dodoi{10.3847/1538-4357/abb6fe}

\bibitem[{{Lee} {et~al.}(2021){Lee}, {Johnstone}, {Lee}, {Herczeg}, {Mairs},
  {Contreras-Pe{\~n}a}, {Hatchell}, {Naylor}, {Bell}, {Bourke}, {Broughton},
  {Francis}, {Gupta}, {Harsono}, {Liu}, {Park}, {Plovie}, {Moriarty-Schieven},
  {Scholz}, {Sharma}, {Teixeira}, {Wang}, {Aikawa}, {Bower}, {Chen}, {Bae},
  {Baek}, {Chapman}, {Chen}, {Du}, {Dutta}, {Forbrich}, {Guo}, {Inutsuka},
  {Kang}, {Kirk}, {Kuan}, {Kwon}, {Lai}, {Lalchand}, {Lane}, {Lee}, {Liu},
  {Morata}, {Pearson}, {Pon}, {Sahu}, {Shang}, {Stamatellos}, {Tang}, {Xu}, \&
  {Yoo}}]{lee21}
---. 2021, arXiv e-prints, arXiv:2107.10750.
\newblock \doarXiv{2107.10750}

\bibitem[{{Lucas} {et~al.}(2017){Lucas}, {Smith}, {Contreras Pe{\~n}a},
  {Froebrich}, {Drew}, {Kumar}, {Borissova}, {Minniti}, {Kurtev}, \&
  {Mongui{\'o}}}]{lucas17}
{Lucas}, P.~W., {Smith}, L.~C., {Contreras Pe{\~n}a}, C., {et~al.} 2017,
  \mnras, 472, 2990, \dodoi{10.1093/mnras/stx2058}

\bibitem[{{MacFarlane} {et~al.}(2019{\natexlab{a}}){MacFarlane}, {Stamatellos},
  {Johnstone}, {Herczeg}, {Baek}, {Chen}, {Kang}, \& {Lee}}]{macfarlane19a}
{MacFarlane}, B., {Stamatellos}, D., {Johnstone}, D., {et~al.}
  2019{\natexlab{a}}, \mnras, 487, 5106, \dodoi{10.1093/mnras/stz1512}

\bibitem[{{MacFarlane} {et~al.}(2019{\natexlab{b}}){MacFarlane}, {Stamatellos},
  {Johnstone}, {Herczeg}, {Baek}, {Vivien Chen}, {Kang}, \&
  {Lee}}]{macfarlane19b}
---. 2019{\natexlab{b}}, \mnras, 1508, \dodoi{10.1093/mnras/stz1570}

\bibitem[{{Mainzer} {et~al.}(2011){Mainzer}, {Bauer}, {Grav}, {Masiero},
  {Cutri}, {Dailey}, {Eisenhardt}, {McMillan}, {Wright}, {Walker}, {Jedicke},
  {Spahr}, {Tholen}, {Alles}, {Beck}, {Brandenburg}, {Conrow}, {Evans},
  {Fowler}, {Jarrett}, {Marsh}, {Masci}, {McCallon}, {Wheelock}, {Wittman},
  {Wyatt}, {DeBaun}, {Elliott}, {Elsbury}, {Gautier}, {Gomillion}, {Leisawitz},
  {Maleszewski}, {Micheli}, \& {Wilkins}}]{mainzer11}
{Mainzer}, A., {Bauer}, J., {Grav}, T., {et~al.} 2011, \apj, 731, 53,
  \dodoi{10.1088/0004-637X/731/1/53}

\bibitem[{{Mainzer} {et~al.}(2014){Mainzer}, {Bauer}, {Cutri}, {Grav},
  {Masiero}, {Beck}, {Clarkson}, {Conrow}, {Dailey}, {Eisenhardt}, {Fabinsky},
  {Fajardo-Acosta}, {Fowler}, {Gelino}, {Grillmair}, {Heinrichsen}, {Kendall},
  {Kirkpatrick}, {Liu}, {Masci}, {McCallon}, {Nugent}, {Papin}, {Rice},
  {Royer}, {Ryan}, {Sevilla}, {Sonnett}, {Stevenson}, {Thompson}, {Wheelock},
  {Wiemer}, {Wittman}, {Wright}, \& {Yan}}]{mainzer14}
{Mainzer}, A., {Bauer}, J., {Cutri}, R.~M., {et~al.} 2014, \apj, 792, 30,
  \dodoi{10.1088/0004-637X/792/1/30}

\bibitem[{{Mairs} {et~al.}(2017{\natexlab{a}}){Mairs}, {Johnstone}, {Kirk},
  {Lane}, {Bell}, {Graves}, {Herczeg}, {Scicluna}, {Bower}, \&
  {Chen}}]{Mairs2017GBS}
{Mairs}, S., {Johnstone}, D., {Kirk}, H., {et~al.} 2017{\natexlab{a}}, The
  Astrophysical Journal, 849, 107, \dodoi{10.3847/1538-4357/aa9225}

\bibitem[{{Mairs} {et~al.}(2017{\natexlab{b}}){Mairs}, {Lane}, {Johnstone},
  {Kirk}, {Lacaille}, {Bower}, {Bell}, {Graves}, {Chapman}, \& {The JCMT
  Transient Team}}]{Mairs2017Cal}
{Mairs}, S., {Lane}, J., {Johnstone}, D., {et~al.} 2017{\natexlab{b}}, The
  Astrophysical Journal, 843, 55, \dodoi{10.3847/1538-4357/aa7844}

\bibitem[{{Mairs} {et~al.}(2021){Mairs}, {Dempsey}, {Bell}, {Parsons},
  {Currie}, {Friberg}, {Jiang}, {Tetarenko}, {Bintley}, {Cookson}, {Li},
  {Rawlings}, {Wouterloot}, {Berry}, {Graves}, {Mizuno}, {Acohido}, {Clark},
  {Cox}, {Fuchs}, {Hoge}, {Kemp}, {Lee}, {Matulonis}, {Montgomerie}, {Silva},
  \& {Smith}}]{mairs21}
{Mairs}, S., {Dempsey}, J.~T., {Bell}, G.~S., {et~al.} 2021, arXiv e-prints,
  arXiv:2107.13558.
\newblock \doarXiv{2107.13558}

\bibitem[{{Ma{\'\i}z Apell{\'a}niz} {et~al.}(2020){Ma{\'\i}z Apell{\'a}niz},
  {Pantaleoni Gonz{\'a}lez}, {Barb{\'a}}, {Garc{\'\i}a-Lario}, \&
  {Nogueras-Lara}}]{maiz-apellaniz20}
{Ma{\'\i}z Apell{\'a}niz}, J., {Pantaleoni Gonz{\'a}lez}, M., {Barb{\'a}},
  R.~H., {Garc{\'\i}a-Lario}, P., \& {Nogueras-Lara}, F. 2020, \mnras, 496,
  4951, \dodoi{10.1093/mnras/staa1790}

\bibitem[{{Manoj} {et~al.}(2013){Manoj}, {Watson}, {Neufeld}, {Megeath},
  {Vavrek}, {Yu}, {Visser}, {Bergin}, {Fischer}, {Tobin}, {Stutz}, {Ali},
  {Wilson}, {Di Francesco}, {Osorio}, {Maret}, \& {Poteet}}]{manoj13}
{Manoj}, P., {Watson}, D.~M., {Neufeld}, D.~A., {et~al.} 2013, \apj, 763, 83,
  \dodoi{10.1088/0004-637X/763/2/83}

\bibitem[{{Manoj} {et~al.}(2016){Manoj}, {Green}, {Megeath}, {Evans}, {Stutz},
  {Tobin}, {Watson}, {Fischer}, {Furlan}, \& {Henning}}]{manoj16}
{Manoj}, P., {Green}, J.~D., {Megeath}, S.~T., {et~al.} 2016, \apj, 831, 69,
  \dodoi{10.3847/0004-637X/831/1/69}

\bibitem[{{McMahon} {et~al.}(2013){McMahon}, {Banerji}, {Gonzalez}, {Koposov},
  {Bejar}, {Lodieu}, {Rebolo}, \& {VHS Collaboration}}]{mcmahon13}
{McMahon}, R.~G., {Banerji}, M., {Gonzalez}, E., {et~al.} 2013, The Messenger,
  154, 35

\bibitem[{{McMullin} {et~al.}(2007){McMullin}, {Waters}, {Schiebel}, {Young},
  \& {Golap}}]{mcmullin2007}
{McMullin}, J.~P., {Waters}, B., {Schiebel}, D., {Young}, W., \& {Golap}, K.
  2007, in Astronomical Society of the Pacific Conference Series, Vol. 376,
  Astronomical Data Analysis Software and Systems XVI, ed. R.~A. {Shaw},
  F.~{Hill}, \& D.~J. {Bell}, 127

\bibitem[{{Megeath} {et~al.}(2012){Megeath}, {Gutermuth}, {Muzerolle},
  {Kryukova}, {Flaherty}, {Hora}, {Allen}, {Hartmann}, {Myers}, {Pipher},
  {Stauffer}, {Young}, \& {Fazio}}]{megeath12}
{Megeath}, S.~T., {Gutermuth}, R., {Muzerolle}, J., {et~al.} 2012, \aj, 144,
  192, \dodoi{10.1088/0004-6256/144/6/192}

\bibitem[{{Mitchell} {et~al.}(2001){Mitchell}, {Johnstone},
  {Moriarty-Schieven}, {Fich}, \& {Tothill}}]{mitchell01}
{Mitchell}, G.~F., {Johnstone}, D., {Moriarty-Schieven}, G., {Fich}, M., \&
  {Tothill}, N.~F.~H. 2001, \apj, 556, 215, \dodoi{10.1086/321574}

\bibitem[{{Motte} {et~al.}(2001){Motte}, {Andr{\'e}}, {Ward-Thompson}, \&
  {Bontemps}}]{motte01}
{Motte}, F., {Andr{\'e}}, P., {Ward-Thompson}, D., \& {Bontemps}, S. 2001,
  \aap, 372, L41, \dodoi{10.1051/0004-6361:20010543}

\bibitem[{{Muzerolle} {et~al.}(2013){Muzerolle}, {Furlan}, {Flaherty}, {Balog},
  \& {Gutermuth}}]{muzerolle13}
{Muzerolle}, J., {Furlan}, E., {Flaherty}, K., {Balog}, Z., \& {Gutermuth}, R.
  2013, \nat, 493, 378, \dodoi{10.1038/nature11746}

\bibitem[{{Nagy} {et~al.}(2020){Nagy}, {Menechella}, {Megeath}, {Tobin},
  {Booker}, {Fischer}, {Manoj}, {Stanke}, {Stutz}, \& {Wyrowski}}]{nagy20}
{Nagy}, Z., {Menechella}, A., {Megeath}, S.~T., {et~al.} 2020, \aap, 642, A137,
  \dodoi{10.1051/0004-6361/201937342}

\bibitem[{{Najita} {et~al.}(1996){Najita}, {Carr}, {Glassgold}, {Shu}, \&
  {Tokunaga}}]{najita96}
{Najita}, J., {Carr}, J.~S., {Glassgold}, A.~E., {Shu}, F.~H., \& {Tokunaga},
  A.~T. 1996, \apj, 462, 919, \dodoi{10.1086/177205}

\bibitem[{{Nomura} {et~al.}(2007){Nomura}, {Aikawa}, {Tsujimoto}, {Nakagawa},
  \& {Millar}}]{nomura07}
{Nomura}, H., {Aikawa}, Y., {Tsujimoto}, M., {Nakagawa}, Y., \& {Millar}, T.~J.
  2007, \apj, 661, 334, \dodoi{10.1086/513419}

\bibitem[{{Oh} {et~al.}(2018){Oh}, {Pyo}, {Koo}, {Yuk}, {Kaplan}, {Lee},
  {Sokal}, {Mace}, {Park}, {Lee}, {Park}, {Hwang}, {Kim}, \& {Jaffe}}]{oh18}
{Oh}, H., {Pyo}, T.-S., {Koo}, B.-C., {et~al.} 2018, \apj, 858, 23,
  \dodoi{10.3847/1538-4357/aabba4}

\bibitem[{{Okoda} {et~al.}(2021){Okoda}, {Oya}, {Francis}, {Johnstone},
  {Inutsuka}, {Ceccarelli}, {Codella}, {Chandler}, {Sakai}, {Aikawa}, {Alves},
  {Balucani}, {Bianchi}, {Bouvier}, {Caselli}, {Caux}, {Charnley}, {Choudhury},
  {De Simone}, {Dulieu}, {Dur{\'a}n}, {Evans}, {Favre}, {Fedele}, {Feng},
  {Fontani}, {Hama}, {Hanawa}, {Herbst}, {Hirota}, {Imai}, {Isella},
  {J{\'\i}menez-Serra}, {Kahane}, {Lefloch}, {Loinard}, {L{\'o}pez-Sepulcre},
  {Maud}, {Maureira}, {Menard}, {Mercimek}, {Miotello}, {Moellenbrock}, {Mori},
  {Murillo}, {Nakatani}, {Nomura}, {Oba}, {O'Donoghue}, {Ohashi},
  {Ospina-Zamudio}, {Pineda}, {Podio}, {Rimola}, {Sakai}, {Segura-Cox},
  {Shirley}, {Svoboda}, {Taquet}, {Testi}, {Vastel}, {Viti}, {Watanabe},
  {Watanabe}, {Witzel}, {Xue}, {Zhang}, {Zhao}, \& {Yamamoto}}]{okoda21}
{Okoda}, Y., {Oya}, Y., {Francis}, L., {et~al.} 2021, \apj, 910, 11,
  \dodoi{10.3847/1538-4357/abddb1}

\bibitem[{{Park} {et~al.}(2021){Park}, {Lee}, {Contreras Pe{\~n}a},
  {Johnstone}, {Herczeg}, {Lee}, {Lee}, {Bhardwaj}, \&
  {Moriarty-Schieven}}]{park21}
{Park}, W., {Lee}, J.-E., {Contreras Pe{\~n}a}, C., {et~al.} 2021, arXiv
  e-prints, arXiv:2107.10751.
\newblock \doarXiv{2107.10751}

\bibitem[{{Phillips} {et~al.}(2001){Phillips}, {Gibb}, \&
  {Little}}]{phillips01}
{Phillips}, R.~R., {Gibb}, A.~G., \& {Little}, L.~T. 2001, \mnras, 326, 927,
  \dodoi{10.1046/j.1365-8711.2001.04502.x}

\bibitem[{{Pilbratt} {et~al.}(2010){Pilbratt}, {Riedinger}, {Passvogel},
  {Crone}, {Doyle}, {Gageur}, {Heras}, {Jewell}, {Metcalfe}, {Ott}, \&
  {Schmidt}}]{pilbratt10}
{Pilbratt}, G.~L., {Riedinger}, J.~R., {Passvogel}, T., {et~al.} 2010, \aap,
  518, L1, \dodoi{10.1051/0004-6361/201014759}

\bibitem[{{Plunkett} {et~al.}(2015){Plunkett}, {Arce}, {Mardones}, {van
  Dokkum}, {Dunham}, {Fern{\'a}ndez-L{\'o}pez}, {Gallardo}, \&
  {Corder}}]{plunkett15}
{Plunkett}, A.~L., {Arce}, H.~G., {Mardones}, D., {et~al.} 2015, \nat, 527, 70,
  \dodoi{10.1038/nature15702}

\bibitem[{{Rebull} {et~al.}(2014){Rebull}, {Cody}, {Covey}, {G{\"u}nther},
  {Hillenbrand}, {Plavchan}, {Poppenhaeger}, {Stauffer}, {Wolk}, {Gutermuth},
  {Morales-Calder{\'o}n}, {Song}, {Barrado}, {Bayo}, {James}, {Hora}, {Vrba},
  {Alves de Oliveira}, {Bouvier}, {Carey}, {Carpenter}, {Favata}, {Flaherty},
  {Forbrich}, {Hernandez}, {McCaughrean}, {Megeath}, {Micela}, {Smith},
  {Terebey}, {Turner}, {Allen}, {Ardila}, {Bouy}, \& {Guieu}}]{rebull14}
{Rebull}, L.~M., {Cody}, A.~M., {Covey}, K.~R., {et~al.} 2014, \aj, 148, 92,
  \dodoi{10.1088/0004-6256/148/5/92}

\bibitem[{{Reipurth}(1989)}]{reipurth89}
{Reipurth}, B. 1989, \nat, 340, 42, \dodoi{10.1038/340042a0}

\bibitem[{{Safron} {et~al.}(2015){Safron}, {Fischer}, {Megeath}, {Furlan},
  {Stutz}, {Stanke}, {Billot}, {Rebull}, {Tobin}, {Ali}, {Allen}, {Booker},
  {Watson}, \& {Wilson}}]{safron15}
{Safron}, E.~J., {Fischer}, W.~J., {Megeath}, S.~T., {et~al.} 2015, \apjl, 800,
  L5, \dodoi{10.1088/2041-8205/800/1/L5}

\bibitem[{{Scholz} {et~al.}(2013){Scholz}, {Froebrich}, \& {Wood}}]{scholz13}
{Scholz}, A., {Froebrich}, D., \& {Wood}, K. 2013, \mnras, 430, 2910,
  \dodoi{10.1093/mnras/stt091}

\bibitem[{{Smith}(1995)}]{smith95}
{Smith}, M.~D. 1995, \aap, 296, 789

\bibitem[{{Spezzi} {et~al.}(2015){Spezzi}, {Petr-Gotzens}, {Alcal{\'a}},
  {J{\o}rgensen}, {Stanke}, {Lombardi}, \& {Alves}}]{spezzi15}
{Spezzi}, L., {Petr-Gotzens}, M.~G., {Alcal{\'a}}, J.~M., {et~al.} 2015, \aap,
  581, A140, \dodoi{10.1051/0004-6361/201425417}

\bibitem[{{Stecklum} {et~al.}(2021){Stecklum}, {Wolf}, {Linz}, {Caratti o
  Garatti}, {Schmidl}, {Klose}, {Eisl{\"o}ffel}, {Fischer}, {Brogan}, {Burns},
  {Bayandina}, {Cyganowski}, {Gurwell}, {Hunter}, {Hirano}, {Kim}, {MacLeod},
  {Menten}, {Olech}, {Orosz}, {Sobolev}, {Sridharan}, {Surcis}, {Sugiyama},
  {van der Walt}, {Volvach}, \& {Yonekura}}]{stecklum21}
{Stecklum}, B., {Wolf}, V., {Linz}, H., {et~al.} 2021, \aap, 646, A161,
  \dodoi{10.1051/0004-6361/202039645}

\bibitem[{{Stutz} {et~al.}(2013){Stutz}, {Tobin}, {Stanke}, {Megeath},
  {Fischer}, {Robitaille}, {Henning}, {Ali}, {di Francesco}, \&
  {Furlan}}]{Stutz2013}
{Stutz}, A.~M., {Tobin}, J.~J., {Stanke}, T., {et~al.} 2013, The Astrophysical
  Journal, 767, 36, \dodoi{10.1088/0004-637X/767/1/36}

\bibitem[{{Szegedi-Elek} {et~al.}(2020){Szegedi-Elek}, {{\'A}brah{\'a}m},
  {Wyrzykowski}, {Kun}, {K{\'o}sp{\'a}l}, {Chen}, {Marton}, {Mo{\'o}r}, {Kiss},
  {P{\'a}l}, {Szabados}, {Varga}, {Varga-Vereb{\'e}lyi}, {Andreas}, {Bachelet},
  {Bischoff}, {B{\'o}di}, {Breedt}, {Burgaz}, {Butterley}, {Carrasco},
  {{\v{C}}epas}, {Damljanovic}, {Gezer}, {Godunova}, {Gromadzki}, {Gurgul},
  {Hardy}, {Hildebrandt}, {Hoffmann}, {Hundertmark}, {Ihanec}, {Janulis},
  {Kalup}, {Kaczmarek}, {K{\"o}nyves-T{\'o}th}, {Krezinger}, {Kruszy{\'n}ska},
  {Littlefair}, {Maskoli{\={u}}nas}, {M{\'e}sz{\'a}ros}, {Miko{\l}ajczyk},
  {Mugrauer}, {Netzel}, {Ordasi}, {Pak{\v{s}}tien{\.{e}}}, {Rybicki},
  {S{\'a}rneczky}, {Seli}, {Simon}, {{\v{S}}i{\v{s}}kauskait{\.{e}}},
  {S{\'o}dor}, {Sokolovsky}, {Stenglein}, {Street}, {Szak{\'a}ts}, {Tomasella},
  {Tsapras}, {Vida}, {Zdanavi{\v{c}}ius}, {Zieli{\'n}ski}, {Zieli{\'n}ski}, \&
  {Zi{\'o}{\l}kowska}}]{szegedi20}
{Szegedi-Elek}, E., {{\'A}brah{\'a}m}, P., {Wyrzykowski}, {\L}., {et~al.} 2020,
  \apj, 899, 130, \dodoi{10.3847/1538-4357/aba129}

\bibitem[{{Takami} {et~al.}(2006){Takami}, {Chrysostomou}, {Ray}, {Davis},
  {Dent}, {Bailey}, {Tamura}, {Terada}, \& {Pyo}}]{takami06}
{Takami}, M., {Chrysostomou}, A., {Ray}, T.~P., {et~al.} 2006, \apj, 641, 357,
  \dodoi{10.1086/500352}

\bibitem[{{Tappe} {et~al.}(2012){Tappe}, {Forbrich}, {Mart{\'\i}n}, {Yuan}, \&
  {Lada}}]{tappe12}
{Tappe}, A., {Forbrich}, J., {Mart{\'\i}n}, S., {Yuan}, Y., \& {Lada}, C.~J.
  2012, \apj, 751, 9, \dodoi{10.1088/0004-637X/751/1/9}

\bibitem[{{Tobin} {et~al.}(2015){Tobin}, {Stutz}, {Megeath}, {Fischer},
  {Henning}, {Ragan}, {Ali}, {Stanke}, {Manoj}, {Calvet}, \&
  {Hartmann}}]{tobin15}
{Tobin}, J.~J., {Stutz}, A.~M., {Megeath}, S.~T., {et~al.} 2015, \apj, 798,
  128, \dodoi{10.1088/0004-637X/798/2/128}

\bibitem[{{Tobin} {et~al.}(2016){Tobin}, {Stutz}, {Manoj}, {Megeath}, {Karska},
  {Nagy}, {Wyrowski}, {Fischer}, {Watson}, \& {Stanke}}]{tobin16}
{Tobin}, J.~J., {Stutz}, A.~M., {Manoj}, P., {et~al.} 2016, \apj, 831, 36,
  \dodoi{10.3847/0004-637X/831/1/36}

\bibitem[{{Tobin} {et~al.}(2020{\natexlab{a}}){Tobin}, {Sheehan}, {Reynolds},
  {Megeath}, {Osorio}, {Anglada}, {D{\'\i}az-Rodr{\'\i}guez}, {Furlan},
  {Kratter}, {Offner}, {Looney}, {Kama}, {Li}, {van't Hoff}, {Sadavoy}, \&
  {Karnath}}]{tobin20hops370}
{Tobin}, J.~J., {Sheehan}, P.~D., {Reynolds}, N., {et~al.} 2020{\natexlab{a}},
  \apj, 905, 162, \dodoi{10.3847/1538-4357/abc5bf}

\bibitem[{{Tobin} {et~al.}(2020{\natexlab{b}}){Tobin}, {Sheehan}, {Megeath},
  {D{\'\i}az-Rodr{\'\i}guez}, {Offner}, {Murillo}, {van 't Hoff}, {van
  Dishoeck}, {Osorio}, {Anglada}, {Furlan}, {Stutz}, {Reynolds}, {Karnath},
  {Fischer}, {Persson}, {Looney}, {Li}, {Stephens}, {Chandler}, {Cox},
  {Dunham}, {Tychoniec}, {Kama}, {Kratter}, {Kounkel}, {Mazur}, {Maud},
  {Patel}, {Perez}, {Sadavoy}, {Segura-Cox}, {Sharma}, {Stephenson}, {Watson},
  \& {Wyrowski}}]{tobin20}
{Tobin}, J.~J., {Sheehan}, P.~D., {Megeath}, S.~T., {et~al.}
  2020{\natexlab{b}}, \apj, 890, 130, \dodoi{10.3847/1538-4357/ab6f64}

\bibitem[{{Tychoniec} {et~al.}(2019){Tychoniec}, {Hull}, {Kristensen}, {Tobin},
  {Le Gouellec}, \& {van Dishoeck}}]{tychoniec19}
{Tychoniec}, {\L}., {Hull}, C. L.~H., {Kristensen}, L.~E., {et~al.} 2019, \aap,
  632, A101, \dodoi{10.1051/0004-6361/201935409}

\bibitem[{{Varricatt} {et~al.}(2010){Varricatt}, {Davis}, {Ramsay}, \&
  {Todd}}]{varricatt10}
{Varricatt}, W.~P., {Davis}, C.~J., {Ramsay}, S., \& {Todd}, S.~P. 2010,
  \mnras, 404, 661, \dodoi{10.1111/j.1365-2966.2010.16356.x}

\bibitem[{{Visser} {et~al.}(2012){Visser}, {Kristensen}, {Bruderer}, {van
  Dishoeck}, {Herczeg}, {Brinch}, {Doty}, {Harsono}, \& {Wolfire}}]{visser12}
{Visser}, R., {Kristensen}, L.~E., {Bruderer}, S., {et~al.} 2012, \aap, 537,
  A55, \dodoi{10.1051/0004-6361/201117109}

\bibitem[{{Wang} \& {Chen}(2019)}]{wang19}
{Wang}, S., \& {Chen}, X. 2019, \apj, 877, 116,
  \dodoi{10.3847/1538-4357/ab1c61}

\bibitem[{{Weingartner} \& {Draine}(2001)}]{weingartner01}
{Weingartner}, J.~C., \& {Draine}, B.~T. 2001, \apj, 548, 296,
  \dodoi{10.1086/318651}

\bibitem[{{Wright} {et~al.}(2010){Wright}, {Eisenhardt}, {Mainzer}, {Ressler},
  {Cutri}, {Jarrett}, {Kirkpatrick}, {Padgett}, {McMillan}, {Skrutskie},
  {Stanford}, {Cohen}, {Walker}, {Mather}, {Leisawitz}, {Gautier}, {McLean},
  {Benford}, {Lonsdale}, {Blain}, {Mendez}, {Irace}, {Duval}, {Liu}, {Royer},
  {Heinrichsen}, {Howard}, {Shannon}, {Kendall}, {Walsh}, {Larsen}, {Cardon},
  {Schick}, {Schwalm}, {Abid}, {Fabinsky}, {Naes}, \& {Tsai}}]{wright10}
{Wright}, E.~L., {Eisenhardt}, P. R.~M., {Mainzer}, A.~K., {et~al.} 2010, \aj,
  140, 1868, \dodoi{10.1088/0004-6256/140/6/1868}

\bibitem[{{Zakri} {et~al.}(2022){Zakri}, {Megeath}, {Fischer}, {Gutermuth},
  {Furlan}, {Hartmann}, {Karnath}, {Osorio}, {Safron}, {Stanke}, {Stutz},
  {Tobin}, {Allen}, {Federman}, {Habel}, {Manoj}, {Narang}, {Pokhrel},
  {Rebull}, {Sheehan}, \& {Watson}}]{zakri22}
{Zakri}, W., {Megeath}, S.~T., {Fischer}, W.~J., {et~al.} 2022, arXiv e-prints,
  arXiv:2201.04647.
\newblock \doarXiv{2201.04647}

\bibitem[{{Zhu} {et~al.}(2007){Zhu}, {Hartmann}, {Calvet}, {Hernandez},
  {Muzerolle}, \& {Tannirkulam}}]{zhu07}
{Zhu}, Z., {Hartmann}, L., {Calvet}, N., {et~al.} 2007, \apj, 669, 483,
  \dodoi{10.1086/521345}

\end{thebibliography}

\pagebreak

\appendix

\section{Astrometric shifts and uncertainties for accurate positional comparisons} \label{appendix-a}

The absolute astrometry for some observations may be unreliable.  In each image listed in Table~\ref{tab:astrometry}, we measure the centroid of a set of objects and then register the absolute pointing of the image based on the positions of those objects in WISE \citep{cutri14}.

Table~\ref{tab:astrometry} and~\ref{tab:midir_img_uncertainty} lists the adopted uncertainties in right ascention and declination in each band. For 2MASS, PACS, and SCUBA-2 450 and 850 $\mu$m images, the uncertainties are measured from the astrometry using the centroids of objects presented in Table~\ref{tab:astrometry}. The uncertainties in the WISE image are adopted from the ALLWISE catalog. Table~\ref{tab:midir_img_uncertainty} lists the uncertainties in mid-infrared images which are estimated from the 2-dimensional Gaussian fit and assuming that the astrometry is accurate.

\begin{table}[!ht]
\caption{Centroid positions}
\label{tab:astrometry}
\begin{center}
\begin{tabular}{llcccc}
Object & Type$^a$ & RA (J2000) & Dec (J2000) & $\Delta$RA ($^{\prime\prime}$)$^b$ & $\Delta$Dec ($^{\prime\prime}$)$^b$ \\
\hline
\multicolumn{6}{c}{WISE Positions}\\
\hline
HOPS 388 & Protostar & 05 46 13.136 & -00 06 04.85 & -- & --\\
LkHA 301 & Disk & 05 46 19.466 & -00 05 20.02 & -- & -- \\
HOPS 321 & Protostar & 05 46 33.184 &  00 00 02.03 & -- & -- \\
HOPS 363 & Protostar & 05 46 43.129 &  00 00 52.28 & -- & -- \\
\hline
\multicolumn{2}{c}{Adopted HOPS 373} & 05 36 30.705 & -00 02 35.23 & -- & -- \\ 
\hline
\hline
\multicolumn{6}{c}{2MASS Positions}\\
\hline
HOPS 388 & Protostar & 05 46 13.135 & -00 06 04.82 & -0.015 & 0.034 \\
LkHA 301 & Disk & 05 46 19.468 & -00 05 19.99 &  0.030 & 0.027 \\
HOPS 321 & Protostar & 05 46 33.188 &  00 00 02.15 &  0.067 & 0.116 \\
HOPS 363 & Protostar & 05 46 43.112 &  00 00 52.30 & -0.246 & 0.022 \\
\hline
\multicolumn{4}{r}{Shift}       & -0.041 & 0.050 \\
\multicolumn{4}{r}{Uncertainty} &  0.141 & 0.044 \\
\hline
\multicolumn{2}{c}{Adopted HOPS 373} & 05 46 30.648 & -00 02 35.02 & -0.855 & 0.210 \\
\hline
\hline
\multicolumn{6}{c}{SCUBA-2 Positions$^c$}\\
\hline
\multicolumn{6}{c}{850 $\mu$m}\\
\hline
LkHa 298 &  Disk  & 05 46 04.618 &   00 04 59.88  & -0.09 & 1.71\\
V1647 Ori & FUor disk  & 05 46 13.140 &  -00 06 04.06  & -0.05 & 0.79\\
LkHa 301 & Disk & 05 46 19.457 & -00 05 18.55 & -0.12 & 1.39 \\
(LkHa 309) & Disk, excluded & 05 47 06.861 &  00 00 48.81 & -1.57 & 1.16 \\
MGM2012 3292 & Disk & 05 46 18.032 &  00 12 12.89 & 0.32 & 0.73 \\
\hline
\multicolumn{6}{c}{450 $\mu$m}\\
\hline
HOPS 315 & Protostar & 05 46 03.604 & -00 14 47.33 & -0.44 & 2.19 \\
HOPS 385 & Protostar & 05 46 04.801 & -00 14 15.20 &  0.24 & 1.49 \\
V1647 Ori & FUor disk & 05 46 13.174 & -00 06 02.74 & 0.57 & 2.11 \\
LkHa 301 & Disk & 05 46 19.530 & -00 05 18.00 & 0.98 & 1.94 \\
\hline
\multicolumn{4}{r}{Shift} & 0.18 & 1.54\\
\multicolumn{4}{r}{Uncertainty} & 0.42 & 0.52\\
\hline
Adopted HOPS 373 & 850 $\mu$m & 05 46 30.913 & -00 02 34.43 & 3.12 & 0.80 \\
Adopted HOPS 373 & 450 $\mu$m & 05 46 30.902 & -00 02 34.17 & 2.96 & 1.06 \\
\hline
\hline
\multicolumn{6}{c}{MIPS 70 $\mu$m Positions}\\
\hline 
HOPS 388 & Protostar & 05 46 13.046 & -00 06 04.64 & -1.359 &  0.207 \\
LkHA 301 & Disk & 05 46 19.473 & -00 05 16.21 &  0.100 &  3.805 \\ 
HOPS 321 & Protostar & 05 46 33.250 &  00 00 03.83 &  0.991 &  1.796 \\ 
HOPS 363 & Protostar & 05 46 43.145 &  00 00 55.14 &  0.247 &  2.857 \\ 
\hline
\multicolumn{4}{r}{Shift} & -0.005 & 2.166 \\
\multicolumn{4}{r}{Uncertainty} & 0.983 & 1.543 \\
\hline
\multicolumn{2}{c}{Adopted HOPS 373} & 05 46 30.916 & -00 02 37.85 & 3.165 & -2.62 \\
\hline
\hline
\multicolumn{6}{c}{PACS 70 $\mu$m Positions} \\
\hline 
HOPS 388 & Protostar & 05 46 13.128 & -00 06 04.84 & -0.128 &  0.013 \\
LkHA 301 & Disk & 05 46 19.432 & -00 05 19.64 & -0.512 &  0.372 \\ 
HOPS 321 &Protostar & 05 46 33.152 &  00 00 01.45 & -0.485 & -0.581 \\ 
HOPS 363 &Protostar & 05 46 43.084 &  00 00 51.93 & -0.665 & -0.347 \\ 
\hline
\multicolumn{4}{r}{Shift} & -0.298 & -0.136 \\
\multicolumn{4}{r}{Uncertainty} & 0.233 & 0.417 \\
\hline 
\multicolumn{2}{c}{Adopted HOPS 373} & 05 46 30.859 & -00 02 35.31 &  2.310 & -0.08 \\
\hline
\hline
\multicolumn{6}{c}{PACS 160 $\mu$m Positions} \\
\hline
HOPS 388 &Protostar & 05 46 13.150 & -00 06 05.08 &  0.207 & -0.233 \\
LkHA 301 & Disk & 05 46 19.450 & -00 05 19.27 & -0.243 &  0.751 \\ 
HOPS 321 &Protostar & 05 46 33.291 &  00 00 00.62 &  1.609 & -1.405 \\ 
HOPS 363 &Protostar & 05 46 43.081 &  00 00 51.08 & -0.712 & -1.195 \\ 
\hline
\multicolumn{4}{r}{Shift} & 0.215 & -0.512 \\
\multicolumn{4}{r}{Uncertainty} & 1.002 & 0.989 \\
\hline
\multicolumn{2}{c}{Adopted HOPS 373} &  05 46 30.855 & -00 02 34.87 & 2.250  & 0.36 \\
\hline
\hline
\multicolumn{6}{l}{$^a$Classifications from \citet{megeath12}.}\\
\multicolumn{6}{l}{$^b$Offset between {\it WISE} position and measured position.}\\
\multicolumn{6}{l}{$^c$SCUBA-2 450 and 850 $\mu$m images have same pointing solution.}\\
\end{tabular}
\end{center}
\end{table}

\begin{table}[!ht]
    \caption{Uncertainty in centroid positions for mid-infrared images}
    \label{tab:midir_img_uncertainty}
    \centering
    \begin{tabular}{cccc}
        Instrument & $\lambda$ & $\Delta$RA (\arcsec) & $\Delta$Dec (\arcsec) \\
        \hline
        IRAC & 3.6 $\mu$m & 0.012 & 0.009 \\ 
        IRAC & 4.5 $\mu$m & 0.018 & 0.012 \\
        IRAC & 5.8 $\mu$m & 0.020 & 0.014 \\
        IRAC & 8.0 $\mu$m & 0.020 & 0.014 \\
        IRAC & 8.0 $\mu$m & 0.020 & 0.014 \\
        MIPS & 24 $\mu$m & 0.072 & 0.063 \\
        \hline 
    \end{tabular}
\end{table}

\section{Channel maps for CO emission from HOPS 373}

\begin{figure*}
    \includegraphics[width=0.99\textwidth]{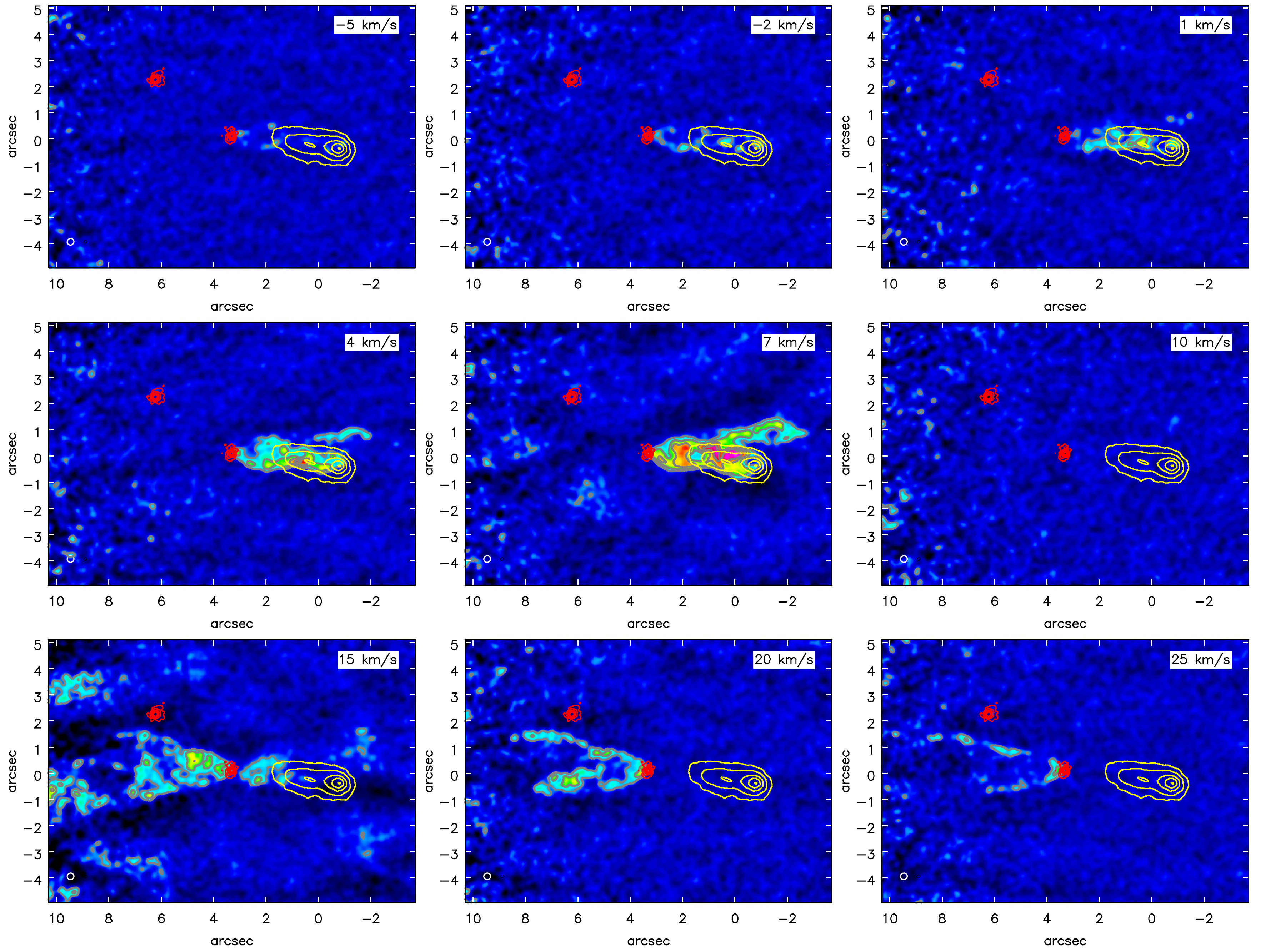}
    \caption{The $^{12}$CO (3-2) channel map. The central velocity is 10.3 km s$^{-1}$. The grey contours show the flux density levels of $4\sigma\times$(1, 2, 3, 4 5) with $1\sigma=2.74\times 10^{-2}$ Jy beam$^{-1}$. The 0.89 mm continuum is superposed on the map in red contours with levels of 5$\sigma\times$(1, 2, 4, 6, 10, 20) and 1$\sigma$=4.0$\times$10$^{-4}$ Jy beam$^{-1}$. The GNIRS K-band emission is overlaid onto the channel map with the yellow contours, in which the contour levels are $20\sigma$, $50\sigma$, $100\sigma$, $150\sigma$, and $200\sigma$ with 1$\sigma$=8.4 in arbitrary units. The beam size is presented in bottom left in each panel. }
    \label{fig:CO-chmap}
\end{figure*}

Figure~\ref{fig:CO-chmap} presents the channel maps for $^{12}$CO 3-2 emission from HOPS 373. The blueshifted emission from the central velocity of 10.3 km s$^{-1}$ is distributed along with the elongated feature of the GNIRS Ks emission. On the other hand, the redshifted emission shows a limb-brightened structure that traces the outflow cavity wall.

\end{CJK*}{UTF8}{gbsn}

\end{document}